\documentclass[useAMS,usenatbib]{mn2e}
\usepackage{float}				
\usepackage{graphicx}		
\usepackage{amsmath}			
\usepackage{amssymb}

\newcommand{\reffig}[1]{Fig.~\ref{#1}}

\newcommand{\refsec}[1]{Sect.~\ref{#1}}
\newcommand{\refappendix}[1]{Appendix~\ref{#1}}
\newcommand{\refpar}[1]{\S\ref{#1}}
\newcommand{\refeq}[1]{Eq.(\ref{#1})}

\title{Do \textit{Fermi}-LAT observations imply very large Lorentz factors in GRB outflows ?}
\author[R.~Hasco\"et, F.~Daigne, R.~Mochkovitch and V.~Vennin]{R.~Hasco\"et\thanks{E-mail: hascoet@iap.fr}
, F.~Daigne\thanks{Institut Universitaire de France}
, R.~Mochkovitch
 and V.Vennin\\
UPMC-CNRS, UMR7095, Institut d'Astrophysique de Paris, F-75014, Paris, France}
\begin{document}

\date{Accepted **.**.**. Received **.**.**; in original form **.**.**}

\pagerange{\pageref{firstpage}--\pageref{lastpage}} \pubyear{2011}

\maketitle

\label{firstpage}

\begin{abstract}
 Recent detections of GeV photons in a few gamma-ray bursts (hereafter GRBs) by \textit{Fermi}-LAT imply huge bulk Lorentz factors to avoid a large $\gamma\gamma$ optical depth at high energy. Estimates can be as high as $\Gamma\simeq 10^3$ in the most extreme cases. This puts severe constraints on models of the central engine and the jet acceleration in GRBs. These estimates are however obtained from a simplified single zone model. We present here a more realistic calculation of the $\gamma\gamma$ opacity which takes into account the time, space and direction dependent photon field existing in an outflow with several relativistically moving emitting zones. The formalism is very general and can be applied to many models of the prompt GRB emission. We present results obtained for a numerical implementation in the framework of the internal shock model. We show that (i) the minimum Lorentz factor $\Gamma_\mathrm{min}$ in bright \textit{Fermi}-LAT GRBs is reduced by a factor $\simeq 2-3$ compared to previous estimates if the GeV and MeV emission are produced in the same region, and by an additional factor $\simeq 2-8$ if the GeV emission is produced at larger radii. We provide an improved approximate formula for $\Gamma_\mathrm{min}$ which is in good agreement with our numerical results and which can be directly applied to \textit{Fermi}-LAT GRB data; (ii) a delayed onset of the GeV emission can be due to the time evolution of the opacity in a GRB outflow. As an illustration of these first two results, we present a synthetic GRB that reproduces most features of GRB 080916C with a mean Lorentz factor of $\simeq 340$, an optically thin regime for $\gamma\gamma$ opacity at 3 GeV in time bin 'b', a variability timescale of $\simeq 0.5$ s in the MeV lightcurve and a delayed onset of $\simeq 5$ s of the GeV emission; (iii) the $\gamma\gamma$ opacity can smooth the short timescale variability in the GeV lightcurve. This last result implies that the observed variability at high energy is not necessarily a good test to distinguish between an internal and an external origin for the GeV emission in GRBs.
\end{abstract}

\begin{keywords}
gamma-ray burst: general; gamma-ray burst: individual: GRB 080916C; radiative transfer; radiation mechanisms: non-thermal.
\end{keywords}

\vspace*{-14ex}

\section{Introduction}
The combination of  a short timescale variability with a huge radiated gamma-ray energies in gamma-ray bursts 
leads to the well-known compactness problem: $\gamma$-ray photons should not escape due to $\gamma\gamma$ annihilation. This contradiction is solved by assuming that the emitting region has an ultra-relativistic motion \citep{rees:1966}. This leads to an important constraint on the minimum Lorentz factor of the outflow in cases where the GRB spectrum does not show any evidence for a high-energy cutoff \citep[see e.g.][]{baring:1997,lithwick:2001}, or to a direct measurement of the Lorentz factor if such a cutoff is identified, as possibly in GRB 090926A \citep{ackermann:2011}.\\

Since the launch of \textit{Fermi} in June 2008, the LAT instrument has detected high energy photons above 10 GeV in a few GRBs. The observed $\gamma$-ray spectrum often remains consistent with a Band function covering the GBM and LAT spectral ranges without any evidence of a high energy cutoff which could be identified as a signature of $\gamma\gamma$ annihilation 
(with possibly the exception of GRB 090926A, \citealt{ackermann:2011}). This extension by \textit{Fermi} of the observed spectral range upper bound from a few MeV 
to a few tens of GeV implies constraints on $\Gamma_{\mathrm{min}}$ which are much more severe than the ones obtained previously \citep[see e.g.][]{racusin:2011}. In a few cases $\Gamma_{\mathrm{min}}$ has been estimated to be of the order of 1000. For instance, in the four brightest \textit{Fermi}-LAT GRBs, the estimated minimum values are $\Gamma_{\mathrm{min}}\simeq 887$ in GRB 080916C \citep{abdo:2009a},  $\simeq 10^3$ in GRB 090902B \citep{abdo:2009b}, $\simeq 1200$ in GRB 090510  \citep{ackermann:2010}), $\simeq 720$ in GRB 090926A \citep{ackermann:2011}. These extreme values put severe constraints on the physics of the central engine which should be able to strongly limit the baryon load in the outflow. However these minimum Lorentz factors were obtained from a simplified ``single zone'' model where the spatial and temporal dependencies of the radiation field are averaged out. This calls for more realistic calculations of the $\gamma\gamma$ opacity in GRB outflows. We present here a detailed approach taking into account the exact time, space and direction dependent radiation field in outflows with several relativistically moving emitting regions, such as internal shocks. \\

The paper is organized as follows. In \refsec{section_calculation_tgg}, we derive the exact formula for the $\gamma\gamma$ opacity in complex and variable GRB outflows, we describe its dependency on several physical parameters such as the emission radius or the Lorentz factor and we compare the exact calculation to the approximate formula which is broadly used in the literature. Finally we validate our approach by a comparison with the semi-analytical study of \citet{granot:2008}. \refsec{section_internal_shocks} is devoted to the investigation of several consequences of $\gamma\gamma$ annihilation in GRBs in the framework of the internal shock model.
We discuss successively the constraints on the minimum Lorentz factor, 
with a focus on 
GRB 080916C, the spectral shape of the high-energy cutoff, the delayed onset of the GeV emission and the smoothing of the short timescale variability in GeV lightcurves. We compare also in this section the  $\gamma\gamma$ opacity with other expected sources of opacities. \refsec{sec_2zones_model}
 presents a discussion of models where the GeV emission is produced at a larger radius than the MeV main component. Finally, \refsec{sec_conclusions} is the conclusion.


\section{Calculation of the $\gamma\gamma$ optical depth}
\label{section_calculation_tgg}

\subsection{General $\gamma\gamma$ opacity formula.} 

The $\gamma\gamma$ opacity $\tau_{\gamma \gamma}$ seen by a photon of high energy $E_\mathrm{HE}$ propagating in a relativistic outflow is given by: 
\begin{equation}
 \tau_{\gamma \gamma} (E_\mathrm{HE}) = 
 \! \! \int 
  \! \! d\ell \int 
 \! \! d \Omega \int_{E_\mathrm{c}(E_\mathrm{HE}, \psi)}^{\infty} 
\! \! \! \! \! \! \! \! \! \! \! \! \! \! \! \! \! \! \! \! \! \! \!  
 dE \ n_{\Omega}(E) \sigma_{\gamma \gamma}(E_\mathrm{HE},E,\psi) \left(1-\cos{\psi}\right)\, .
\label{eqn_taugamma_general}
\end{equation}
All the physical quantities are measured 
in the fixed frame associated to the central source (``laboratory frame", hereafter the source frame):
$E$ is the energy of the interacting field photon, 
$\psi$ represents the interaction angle between the HE photon and the interacting photon. The energy threshold $E_\mathrm{c}$ of the field photon above which $\gamma\gamma$ annihilation can occur is
\begin{equation}
E_\mathrm{c}\left(E_\mathrm{HE},\psi\right) = \frac{2(m_\mathrm{e} c^2)^2}{ E_\mathrm{HE}\left(1-\cos{\psi}\right)}\, .
\label{eqn_threshold_energy}
\end{equation}
and the $\gamma\gamma$ cross section $\sigma_{\gamma \gamma}$ is  given by
\begin{equation}
\sigma_{\gamma \gamma}(y) = \sigma_\mathrm{T}\,  g(y) \, ,
\label{eqn_cross_section}
\end{equation}
where
\begin{equation}
g(y) = \frac{3}{16} (1-y^2) \left[ (3-y^4) \ln{\frac{1+y}{1-y}} - 2 y (2-y^2) \right]\, ,
\label{eqn_g_function}
\end{equation}
and $y$ is defined by 
\begin{equation}
     y = \sqrt{ 1 - \frac{2(m_\mathrm{e} c^2)^2} {E_\mathrm{HE}E (1-\cos{\psi})}}=\sqrt{1-\frac{E_\mathrm{c}}{E}} 
\end{equation}
for $E \ge E_\mathrm{c}$ and $y=0$ for $E \le E_\mathrm{c}$. Finally $n_{\Omega}(E)$ is the photon field distribution $[\mathrm{ph} \cdot \mathrm{cm}^{-3} \cdot \mathrm{keV}^{-1} \cdot \mathrm{sr}^{-1}]$ at a given location and time.\\

\refeq{eqn_taugamma_general} is made of a triple integral: the $d\ell$-integration is done over the path of the GeV photon from its emission location to the observer, the $d \Omega$-integration is done over the solid angle distribution of the interacting photon field surrounding the GeV photon whereas the $d E$-integration is done over its energy distribution. \refeq{eqn_taugamma_general} is general and can be applied to any photon emitted at a given location and time with a given direction of propagation in the GRB outflow.

\subsection{$\gamma\gamma$ opacity created by a ``spherical flash''.} 
\label{sec:flash}
\subsubsection{Calculation of the $\gamma\gamma$ opacity}
\begin{figure}
\includegraphics[width=0.45\textwidth]{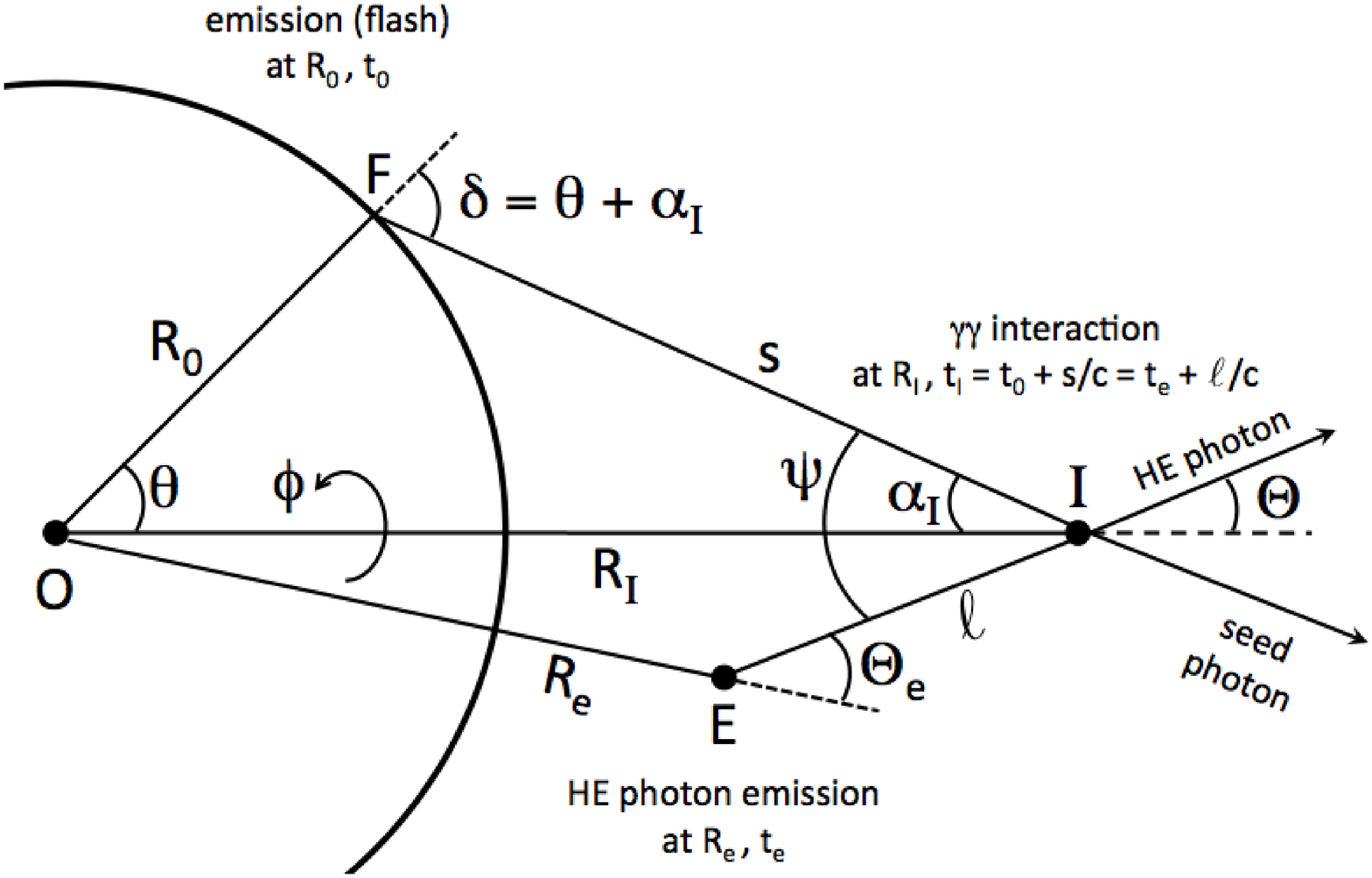} 
\caption{\textbf{Geometry of the $\gamma\gamma$ interaction of a high energy photon with a spherical flash.} All quantities are defined in the source frame. 
A high energy photon is emitted at radius $R_\mathrm{e}$ and time $t_\mathrm{e}$ (point E) in a direction  $\Theta_\mathrm{e}$ relative to the radial direction, and interacts with a  flash emitted at radius $R_0$ and time $t_0$. 
The figure describes the interaction geometry at a given time $t_\mathrm{I}>t_0$: F is the emission point of the low-energy seed photon and I the interaction point. The angles $\theta$, $\phi$ and $\delta$ are respectively the emission colatitude, longitude and Doppler angle of the seed photon. 
Note that the path EI (high energy photon) is contained in the plane of the figure, which is not the case for the path FI (seed photon).
The radius of the interaction point I is $R_I$ and, from I, the arrival direction of the seed photon and of the HE energy photon make respectively an angle $\alpha_\mathrm{I}$ and $\Theta$ with the radial direction. 
The distances from F to I and from E to I are respectively $s$ and $\ell$, leading to $t_\mathrm{I}=t_\mathrm{e}+\ell/c=t_0+s/c$. Finally $\psi$ is the angle between the directions of the two interacting photons (hereafter the interaction angle). Note that, at a given time $t_\mathrm{I}$, all the seed photons come from the same emission colatitude $\theta$, leading to the same values of $\alpha_\mathrm{I}$, $s$ and $\delta=\theta+\alpha_\mathrm{I}$. However, the angle $\psi$ also depends on $\phi$.}
\label{fig_flash_photon}
\end{figure}
The kernel of our study is the calculation of the $\gamma\gamma$ opacity created by a spherical flash. Using this kernel, it is possible to model the case of a propagating radiating spherical front (representing for example a shock wave) by the succession of many spherical flashes. One critical step is the exact calculation of the photon density $n_{\Omega}$  (see \refappendix{sec_calc}). The emissivity $[\mathrm{erg}\cdot \mathrm{cm}^{-3} \cdot \mathrm{s}^{-1} \cdot \mathrm{Hz}^{-1} \cdot \mathrm{sr}^{-1}]$ in the comoving frame\footnote{Physical quantities with a prime are measured in the  comoving frame.} of a spherical flash occurring at a radius $R_0$ and time $t_0$ (source frame) is given by :
\begin{equation}
j'_{\nu'} = \frac{1}{4\pi} \frac{1}{4\pi R_0^2} \frac{\mathcal{E}_\mathrm{rad}}{\Gamma_0}   \frac{1}{\nu'_{p,0}} \mathcal{B}\left( \frac{\nu'}{\nu'_{p,0}} \right) \delta(r-R_0) \delta(t-t_0)\, ,
\label{eqn_emissivity_comov}
\end{equation}
where $\mathcal{E}_\mathrm{rad}$ is the total radiated energy, $\Gamma_{0}=1/\sqrt{1-\beta_0^2}$ is the Lorentz factor of the material moving with velocity $\beta_0 c$ and emitting the spherical flash, 
$\nu'$ is the photon frequency in  the comoving frame, $\nu'_{p,0}$ is the comoving peak frequency of the spectrum, and $\mathcal{B}$ is the normalized spectral shape  ($\int_0^\infty\mathcal{B}(x)dx=1$). \\

The geometry associated to a spherical flash is described in \reffig{fig_flash_photon} and the calculation of the created photon field $n_{\Omega}(E)$  is given in \refappendix{sec_calc}. 
The resulting general expression of the $\gamma\gamma$ opacity created by a spherical flash is
\begin{eqnarray}
\tau_{\gamma \gamma} (E_\mathrm{HE})\!\!\! 
& = &  \int d\ell\, \frac{\mathcal{E}_\mathrm{rad}}{4\pi R_0^2 \Gamma_0}\, \frac{R_0}{s R_\mathrm{I}}\, \frac{\mathcal{D}^2}{E'_\mathrm{p,0}} \int_0^{2\pi}\frac{d\phi}{4\pi}\left(1-\cos{\psi}\right)
\nonumber\\ 
 & & \times 
\int_{E_\mathrm{c}\left(E_\mathrm{HE},\psi\right)}^{+\infty}\frac{dE}{E}\sigma_{\gamma\gamma}\left(E_\mathrm{HE},E,\psi\right)\mathcal{B}\left(\frac{E}{\mathcal{D}E'_\mathrm{p,0}}\right)\nonumber\\
& = & \tau_0 \int d\ell \frac{R_0}{s R_\mathrm{I}}\mathcal{D}^2\int_0^{2\pi}\frac{d\phi}{2\pi}\left(1-\cos{\psi}\right)
\nonumber\\ 
& & \times 
\int_0^{1}dy \frac{y g(y)}{1-y^2}\,\mathcal{B}\left(\frac{E_\mathrm{c}\left(E_\mathrm{HE},\psi\right)}{\mathcal{D}E'_\mathrm{p,0}(1-y^2)}\right)\, ,
\label{eqn_taugamma_flash}
\end{eqnarray} 
where $E'_\mathrm{p,0}=h \nu'_\mathrm{p,0}$ is the comoving peak energy, $\psi$ is the interaction angle given by
\begin{equation}
\cos{\psi} = \cos{\Theta} \cos{\alpha_I} + \sin{\Theta} \sin{\alpha_I} \cos{\phi}\, ,
\label{eqn_interaction_angle}
\end{equation}
$\mathcal{D}$ is the Doppler factor
\begin{equation}
\mathcal{D}=\frac{1}{\Gamma_0\left(1-\beta_0\cos{\delta}\right)}\, ,
\end{equation}
and the dimensionless constant $\tau_0$ is defined by
\begin{equation}
\tau_0 = \frac{\sigma_\mathrm{T}\mathcal{E}_\mathrm{rad}}{4\pi R_0^2 \Gamma_0 E'_\mathrm{p,0}}\, .
\label{eq_tau0}
\end{equation}

If the high energy photon is emitted at radius $R_\mathrm{e}$ and time $t_\mathrm{e}$ in a direction $\Theta_\mathrm{e}$ relative to the radial direction (see \reffig{fig_flash_photon}), the quantities appearing in \refeq{eqn_taugamma_flash} are given as a function of the length $\ell$ traveled by the high energy photon from its emission point by
\begin{eqnarray}
R_\mathrm{I}^2 & = & R_\mathrm{e}^2 + \ell^2 + 2 R_\mathrm{e} l \cos{\Theta_\mathrm{e}}\, ,\\
\sin{\Theta} & = & \frac{R_\mathrm{e}}{R_\mathrm{I}}\sin{\Theta_\mathrm{e}}\, ,\\
s & = & c \left(t_\mathrm{I}-t_0\right) = c \left( t_\mathrm{e}+\frac{\ell}{c}-t_0\right)\, ,\\
\cos{\alpha_\mathrm{I}} & = & \frac{R_\mathrm{I}^2+s^2-R_0^2}{2 R_\mathrm{I} s}\, ,\\
\cos{\delta} & = & \frac{R_\mathrm{I}^2-R_0^2-s^2}{2 R_0 s }\, ,\\
\mathcal{D} & = & \frac{1}{\Gamma_0\left(1-\frac{v_0}{c}\cos{\delta}\right)}\, .
\end{eqnarray}
In \refeq{eqn_taugamma_flash}, the $d\ell$-integration is limited to the range of $\ell$ allowing an interaction, i.e. fulfilling the condition
\begin{equation}
\left| R_0-R_\mathrm{I}\right| \le s \le R_0+R_\mathrm{I}\, .
\label{eq_cond_s}
\end{equation}

\noindent\textbf{--~Case of a photon emitted on axis.}  
If the high energy photon is emitted on axis ($\Theta_\mathrm{e}=\Theta = 0$ and $\psi = \alpha_I$) the axial symmetry gives a simplified formula :
\begin{eqnarray}
\tau_{\gamma \gamma} (E_\mathrm{HE})
& = & \tau_0 \int d\ell \frac{R_0 \left(1-\cos{\psi}\right)}{\left(R_\mathrm{e}+\ell\right) \left(c\left(t_\mathrm{e}-t_0\right)+\ell\right) } \mathcal{D}^2 
\nonumber\\ 
& & \times 
\int_0^{1}dy \frac{y g(y)}{1-y^2}\,\mathcal{B}\left(\frac{E_\mathrm{c}\left(E_\mathrm{HE},\psi\right)}{\mathcal{D}E'_\mathrm{p,0}(1-y^2)}\right)\, ,
\label{eqn_taugamma_onaxis}
\end{eqnarray} 
with
\begin{eqnarray}
\cos{\psi} & = & \frac{\left(R_\mathrm{e}+\ell\right)^2+\left(c\left(t_\mathrm{e}-t_0\right)+\ell\right)^2-R_0^2}{2 \left(R_\mathrm{e}+\ell\right)  \left(c\left(t_\mathrm{e}-t_0\right)+\ell\right)}\, ,\\
\cos{\delta} & = & \frac{\left(R_\mathrm{e}+\ell\right)^2-R_0^2-\left(c\left(t_\mathrm{e}-t_0\right)+\ell\right)^2}{2 R_0 \left(c\left(t_\mathrm{e}-t_0\right)+\ell\right) }\, .
\end{eqnarray}
\vspace*{1ex}

\noindent \textbf{--~Case of a flash having a power-law spectrum.} 
If the spectrum of the flash emission can be approximated by a single power-law of photon index $\beta$, the spectral shape function has the following form $\mathcal{B}(x) = A_0 x^{1+\beta}$ for $x\ge 1$ with $A_0=-(2+\beta)$, and $\mathcal{B}(x)=0$ for $x< 1$. Then \refeq{eqn_taugamma_flash} can be further simplified:
\begin{eqnarray}
\tau_{\gamma \gamma} (E_\mathrm{HE})\!\!\! 
& = & \mathcal{I}(\beta) \tau_0 A_0 \left(\frac{2 \left(m_\mathrm{e}c^2\right)^2}{E_\mathrm{HE} E'_\mathrm{p,0}}\right)^{1+\beta}
\nonumber\\ 
& & \times 
\int d\ell \frac{R_0}{s R_\mathrm{I}}\mathcal{D}^{1-\beta}\int_0^{2\pi}\frac{d\phi}{2\pi}\left(1-\cos{\psi}\right)^{-\beta}\, ,
\label{eqn_taugamma_flash_pl}
\end{eqnarray} 
where $\mathcal{I}(\beta)$ is a finite number 
 for $\beta<0$ :
\begin{equation} 
\mathcal{I}(\beta) = \int_{0}^{1} \frac{y}{(1-y^2)^{2+\beta}} g(y) \ dy\, .
\label{eqn_i_beta}
\end{equation}
For $\beta=-2$, $-2.2$, $-2.3$, $-2.5$ and $-3$, this integral equals respectively  $\mathcal{I}(\beta)\simeq 0.091$, $0.078$, $0.072$, $0.063$ and $0.046$.\\

\noindent \textbf{--~Case of a photon emitted on axis interacting with a flash having a power-law spectrum.}  
Finally if the two simplifications are combined, the $\tau_{\gamma \gamma}$ formula becomes:
\begin{eqnarray}
\tau_{\gamma \gamma} (E_\mathrm{HE})\!\!\! 
& = & \mathcal{I}(\beta) \tau_0 A_0 \left(\frac{2 \left(m_\mathrm{e}c^2\right)^2}{E_\mathrm{HE} E'_\mathrm{p,0}}\right)^{1+\beta}
\nonumber\\ 
& & \times 
\int d\ell \frac{R_0 \left(1-\cos{\psi}\right)^{-\beta}}{\left(R_\mathrm{e}+\ell\right) \left(c\left(t_\mathrm{e}-t_0\right)+\ell\right) }\mathcal{D}^{1-\beta}\, .
\label{eqn_taugamma_flash_pl_onaxis}
\end{eqnarray} 
\subsubsection{Approximate formula}
\label{sec_approx_formula}
Most estimates of the minimum Lorentz factor in GRB outflows are based on approximate formulae corresponding to the case of a high-energy photon emitted on axis and interacting 
with an isotropic radiation field
over a length $\ell' \sim R/\Gamma_0$ in the comoving frame of the emitting material assumed to move with a constant Lorentz factor $\Gamma_0$ 
\citep[see e.g.][]{lithwick:2001,gupta:2008,abdo:2009a,zhang:2009}.
In many GRB models, such as internal shocks, a more realistic approximation corresponds to the situation where the seed photons are produced by a flash emitted at radius $R_0$ by a single thin shell moving with constant velocity $\beta_0$, and where the high-energy photon is emitted on-axis at $R_\mathrm{e}$ and interacts with the resulting anisotropic radiation field.
  Then, the law of motion of the thin shell is $R_\mathrm{e}-R_0 = \beta_0 c \left(t_\mathrm{e}-t_0\right)$ and the 
 high energy photon can interact with the seed photons from the flash only if
\begin{equation}
R_0 < R_\mathrm{e} < \frac{2 R_0}{1-\beta_0}  \simeq 4\Gamma_0^2 R_0\, .
\end{equation}
In this case, an approximate formula for the $\gamma\gamma$ opacity given by \refeq{eqn_taugamma_flash_pl_onaxis} can be obtained (see \refappendix{sec_calc_approx}) :
\begin{equation}
\tau_{\gamma\gamma}\simeq  \left(\frac{2\left(m_\mathrm{e}c^2\right)^2}{E_\mathrm{HE}E'_\mathrm{p,0}}\right)^{1+\beta} \frac{2^{\beta}\mathcal{I}(\beta) \tau_0 A_0\Gamma_0^{\beta+1}}{\left[\left(1+\frac{1}{2}\frac{R_\mathrm{e}-R_0}{R_0}\right)\left(1+\frac{R_\mathrm{e}-R_0}{R_0}\right)\right]^{1-\beta}}\, .
\label{equation_tgg_flash_approximate}
\end{equation}
To compare with observations, the radius $R_0$ is usually estimated by $\Gamma_0^2 c \Delta t_\mathrm{var}$, where $\Delta t_\mathrm{var}$ is the observed variability timescale, and the isotropic equivalent radiated energy $\mathcal{E}_\mathrm{rad}$ is deduced from the observed photon fluence $\mathcal{F}(E) = \mathcal{F}_0 \left(E/E_\mathrm{p,0}\right)^{\beta}$ 
$[\mathrm{ph} \cdot \mathrm{cm}^{-2} \cdot \mathrm{keV}^{-1}]$, where $E_\mathrm{p,0}=\Gamma_0 E'_\mathrm{p,0}$ is the observed peak energy. This leads to
\begin{equation}
\mathcal{E}_\mathrm{rad}=4\pi D^2 \, \frac{E_\mathrm{p,0}^2 \mathcal{F}_0}{A_0} ,
\label{eq:erad_fluence}
\end{equation}
where $D$ is the distance of the source\footnote{We do not include here the corrections due to
cosmological redshift.}
 Finally the $\gamma\gamma$ opacity reads
\begin{equation}
\tau_{\gamma\gamma}\simeq  K\left(R_\mathrm{e}\right) 
\sigma_\mathrm{T} \left(\frac{D}{c \Delta t_\mathrm{var}}\right)^2 E_\mathrm{p,0}\, \mathcal{F}_0\, \Gamma_0^{-2\left(1-\beta\right)}
\left(\frac{\left(m_\mathrm{e}c^2\right)^2}{E_\mathrm{HE}E_\mathrm{p,0}}\right)^{1+\beta} 
\, ,
\label{eq_tau_approx}
\end{equation} 
where
\begin{equation}
K\left(R_\mathrm{e}\right) = \frac{2^{1+2\beta} \mathcal{I}(\beta)}{\left[\left(1+\frac{1}{2}\frac{R_\mathrm{e}-R_0}{R_0}\right)\left(1+\frac{R_\mathrm{e}-R_0}{R_0}\right)\right]^{1-\beta}}
\end{equation}
is a coefficient that depends on the location where the high-energy photon is emitted. 
Most previous
 studies considered usually the limit $R_\mathrm{e}\to R_0$ (all photons are radiated in the same region). This leads here to a coefficient $K_0 = K\left(R_\mathrm{e}\to R_0\right)$ given by
\begin{equation}
K_0 \simeq 2^{1+2\beta} \mathcal{I}(\beta)\, .
\label{eq_correct_norm}
\end{equation}
Compared to the simple model assuming an isotropic radiation field in the comoving frame of the outflow, the approximate formula derived here for an anisotropic field produced by an impulsive flash shows the same scaling with the parameters of the problem, but a different normalization $K_0$.
\citet{svensson:1987} has derived the exact coefficient for the single zone isotropic model and gives the following accurate approximation (error less than 0.3\% for $-7<\beta<-1$) 
\begin{equation}
K_0^\mathrm{S87}\simeq
\frac{7}{6 (-\beta)^{5/3} (1-\beta)}\, .
\end{equation}
The coefficient 
\begin{equation}
K_0^\mathrm{A09} = \frac{4}{1-\beta} \mathcal{I}(\beta) = \frac{2^{1-2\beta}}{1-\beta} K_0
\end{equation}
used by \citet{abdo:2009a}
differs from $K_0^\mathrm{S87}$ by less than 0.5\%. The coefficient 
\begin{equation}
K_0^\mathrm{LS01}=
-\frac{11}{180}\frac{1}{1+\beta} 
\end{equation}
used by \citet{lithwick:2001} is less accurate, with a difference by a factor 1.5 to 2 in the range $-3 < \beta < -2$.\\

For a typical value $\beta=-2.3$, 
the differences between the two assumptions (isotropic radiation field vs anisotropic field created by a flash) is noticeable. 
We
have $K^\mathrm{A09}/K_0\simeq 14.7$ and $K^\mathrm{LS01}/K_0\simeq 7.3$, already leading to some 
differences in the estimate of the minimum Lorentz factor in GRB outflows. 
Note that in all these single zone models, the opacity is built over a length $\ell\sim R$ in the source frame (see \refappendix{sec_calc_approx} and see also \refpar{sec:grb090816C}), so that the origin of the difference is mainly the geometry of the radiation field (isotropy vs anisotropy in the comoving frame).
However, as described in \refpar{subsection_lfmin}, the estimate of $\Gamma_\mathrm{min}$ is even more affected by 
dynamical effects, that are not included in these single zone models.

\begin{figure*}
\begin{center}
\begin{tabular}{cc}
\includegraphics[scale=0.4]{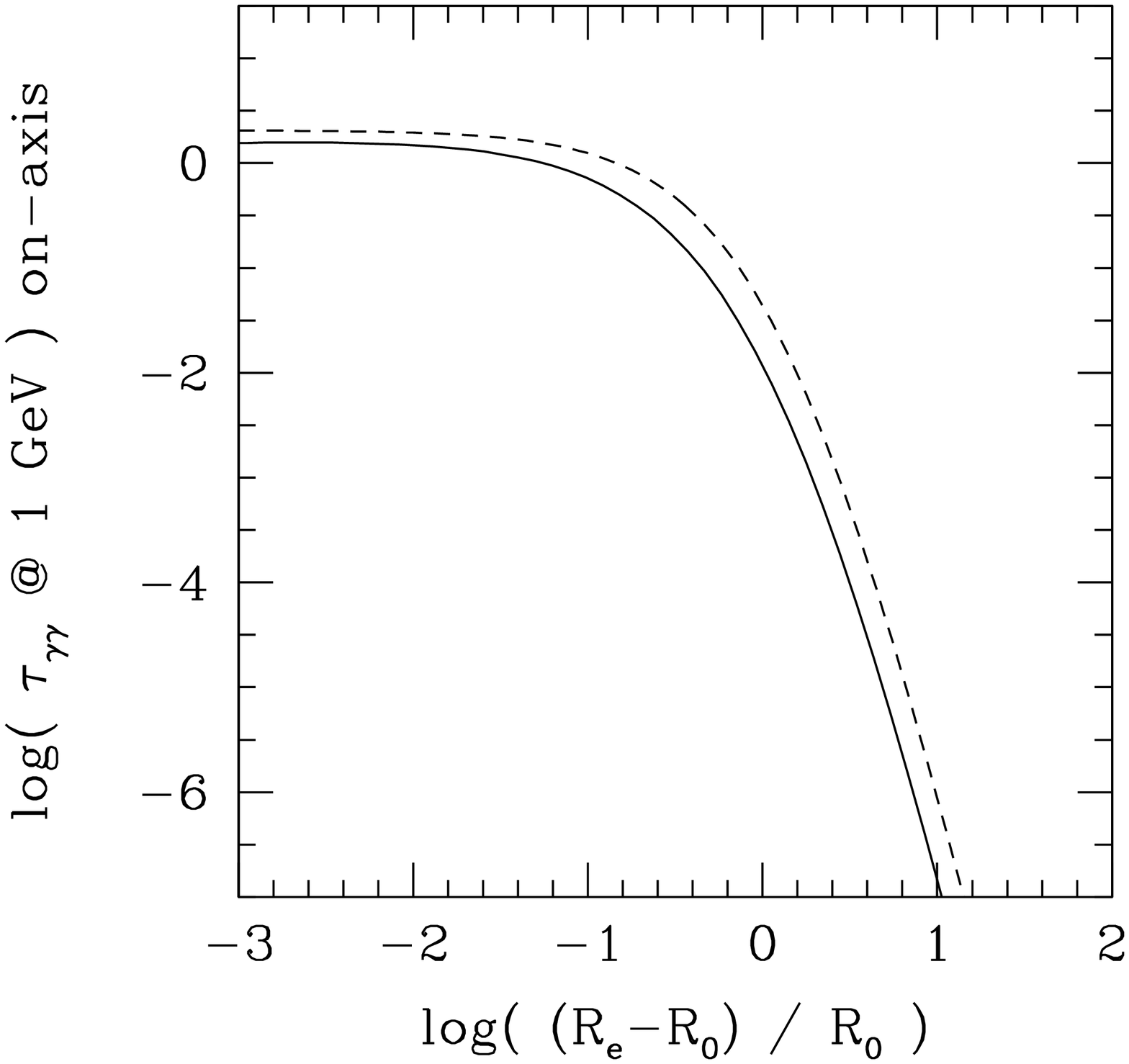} & \includegraphics[scale=0.4]{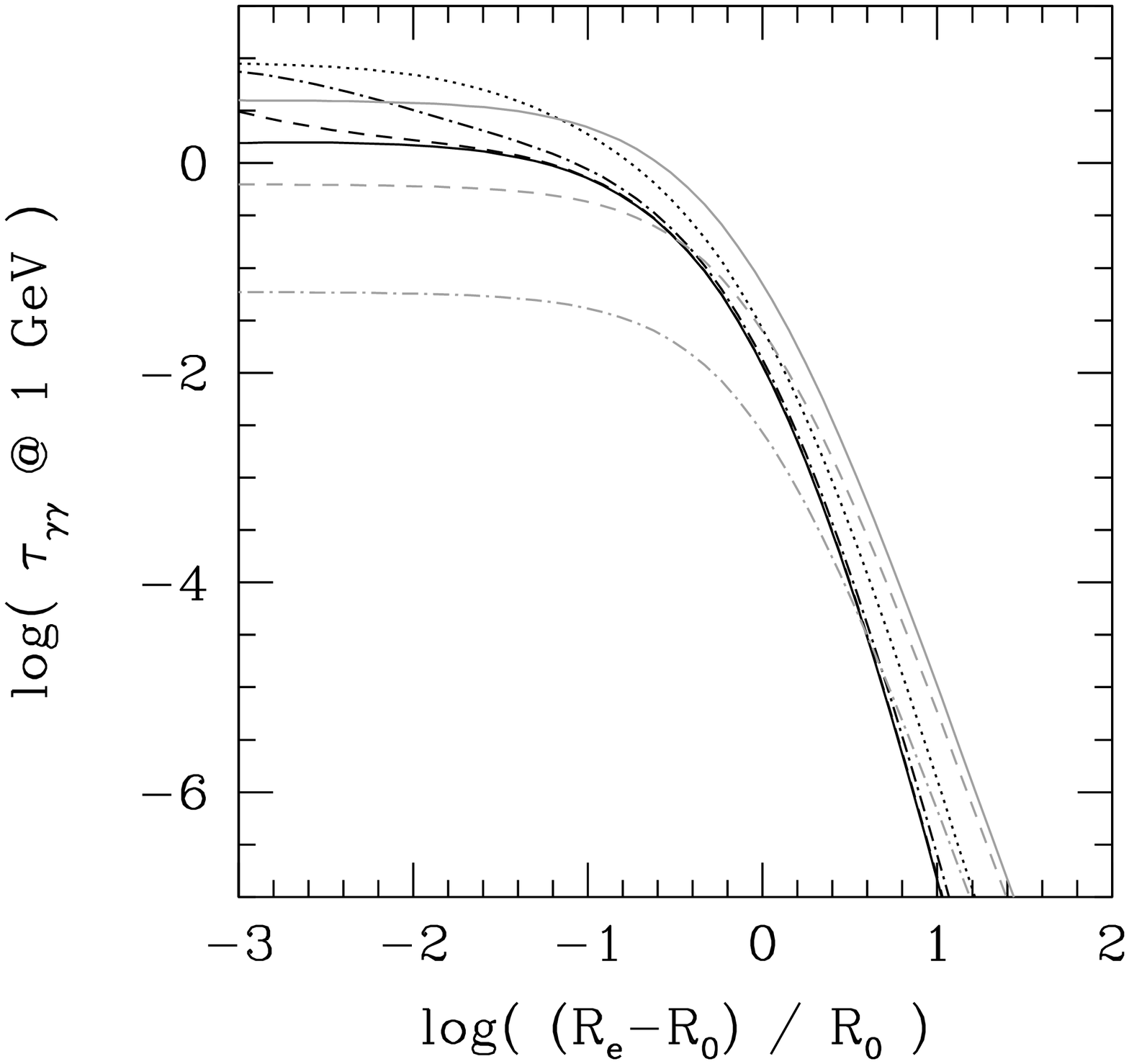} \\ 
\includegraphics[scale=0.4]{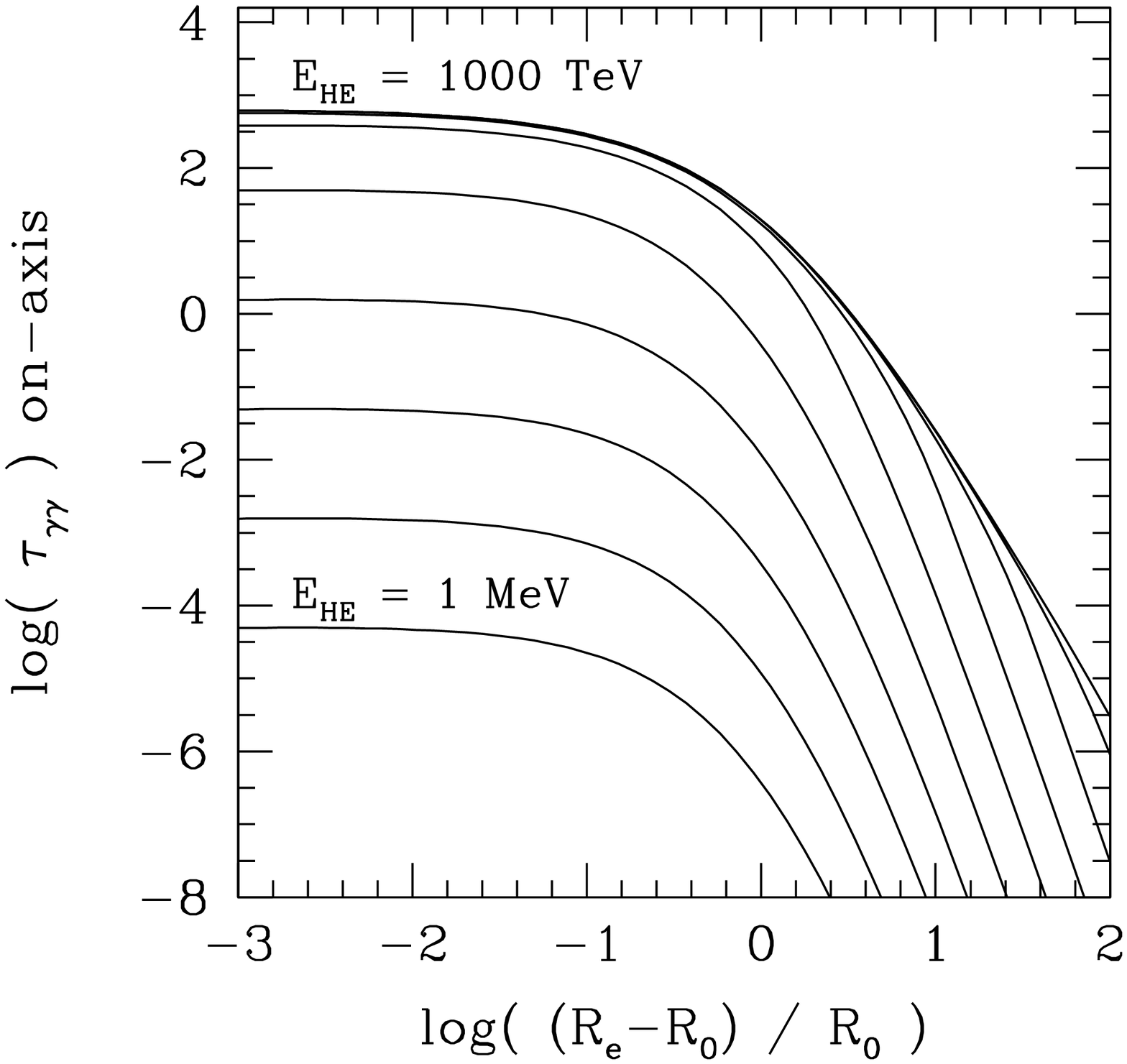}  & \includegraphics[scale=0.4]{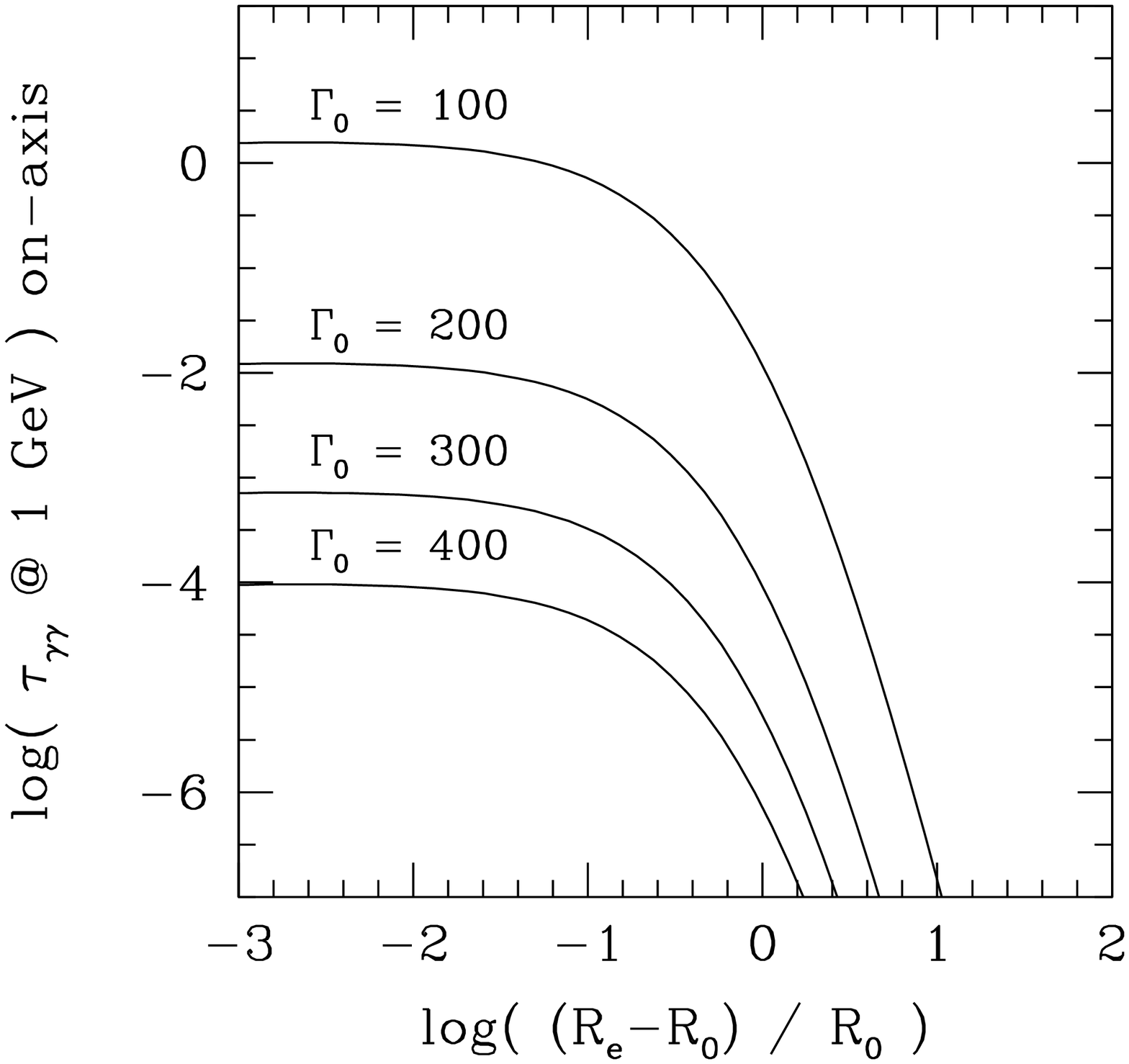} 
\end{tabular}
\end{center}

\caption{\textbf{Opacity created by a spherical flash.} In each panel, the $\gamma\gamma$ opacity is plotted as a function of the high energy photon emission radius.  In the first three panels, the Lorentz factor is $\Gamma_0 = 100$. \textit{Upper  left panel:} opacity seen by a 1 GeV photon emitted on axis. The numerical calculation  is plotted in solid line and the approximate formula given by \refeq{equation_tgg_flash_approximate} in dashed line. \textit{Upper right panel:} opacity seen by a 1 GeV photon 
for different emission colatitudes $\Theta_\mathrm{e}\Gamma_0 = 0$ (black, solid line); $0.03$ (black, dashed line); $0.1$ (black, dashed dotted line); $0.3$ (black, dotted line); $1$ (gray, solid line); $3$ (gray, dashed line); $10$ (gray, dashed dotted line). \textit{Bottom left:} opacity seen by a high energy photon emitted on axis for
$E_\mathrm{HE} = 1$; $10$; $100$ MeV; $1$; $10$; $100$ GeV ; $1$; $10$; $100$ ; $1000$ TeV. 
\textit{Bottom right:} opacity seen by 1 GeV photons emitted on axis for different Lorentz factors $\Gamma_0 = 100$; $200$; $300$; $400$.}
\label{fig_case_flash}
\end{figure*}

\subsubsection{A case study}
\label{subsection_flash_case_study}

To better understand the different effects at stake when studying the $\gamma\gamma$ opacity in relativistic outflows, it is worth to  focus first on the case of a single spherical flash for which we have obtained an approximate estimate in \refpar{sec_approx_formula}. Let us consider a spherical shell expanding at a constant Lorentz factor $\Gamma_0$ 
that emits a photon flash at a radius $R_0 = (\Gamma_0/100)^2\, 10^{14}$ cm. The total radiated energy (source frame) is 
$\mathcal{E}_\mathrm{rad} = 2\cdot 10^{51}$ erg 
and the spectrum is a broken power-law with a comoving peak energy $E'_\mathrm{p,0} = \left(1\, \mathrm{MeV}\right) / \Gamma_0$, a low energy photon index $\alpha = -1$, and a high energy photon index $\beta = -2.5$. 
With these assumptions, the observed pulse has a fixed duration $\Delta t_\mathrm{obs} = R_0/(2 \Gamma_0^2 c) \simeq 0.17\,\mathrm{s}$, a fixed bolometric isotropic equivalent energy $\mathcal{E}_\mathrm{rad}$ and a fixed observed peak energy $\simeq \Gamma_0 E'_\mathrm{p,0}\simeq 1$ MeV. 
In addition, we assume that the same shell, expanding at a constant Lorentz factor $\Gamma_0$, is emitting a high energy photon at a radius $R_\mathrm{e}\ge R_0$ and a colatitude $\Theta_\mathrm{e}$. \refeq{eqn_taugamma_flash} allows  to compute the optical depth for $\gamma\gamma$ annihilation $\tau_{\gamma \gamma}$ seen by 
this high energy photon, as a function of its energy $E_\mathrm{HE}$. 
This photon is detected by a distant observer at time $t_\mathrm{obs}^\mathrm{HE}$ given by
\begin{equation}
t_\mathrm{obs}^\mathrm{e} \simeq  t_\mathrm{obs}^\mathrm{flash} + \frac{R_\mathrm{e}-R_{0}}{2 \Gamma_0^{2} c} + \left(1-\cos{\Theta_\mathrm{e}}\right) \frac{R_\mathrm{e}}{c}\, ,
\label{eqn_tobs_gev}
\end{equation}
where $t_\mathrm{obs}^\mathrm{flash}$ is the time at which the first photon from the flash is observed.\\
The four panels of \reffig{fig_case_flash} give an overview of the obtained results showing the different dependencies of $\tau_{\gamma \gamma}$.\\

\noindent \textbf{Dependence on the emission radius.}
The top left panel of \reffig{fig_case_flash} gives the evolution of $\tau_{\gamma \gamma}$ as a function of $R_\mathrm{e}$ 
for a photon emitted on axis ($\Theta_\mathrm{e}=0$). The opacity $\tau_{\gamma \gamma}$ remains constant for 
\begin{equation}
\frac{R_\mathrm{e}}{R_{0}}-1= \frac{t_\mathrm{obs}^\mathrm{HE} -  t_\mathrm{obs}^\mathrm{flash}}{\Delta t_\mathrm{obs}}\la 0.2
\end{equation}
and then decreases as $\tau_{\gamma\gamma}\propto R_\mathrm{e}^{-2(1-\beta)}$, in excellent agreement with the approximate formula given by \refeq{eq_tau_approx}. The top right panel of \reffig{fig_case_flash} shows the same curves $\tau_{\gamma\gamma}(R_\mathrm{e})$, but for different emission colatitudes $\Theta_\mathrm{e}$.
The evolution is more complicated but remains dominated by a transition from a plateau when $R_\mathrm{e}$ is still close to $R_0$, to a steep decay at large radii. As explained in \refappendix{sec_calc_approx}, the plateau is due to a favorable configuration when $R_\mathrm{e}\to R_0$: both the Doppler angle $\delta$ and the interaction angle $\psi$ are close to $1/\Gamma_0$ at the beginning of the propagation of the high-energy photon; at larger emission radii $R_\mathrm{e}$, the Doppler angle increases, leading to a reduction of the specific intensity, and the interaction angle decreases, leading to a higher energy threshold for the seed photons and a lower effective cross section $\left(1-\cos{\psi}\right)\sigma_{\gamma\gamma}$.\\

\noindent \textbf{Dependence on the emission latitude.}
As seen in the top right panel of \reffig{fig_case_flash}, the value of the opacity in the plateau discussed in the previous paragraph depends on the emission colatitude $\Theta_\mathrm{e}$ of the high energy photon. When $\Theta_\mathrm{e}$ is increasing, the dominant effect is initially an increase of the interaction angle $\Psi$ at the beginning of the propagation, which leads to an increase of the opacity. However, when the colatitude $\Theta_\mathrm{e}$ increases even more, this effect is reduced by an increasing Doppler angle : the opacity reaches a maximum for $\Theta_\mathrm{e}\simeq 0.3/\Gamma_0$ and then decreases. At very high colatitudes, the opacity is strongly reduced compared to the on-axis photon: e.g. by  a factor $\simeq 30$ for $\Theta_\mathrm{e}\simeq 10/\Gamma_0$.\\
 
\noindent \textbf{Dependence on the photon energy.}
The bottom left panel of \reffig{fig_case_flash} gives the evolution of $\tau_{\gamma \gamma}$ as a function of $R_\mathrm{e}$ for different photon energies $E_\mathrm{HE}$. The curves are similar to the on-axis case discussed above. However the value of $\tau_{\gamma\gamma}$ in the initial plateau  evolve with $E_\mathrm{HE}$. For $\Psi\simeq 1/\Gamma_0$ (low emission radius $R_\mathrm{e}$), the threshold for $\gamma\gamma$ annihilation is
\begin{equation}
E_\mathrm{c}\simeq\frac{\left(2\Gamma_0 m_\mathrm{e}c^{2}\right)^2}{E_\mathrm{HE}}\, .
\end{equation}
Then, for $E_\mathrm{HE}<  \left(2\Gamma_0 m_\mathrm{e} c^2\right)^2/E_\mathrm{p,0} \simeq 10\, \mathrm{GeV}$, the threshold energy $E_\mathrm{c}$ remains in the high energy branch of the 
spectrum and the dependency is simply $\tau_{\gamma \gamma} \propto E_\mathrm{HE}^{-(1+\beta)}$ with $-(1+\beta) = 1.5$ for the present case. Then, for $E\ga 10$ GeV
the threshold
$E_\mathrm{c}$ begins to enter the low-energy branch of the spectrum and the dependency progressively evolves from $\tau_{\gamma \gamma} \propto E_\mathrm{HE}^{-(1+\beta)}$ to $\tau_{\gamma \gamma} \propto E_\mathrm{HE}^{-(1+\alpha)}$ because photons contributing mostly to $\tau_{\gamma \gamma}$ have an energy close to $E_\mathrm{c}$. Due to the particular choice of $\alpha=-1$, we have $-(1+\alpha) = 0$ and the $\gamma\gamma$ opacity $\tau_{\gamma \gamma}$ becomes independent of $E_\mathrm{HE}$ at very high energies (see the saturation effect observed in the middle panel of \reffig{fig_case_flash} when $E_\mathrm{HE}$ is increasing).\\

\noindent \textbf{Dependence on the Lorentz factor.}
The bottom right panel of \reffig{fig_case_flash} gives the evolution of $\tau_{\gamma \gamma}$ as a function of $R_\mathrm{e}$ for different Lorentz factors $\Gamma_0$. In this case the scaling is simply $\tau_{\gamma \gamma} \propto \Gamma_0^{2(\beta-1)}$. As discussed in \refappendix{sec_calc_approx}, this scaling comes from the combination of three effects: (i) the threshold energy goes like $E_\mathrm{c} \propto \Gamma_0^2$ so that $\tau_{\gamma \gamma}$ is modified by a factor $\Gamma_0^{2(1+\beta)}$ (see \refeq{eqn_taugamma_flash_pl_onaxis}) (ii) $\Delta t_{obs}$ is fixed and $R_{0} \propto \Gamma_0^2$, which modifies $\tau_{\gamma \gamma}$ by another factor $\Gamma_0^{-2}$ : the specific intensity is diluted by a factor $R_0^{-2}$
(see \refeq{eq_tau0}) but the opacity is built on a typical length $\ell\simeq R_0$ (see \refappendix{sec_calc_approx}) ; (iii) the geometrical factor $\left(1-\cos{\psi}\right)\simeq 1/2\Gamma_0^2$  
 brings another factor $\Gamma^{-2}$. Finally, the dependency is $\tau_{\gamma\gamma}\propto \Gamma_0^{2(1+\beta)-2-2}=\Gamma_0^{2(\beta-1)}$.

\subsection{Towards a complex photon field geometry}

Using the present approach, it is possible to treat any complex photon field configuration as long as it can be modeled as a collection of spherical photon flashes. 
The contribution from a flash
emitted at $R_0$, $t_0$ by a expanding sphere with Lorentz factor $\Gamma_0$,
with a spectrum defined by the comoving peak energy $E'_\mathrm{p,0}$ and the shape $\mathcal{B}$ and with a total radiated energy $\mathcal{E}_\mathrm{rad}$,
to the $\gamma\gamma$ opacity seen by a high-energy photon of energy $E_\mathrm{HE}$ emitted at $R_\mathrm{e}$, $t_\mathrm{e}$ in direction $\Theta_\mathrm{e}$ can be formally written as
\begin{eqnarray}
\tau_{\gamma\gamma}\left(E_\mathrm{HE},R_\mathrm{e},t_\mathrm{e},\Theta_\mathrm{e}\right)
 & = & 
\frac{\sigma_\mathrm{T}\mathcal{E}_\mathrm{rad}}{4\pi R_0^2 \Gamma_0 E'_\mathrm{p,0}}
\nonumber\\ 
& & 
\!\!\!\!\!\!\!\!\!\!\!\!\!\!\!\!\!\!\!\!\!\!\!\!\!\!\!\!\!\!\times 
\int d\ell\,
\mathcal{F}\left[\ell\, ;\, E_\mathrm{HE},R_\mathrm{e},t_\mathrm{e},\Theta_\mathrm{e}\, ;\, R_0, t_0, E'_\mathrm{p,0}, \mathcal{B} \right]
\, ,
\nonumber\\ 
\end{eqnarray}
where the formal function $\mathcal{F}$ can be deduced from \refeq{eqn_taugamma_flash}, and \refeq{eqn_interaction_angle} to \refeq{eq_cond_s}. In a more general situation, the
$\gamma \gamma$ opacity seen by a high energy photon is given by 
\begin{eqnarray}
\tau_{\gamma\gamma}\left(E_\mathrm{HE},R_\mathrm{e},t_\mathrm{e},\Theta_\mathrm{e}\right)
\!\!\!\! & = & \!\!\!\! 
\sum_{i}
\int dt\,
\frac{\sigma_\mathrm{T}\mathcal{L}_{\mathrm{rad},i}(t)}{4\pi R_i^2(t) \Gamma_i(t) E'_{\mathrm{p},i}(t)}
\nonumber\\ 
 & & \!\!\!\!\!\!\!\!\!\!\!\!\!\!\!\!\!\!\!\!\!\!\!\!\!\!\!\!\!\!\!\!\!\!\!\!\!\!\!\!\!\!\!\!\!\!\!\!\!\!\times 
\int d\ell\,
\mathcal{F}\left[\ell\, ;\, E_\mathrm{HE},R_\mathrm{e},t_\mathrm{e},\Theta_\mathrm{e}\, ;\, R_i(t), t, E'_{\mathrm{p},i}(t), \mathcal{B}_i (t)\right]
\, ,
\nonumber\\ 
\end{eqnarray}
where the sum on $i$ is on all emitting regions in the outflow (such as the shocked region behind a propagating shock wave), the $dt$-integral is computed over the whole propagation of the considered emitting region ($t$ being measured in the source frame), and the elementary contribution to the opacity of the propagating emitting region $i$ at time $t$ is given by the same formal expression as for a spherical flash, 
replacing the radiated energy $\mathcal{E}_\mathrm{rad}$ by the radiated luminosity $\mathcal{L}_{\mathrm{rad},i}(t)$ at time $t$, and considering all other properties (radius, Lorentz factor, emitted spectrum) at time $t$.
Finally synthetic light-curves and spectra seen by a distant observer can be computed by integrating the flux from  each emitting region on equal-arrival time surfaces, including at high energy the effect of  $\tau_{\gamma \gamma}$. 

\subsection{Validation of the model} 
\label{sec:model_granot}
\begin{figure}
\centerline{\includegraphics[width=0.5\textwidth]{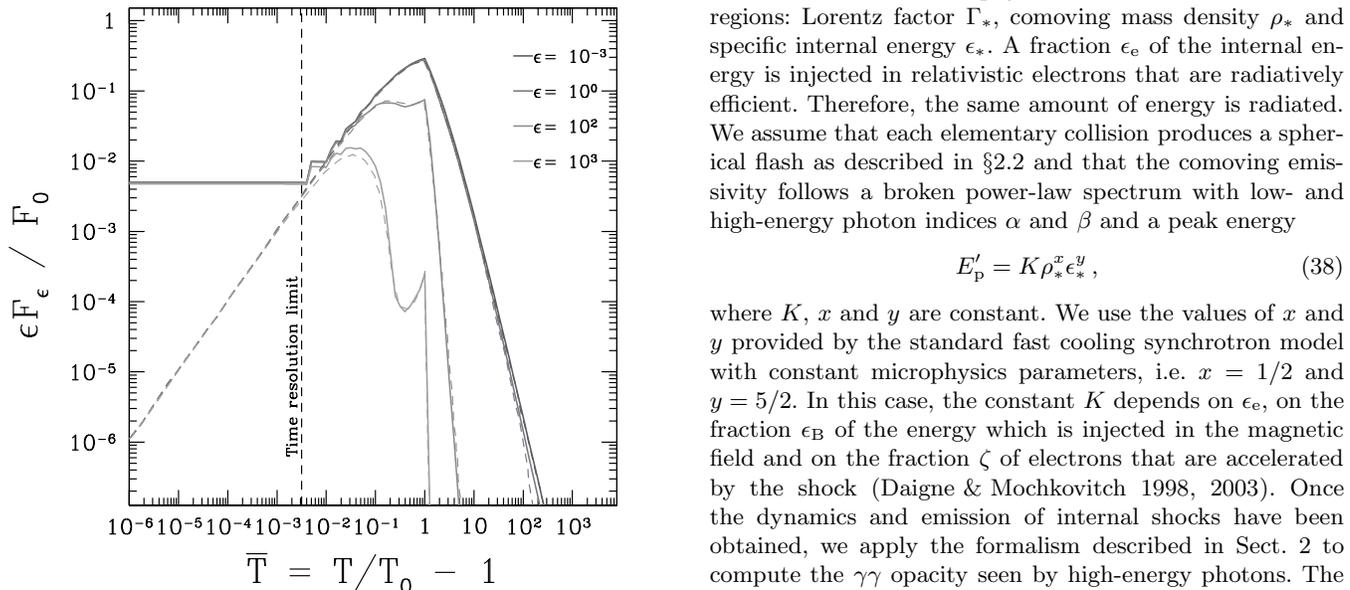}}

\caption{\textbf{Opacity in a single pulse: comparison with the semi-analytical work of \citet{granot:2008}.} 
$\gamma\gamma$ absorbed lightcurves at four different energies $\epsilon = E_\mathrm{HE}/ m_\mathrm{e}c^2$ are plotted as a function of the observer time for a single pulse using the prescriptions and model parameters corresponding to  the middle panel of Fig.~9  in \citet{granot:2008}. 
Our numerical calculation (solid line) is compared to the semi-analytical result 
(dashed line). Notations are the same (the observer time $\overline{T}$ and observed fluxes $\epsilon F_\epsilon / F_0$  have normalized values). The agreement is excellent except for 
$\overline{T} < 10^{-2}$, where the discrepancy  is due to numerical resolution limitations (this corresponds to a true observer time $t_\mathrm{obs} < 0.1$ ms).}
\label{fig_granot}
\end{figure}

\noindent Before dealing with more complex dynamical configurations within the internal shock framework, the validity of our model has been tested on a simple single-shock case with a comparison to the previous semi-analytic study by \cite{granot:2008}. In this study $\gamma$-ray light curves produced by an emitting thin  shell in spherical expansion are computed at different energies taking into account the effect of the $\gamma\gamma$ opacity.  Using our model and adopting the same prescriptions for the dynamics of the expanding shell and the properties of its emission (luminosity and spectrum), we reproduce the results of \cite{granot:2008}. The excellent agreement between the two approaches is illustrated  in \reffig{fig_granot}, corresponding to the example considered in the middle panel of Fig. 9 in \citet{granot:2008}. The only difference occurs at very early times (first 0.1 ms) due to the limitation of the discretization used in our numerical method. The semi-analytical approach used by \citet{granot:2008} does not suffer such a limitation but cannot be easily adapted to more complex outflows such as presented later in this paper. In this example, the $\gamma\gamma$ opacity obtained by \citet{granot:2008} and confirmed by our numerical calculation is found to be a factor $92$ smaller than the prediction from the single zone isotropic model, using the normalization factor $K_0^\mathrm{A09}$ from \citet{abdo:2009a} (see \refpar{sec_approx_formula}). This leads for instance to a cutoff energy in the time-integrated spectrum, defined as the energy $E_\mathrm{HE}$ where $\tau_{\gamma\gamma}\simeq 1$, which is $\approx 45\, m_\mathrm{e}c^2$ instead of $0.49\, m_\mathrm{e}c^2$. This strong discrepancy between single zone models and detailed calculations confirms how crucial it is to take into account the precise geometry and variations of the radiation field when computing the $\gamma\gamma$ opacity.  

\section{The $\gamma\gamma$ opacity in Internal Shocks}
\label{section_internal_shocks}

In this section the model is applied to the dynamical evolution expected in the internal shock framework, where the whole prompt $\gamma$-ray emission is produced by electrons accelerated by shock waves propagating within a relativistic variable outflow \citep{rees:1994,kobayashi:1997,daigne:1998}. A previous study of the $\gamma\gamma$ opacity in internal shocks was made by \cite{aoi:2010}. However the prescription used to compute $\tau_{\gamma \gamma}$ was still approximate, using the local physical conditions of the outflow where the high energy photon is emitted and applying them to an average formula of $\tau_{\gamma \gamma}$ as 
discussed in \refpar{sec_approx_formula}, using the same normalization as in \citet{lithwick:2001}.

\subsection{Description of the internal shock model}
We assume that the ejection by the central engine lasts for a duration $t_\mathrm{w}$ and we
consider the relativistic outflow at the end of the  acceleration process, where it is assumed that the energy content is dominated by the kinetic energy. It is then described by the distribution of the Lorentz factor $\Gamma(t)$ and of the kinetic power $\dot{E}_\mathrm{kin}(t)$. Both quantities can vary on a timescale $\Delta t_\mathrm{var}$ (variability timescale of the central engine), which leads to the formation of internal shocks at larger distance. 
We model the dynamics of these internal shocks via a multiple shell model where the successive collisions between shells mimic the propagation of shock waves within the relativistic outflow \citep{daigne:1998,daigne:2000}. This allows to follow the physical conditions in the shocked regions: 
Lorentz factor $\Gamma_*$, comoving mass density $\rho_*$ and specific internal energy $\epsilon_*$. A fraction $\epsilon_\mathrm{e}$ of the internal energy is injected in relativistic electrons that are radiatively efficient. Therefore, the same amount of energy is radiated. We assume that each elementary collision produces a spherical flash as described in \refpar{sec:flash} and that the comoving emissivity follows a broken power-law spectrum with low- and high-energy photon indices $\alpha$ and $\beta$ and a peak energy 
\begin{equation}
E'_\mathrm{p} = K \rho_*^{x} \epsilon_*^{y}\, ,
\label{eqn_peak_energy}
\end{equation}
where $K$, $x$ and $y$ are constant. We use the values of $x$ and $y$ provided by the
standard fast cooling synchrotron model with constant microphysics parameters, i.e. 
$x=1/2$ and $y=5/2$. In this case, the constant $K$
depends on $\epsilon_\mathrm{e}$, on the fraction $\epsilon_\mathrm{B}$ of the energy which is injected in the magnetic field and on the fraction $\zeta$ of electrons that are accelerated by the shock
\citep{daigne:1998,daigne:2003}.
 Once the dynamics and emission of internal shocks have been obtained, we apply the formalism described in \refsec{section_calculation_tgg} to compute the $\gamma\gamma$ opacity seen by high-energy photons. 
The assumption of a geometrically thin emitting region for each shock wave is well justified\footnote{We have tested in the single pulse burst presented in Fig. 4 that the contribution of a local isotropic radiation field associated with the finite size of the emitting region is negligible as long as its width is smaller than $\sim c t'_\mathrm{dyn}/100$. In the realistic case where this width is fixed by the synchrotron timescale, the additional contribution to the opacity is about five orders of magnitude below the main contribution.
}, as electrons are in the fast cooling regime, limiting the size of the emitting region at the shock front to be of the order of $c\times t'_\mathrm{rad}$, where $t'_\mathrm{rad}$ is the radiative timescale and is much shorter than the dynamical timescale $t'_\mathrm{dyn}$.
 We have tested that the results presented in this paper are independent of the choice for the time discretization of the dynamical calculation, as long as the time step is chosen to be short compared to the variability timescale $\Delta t_\mathrm{var}$ in the outflow.
 The results described hereafter in the paper have been obtained using a time step $t_\mathrm{w}/3000$, except for the example shown in \reffig{fig_smoothing} where the time step has been decreased to $t_\mathrm{w}/10000$ because of the short timescale variability in this particular case.\\
 
\begin{figure}
\centerline{\includegraphics[scale=0.42]{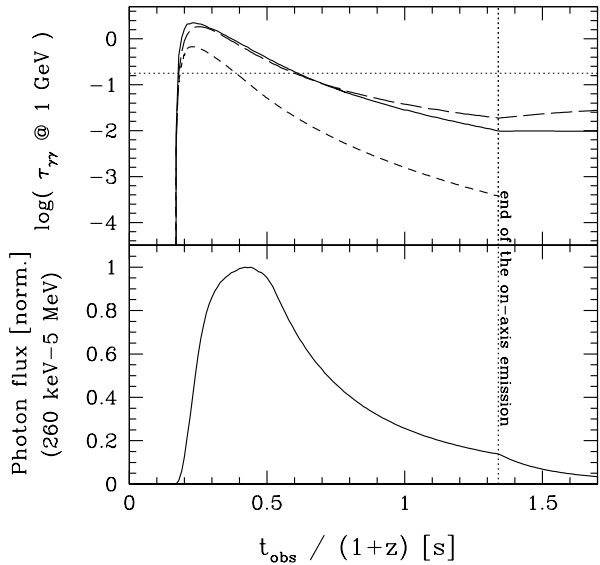}}

\caption{\textbf{$\gamma\gamma$ on-axis opacity as a good proxy.} Same single pulse burst as in \reffig{fig_delay} (see \reffig{fig_delay} for the initial conditions in the outflow).
\textit{Bottom panel:} normalized photon flux in the 260 keV - 5 MeV energy range
as a function of observer time;
\textit{Top panel:} $\gamma\gamma$ opacity seen by 1 GeV photons  (source frame) as a function of the observer time. The solid line shows the flux-averaged value of the exact opacity (latitude dependent), as given by \refeq{eq:taulatitudeaveraged}; the short dashed line shows the opacity seen by photons emitted on-axis; the long dashed line shows the flux-averaged value of the approximate opacity defined by \refeq{eq:proxy} using a constant factor $k=3$.
Note that the curve for the on-axis opacity stops at $t_\mathrm{obs}/(1+z)\simeq 1.33$ s  when the on-axis internal shock emission stops, whereas the flux averaged opacities are still well defined after this point, when the observed flux is only due to the high-(co-)latitude emission.
The approximate value of $\tau_{\gamma\gamma}$ (single zone model) is plotted as an horizontal thin dotted line using 
\refeq{eq:C1}. Note that this approximate formula has been calibrated using time-integrated spectra (see \refpar{sec_C1}).}
\label{fig:proxy}
\end{figure}

\subsection{Use of the on axis $\gamma\gamma$ opacity as a good proxy}
\label{sec:proxy}
From this point until the end of \refsec{section_internal_shocks} the $\gamma \gamma$ opacity computed on-axis will be used as a proxy for the complete (latitude dependent) calculation. More precisely for a given flash, the approximate opacity 
\begin{equation}
\tilde{\tau}_{\gamma\gamma}\left(E_\mathrm{HE}\right)=k\times \tau_{\gamma\gamma}\left(E_\mathrm{HE};\Theta_\mathrm{e}=0\right)
\label{eq:proxy}
\end{equation} 
will be applied to all high energy photons of energy $E_\mathrm{HE}$  radiated 
at the same radius, even those that are emitted off-axis. The constant factor $k$ is determined below, from a detailed comparison with the exact calculation.
This approximation has the advantage of  reducing the computing time without affecting the qualitative discussion. 
Even if the on-axis opacity 
(see \refeq{eqn_taugamma_onaxis}) and the exact 
opacity $\tau_{\gamma \gamma}$
 (see \refeq{eqn_taugamma_flash}) have different values, the ratio between the two 
 remains fairly unchanged during the whole 
 simulation as long as the on-axis emission is active. It is only at the very end of the prompt emission that the observed flux is dominated by the high-(co-)latitude emission and that this approximation is not appropriate anymore.\\

To illustrate the validity of this approximation, \reffig{fig:proxy} shows, in the case of a simple single pulse burst\footnote{
In the examples presented in Figs.~\ref{fig:proxy}, \ref{fig_scattering}, \ref{fig_delay}, \ref{fig_smoothing} and \ref{fig_thomson}, high values of the kinetic power have been chosen as \textit{Fermi}-LAT GRBs are among the brightest GRBs ever detected. For a jet opening angle from $1$ to $10^\circ$, an isotropic equivalent kinetic power of $10^{55}\,\mathrm{erg.s^{-1}}$ corresponds to a true power of $1.5\cdot 10^{51}$ to $1.5\cdot 10^{53}\,\mathrm{erg.s^{-1}}$. In addition, the 
microphysics parameters are fixed to have an observed time-integrated peak energy $E_\mathrm{p,0}= 1\, \mathrm{MeV}/(1+z)$ and the low- and high-energy slopes are fixed to the typical values $\alpha=-1$ and $\beta=-2.2$ \citep{preece:00} in these five figures.}
, the evolution 
of the flux-averaged value of the exact opacity (latitude dependent) 
 $\overline{\tau}_{\gamma \gamma}$ obtained by 
 \begin{equation}
 e^{-\overline{\tau}_{\gamma \gamma}(t_\mathrm{obs},E_\mathrm{HE})} = \frac{\sum_i F_i(t_\mathrm{obs},{E_\mathrm{HE}}) \times e^{-\tau_{\gamma\gamma}(t_\mathrm{obs},E_\mathrm{HE},\Theta_{\mathrm{e},i})}}{\sum_i F_i(t_\mathrm{obs},{E_\mathrm{HE}})}\, ,
 \label{eq:taulatitudeaveraged}
 \end{equation}
 where $t_\mathrm{obs}$ is the observer time and the summation is done over all collisions contributing to the observed flux at $t_\mathrm{obs}$: for each $i$, $F_i(t_\mathrm{obs},{E_\mathrm{HE}})$ is the contribution to the flux of collision $i$ and $\tau_{\gamma\gamma}(t_\mathrm{obs},\Theta_{\mathrm{e},i})$ is the $\gamma\gamma$ opacity computed for collision $i$ taking into account the emission colatitude $\Theta_{\mathrm{e},i}$ at time $t_\mathrm{obs}$.  
 It is compared to the evolution of the opacity for on-axis photons 
 and to the evolution 
 of the flux-averaged value of the approximate opacity $\tilde{\tau}_{\gamma\gamma}$ defined by \refeq{eq:proxy} using $k=3$. 
In this example, the $\gamma \gamma$ opacity evolves over more than 2 orders of magnitude whereas the ratio $\overline{\tilde{\tau}}_{\gamma \gamma}/\overline{\tau}_{\gamma \gamma}$ remains very close to unity, except at the end of the pulse, especially after the end of the on-axis emission. This small discrepancy is negligible in complex lightcurves as long as the central source is active.

\subsection{Minimum Lorentz factor in GRB outflows}
\label{subsection_lfmin}
\begin{figure*}
\begin{center}
\begin{tabular}{cc}
\includegraphics[scale=0.4]{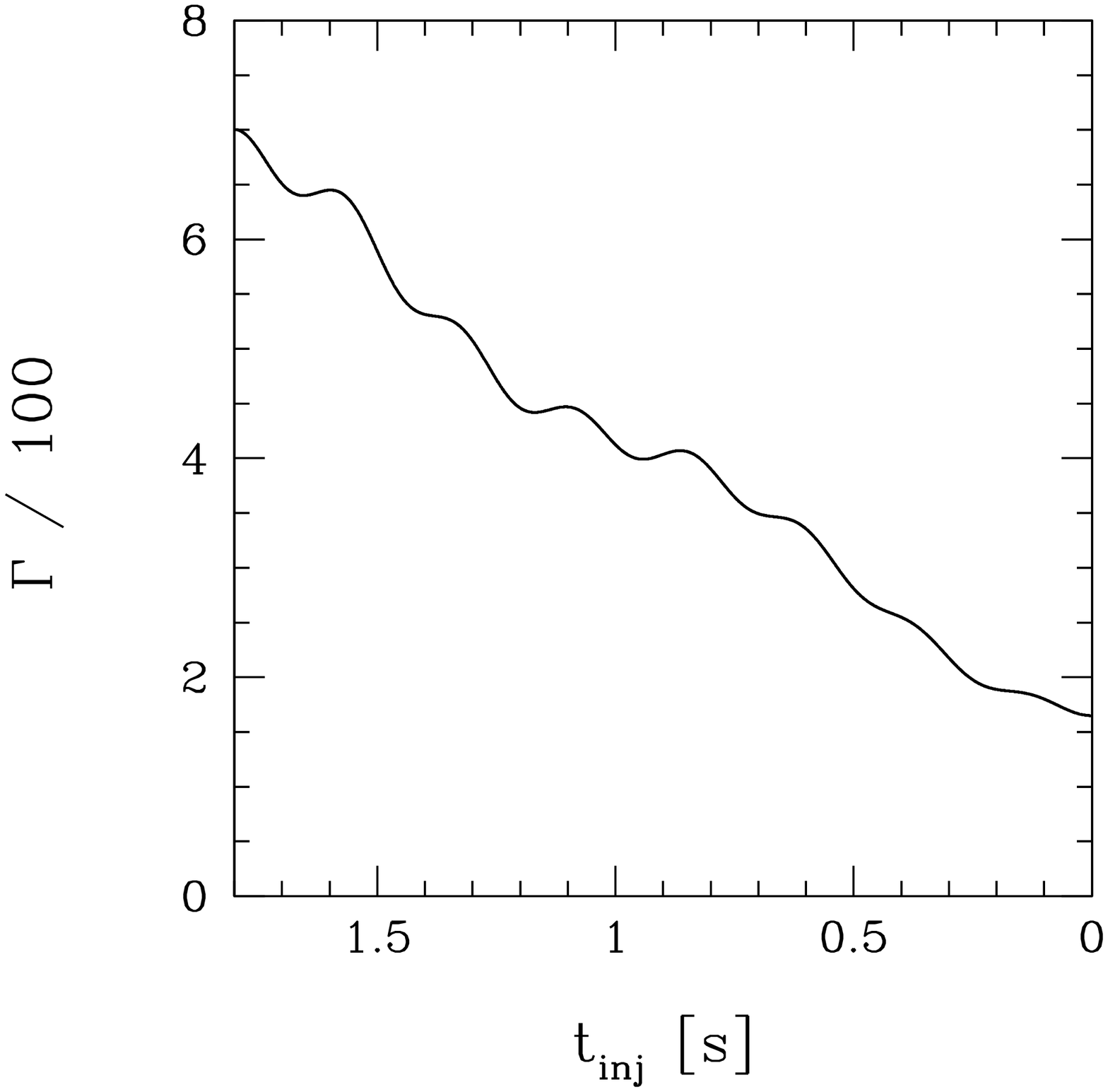} &  \includegraphics[scale=0.4]{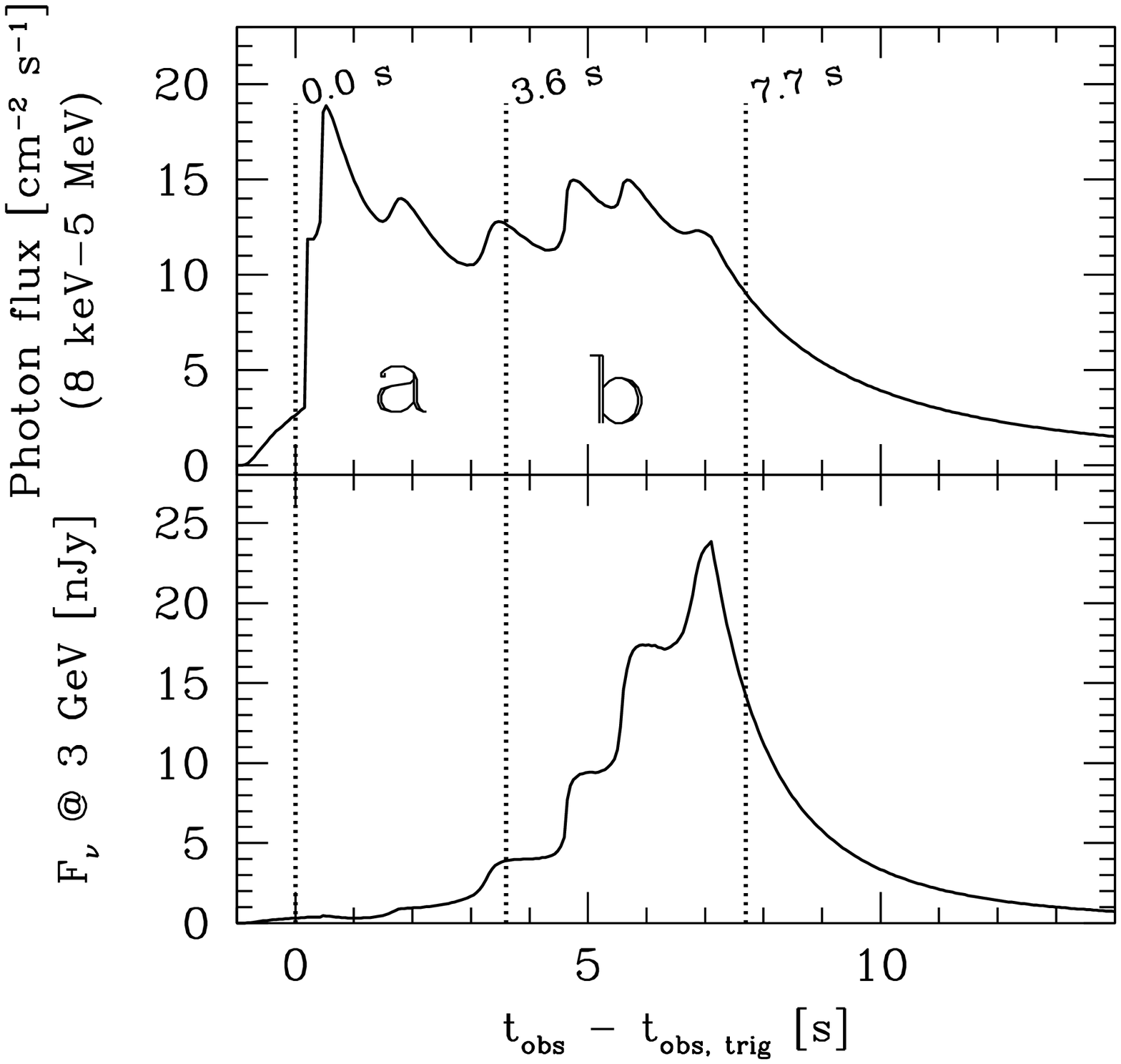} \\
\includegraphics[scale=0.4]{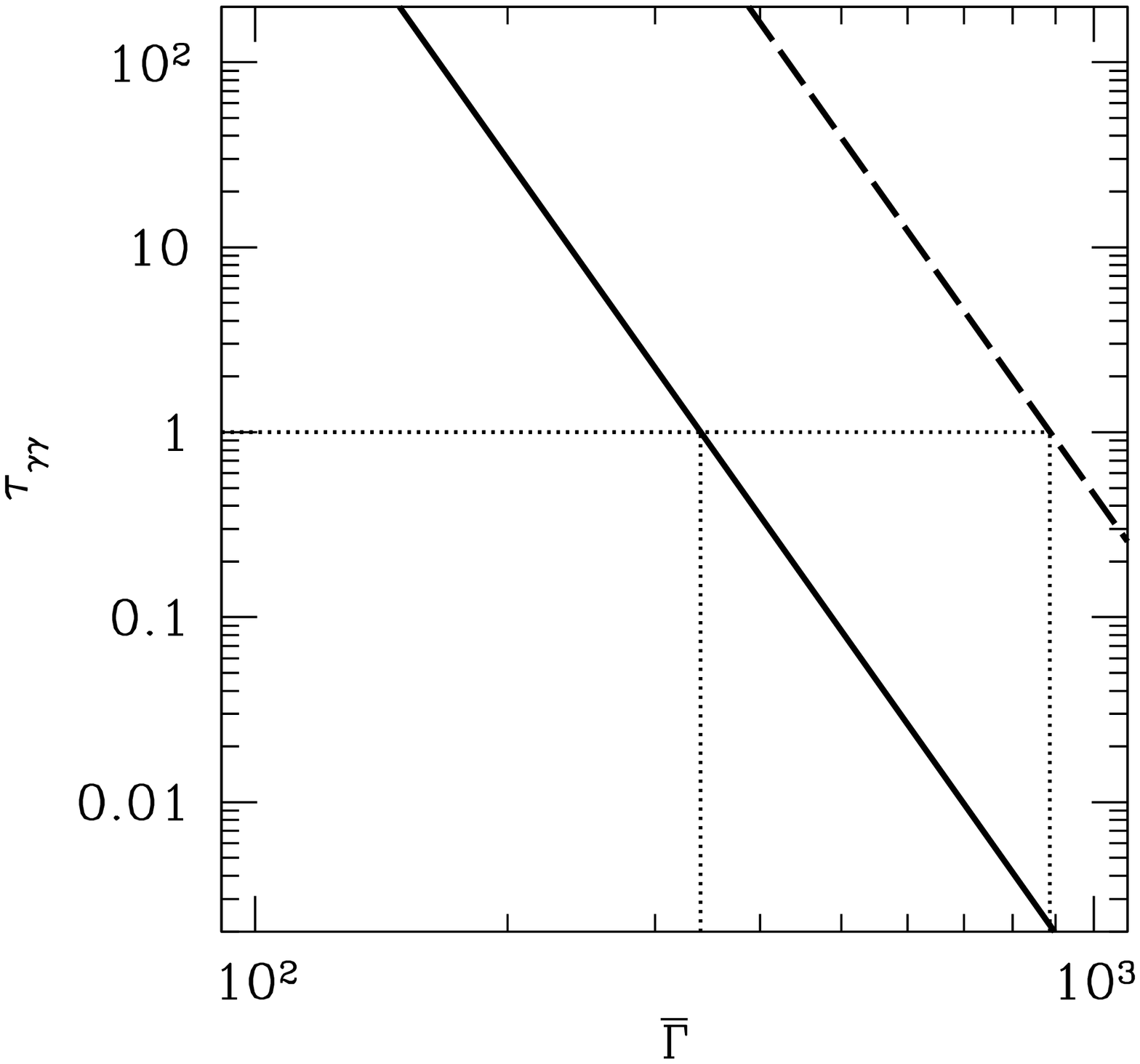} &  \includegraphics[scale=0.4]{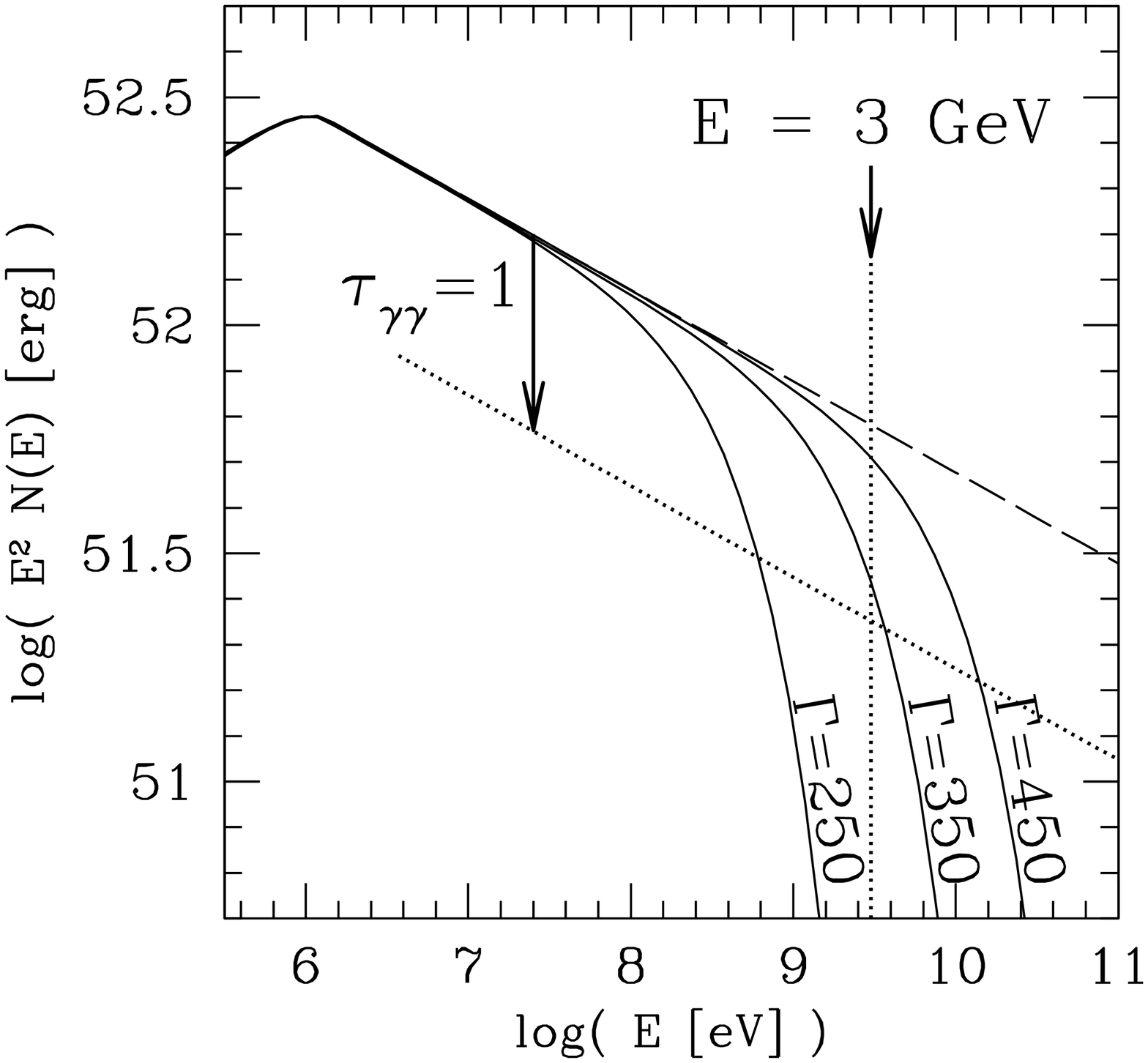}
\end{tabular}
\end{center}

\caption{\textbf{Minimum Lorentz factor for GRB 080916C.}  The two first panels are plotted for the critical case leading to $\tau_{\gamma\gamma}(3\,\mathrm{GeV})=1$ in time bin 'b', i.e. for a mean Lorentz factor $\overline{\Gamma} = \overline{\Gamma}_\mathrm{min}=340$. \textit{Upper left panel:} initial Lorentz factor distribution in the outflow. \textit{Upper right panel:} $\gamma$-ray lightcurves in the GBM/BGO band ($8$ keV -- $5$ MeV, top) and at 3 GeV (bottom). The lightcurves are plotted as a function of $t_\mathrm{obs}-t_\mathrm{obs,trig}$, where $t_\mathrm{obs,trig}$ is the observer time of the first detected photons. \textit{Lower left panel:} evolution of $\tau_{\gamma \gamma}$ at $E_\mathrm{HE}=3$ GeV against the mean Lorentz factor in the outflow $\overline{\Gamma}$, following our detailed modeling (solid line) and using the single zone isotropic model from \citet{abdo:2009a} (dashed line). \textit{Lower right panel:} time integrated spectrum over time bin 'b' for different mean Lorentz factors  (the relative shape of the initial Lorentz factor distribution is kept the same) and reference spectrum without $\gamma\gamma$ annihilation (dashed line). }
\label{fig_gmin}
\end{figure*}

\subsubsection{The case of GRB 080916C}
\label{sec:grb090816C}
The first natural application of our model is the estimate of the minimum bulk Lorentz factor $\Gamma_\mathrm{min}$ in GRB outflows, obtained from the constraint $\tau_{\gamma\gamma}(E_\mathrm{HE,max})\simeq 1$, where $E_\mathrm{HE,max}$ is the highest photon energy detected in the burst. 
To illustrate this aspect, 
we applied our approach to the case of one of the four brightest GRBs detected in the GeV range by \textit{Fermi}, GRB 080916C. The results are shown in \reffig{fig_gmin}. Using our numerical model, a synthetic GRB is generated, which reproduces the main observational features: 
the total radiated isotropic $\gamma$-ray energy ($E_\mathrm{iso} = 8.8 \times 10^{54}$ ergs between 10 keV and 10 GeV), the spectral properties ($E_\mathrm{p}$, $\alpha$, $\beta$ parameters of the Band function\footnote{We use the values given by \citet{abdo:2009a} for time bin 'b'.}), the envelop of the light curve and a short time-scale variability of 0.5 s in the observer frame. The study is focused on the most constraining time bin, bin 'b',
during which the highest observed photon energy was $E_\mathrm{HE,max}=3$ GeV (16 GeV in the source rest frame): for this reason, only time bins 'a' and 'b' are reproduced in the synthetic GRB. These two intervals correspond to 32~\%
of the total radiated 
energy. 
The ejection by the central engine lasts for $t_\mathrm{w}=1.8\,\mathrm{s}$, the injected kinetic power is taken constant and equals $\dot{E}_\mathrm{kin}(t)=8\cdot 10^{55}\,\mathrm{erg.s^{-1}}$ (isotropic equivalent value)
and the assumed initial distribution of the Lorentz factor $\Gamma(t)$ is plotted in the top-left panel of \reffig{fig_gmin}. This initial distribution leads to the formation of two shock waves. The first shock wave, propagating forwards, is short-lived as it reaches rapidly the front edge of the outflow. The second shock wave, propagating backwards, is the main source of radiation. Due to the initial shape of $\Gamma(t)$, the physical conditions in the shocked region vary during the propagation, leading to two main pulses in the lightcurve, with additional variability on shorter timescales (see \reffig{fig_gmin}, top-right panel). Despite its apparent complexity, this case remains relatively simple as there is only one propagating radiating front in the outflow. We find that the $\gamma\gamma$ opacity seen by high-energy photons is dominated by a local contribution, i.e. by the interaction with seed photons emitted almost at the same radius as the high-energy photon. Typically, we find that the opacity $\tau_{\gamma\gamma}\left(E_\mathrm{HE};\Theta_\mathrm{e}=0\right)$ of a high-energy photon emitted on-axis at radius $R_\mathrm{e}$ is built over a distance $\ell\simeq \left(0.1\to 1\right) R_\mathrm{e}$. In addition, in this simple case, the geometry makes impossible head-on collisions: the interaction angle is necessary in the range $0\le \Psi \la 1/\Gamma$, resulting in a seed radiation field which is highly anisotropic. Therefore, the situation considered to model time bins 'a' and 'b' in  GRB 080916C is not very far from the simplest case of a spherical flash, and, as discussed below, the approximate formulae derived in \refpar{sec_approx_formula} remains relevant even in this more complex case.   \\

The minimum mean Lorentz factor $\overline{\Gamma}_\mathrm{min}$ is obtained by requiring that  $\tau_{\gamma \gamma}\left(E_\mathrm{HE,max}\right) \la 1$ (see \reffig{fig_gmin}, lower panel). 
With the detailed calculation,  we find a minimum mean Lorentz factor $\overline{\Gamma}_\mathrm{min} = 340$, i.e. a factor $2.6$ lower than the value $\overline{\Gamma}_\mathrm{min}=887$, which was obtained from an approximate single zone isotropic model \citep{abdo:2009a}. Even more remarkable, the whole initial distribution of the Lorentz factor used in this model of GRB 080916C (from 170 to 700) remains below the ``minimum'' value of the Lorentz factor derived from single zone models (see  \reffig{fig_gmin}, upper left panel).\\

Compared to GRB 080916C, the case of GRB 090926A \citep{ackermann:2011} is more complex : the time-integrated high-energy spectrum shows a cutoff at about 1.4 GeV. If this cutoff is interpreted as due to $\gamma\gamma$ annihilation, it allows a direct measurement of the Lorentz factor instead of a minimum value. It is however difficult to confirm this interpretation from the spectral shape of the cutoff  and other physical explanations are possible (see \refpar{subsection_scattering}). Another source of complexity is due to the presence of an additional component at high-energy, which is not necessarily produced in the same region than the soft gamma-ray component (see the discussion in \refsec{sec_2zones_model}). Nevertheless, interpreting the cutoff as a signature of $\gamma\gamma$ annihilation, \citet{ackermann:2011} obtain a Lorentz factor of $\Gamma \approx 720\pm 76$  with the single zone model, and a value $\Gamma \approx 220$ (i.e. a factor $\sim 3$ lower) with a more realistic approach based on the formalism developed by \citet{granot:2008} (see \refpar{sec:model_granot} and \refpar{sec_C1}). 
Using the approximate formula given in \refeq{gmin_gg} below, which has been calibrated on the numerical results presented in this paper (see \refpar{sec_otheropacities}), we obtain a very similar value $\Gamma \approx 250$.\\
 
If the compactness argument imposes a minimum Lorentz factor for the outflow, we note on the other hand that -- 
if the prompt emission has an internal origin --
there is also an upper limit on the Lorentz factor, imposed by the condition that the internal dissipation should occur before 
the deceleration by the external medium becomes significant, i.e. at radii lower than the deceleration radius:
\begin{equation}
R_\mathrm{dec} = \left( \frac{3-s}{4\pi} \frac{E_\mathrm{kin}}{\overline{\Gamma}^2 A c^2} \right)^{\frac{1}{3-s}}\, ,
\label{eqn_deceleration_radius}
\end{equation}
where $\overline{\Gamma}$ is the mean Lorentz factor and $E_\mathrm{kin}$ the total kinetic energy of the outflow, and $A$ and $s$ parametrize the density profile of the external medium which follows 
\begin{equation}
\rho_\mathrm{ext} = \frac{A}{R^s}\, .
\label{eqn_external_density}
\end{equation}
As $R_\mathrm{dec}$ \textit{decreases} as $\overline{\Gamma}^{-\frac{2}{3-s}}$, this condition may become severe for high values of $\overline{\Gamma}_\mathrm{min}$. This is especially true for the internal shock model, where the typical dissipation radius \textit{increases} with the Lorentz factor as $\overline{\Gamma}^2$. 
In the case of the synthetic burst used to model bins 'a' and 'b' of GRB 080916C, a typical efficiency of 5 \%
for the internal shock phase leads to a total kinetic energy of the outflow $E_\mathrm{kin} \simeq 1.4 \cdot10^{56}$ erg.  Therefore, for a  wind-like (resp. uniform) medium with $A = 5 \cdot 10^{11} \mathrm{g\cdot cm^{-1}}$ and $s=2$ (resp. $A=10^3 m_\mathrm{p}\, \mathrm{cm^{-3}}$ and $s=0$) , the deceleration radius equals $R_\mathrm{dec} = 1.9 \cdot 10^{17}\, \left(\overline{\Gamma}/100\right)^{-2} $ cm (resp. $R_\mathrm{dec} = 5.6 \cdot 10^{16}\, \left(\overline{\Gamma}/100\right)^{-2/3} $ cm). On the other hand, internal shocks are found in the simulation to occur over a radial range $R_\mathrm{IS}\simeq \left(8.7\cdot 10^{13}\to 6.1\cdot 10^{14}\right) \left(\overline{\Gamma}/100\right)^2\, \mathrm{cm}$. Then the condition $R_\mathrm{IS} < R_\mathrm{dec}$ imposes $\overline{\Gamma} \la 420$ in the wind-like medium (resp. $\overline{\Gamma} \la 540$ in the uniform medium). 
Assuming dense external media such as expected around a massive star,
previous estimates of $\overline{\Gamma}_\mathrm{min}=887$ were not consistent with this constraint, whereas the more accurate estimate of $\overline{\Gamma}_\mathrm{min}=340$ obtained in the present study fulfills the condition for both types of external medium.

\subsubsection{Minimum Lorentz factor: detailed modeling vs simple estimates}
\label{sec_C1}

The difference in the estimate of $\overline{\Gamma}_\mathrm{min}$ from single zone isotropic models and the detailed approach presented here has two origins:  as shown in \refpar{sec_approx_formula}, 
the normalization obtained with single zone isotropic models is larger than the normalization obtained by considering the interaction of a high-energy photon with the anisotropic field created by a single flash,
leading to a ratio in the estimate of the minimum Lorentz factor  equal to 
 $\left(K_0^\mathrm{A09}/K_0\right)^{\frac{1}{2(1-\beta)}}\simeq 1.5$   \citep{abdo:2009a} (resp. $\left(K_0^\mathrm{LS01}/K_0\right)^{\frac{1}{2(1-\beta)}}\simeq 1.3\to 1.5$, 
 \citealt{lithwick:2001})
 for $\beta=-2\to -3$. However the difference illustrated by the example of GRB 080916C is much larger, due to the fact that the dynamics of the outflow is properly  taken into account in the detailed approach presented here. Such effects are important, see e.g. 
   the dependency on $R_\mathrm{e}$ in \refeq{equation_tgg_flash_approximate}.  \\

 Note that a precise comparison between single zone models and our detailed approach is difficult, as the value of the Lorentz factor is not unique in our case. Indeed, the variations of $\Gamma$ are at the origin of the prompt emission, as we use the internal shock framework. In the example of GRB 080916C, the value $\overline{\Gamma}_\mathrm{min} = 340$ is the mean value for a distribution varying from $\Gamma_\mathrm{inf}=170$ to $\Gamma_\mathrm{sup}=700$. In the approximate formula for $\tau_{\gamma\gamma}$ obtained in \refpar{sec_approx_formula}, the dependency $\tau_{\gamma\gamma}\propto \Gamma_0^{-2(1-\beta)}$ can be seen as the product of $\Gamma_0^{-4}$ coming from the dependency of $\tau_{\gamma\gamma}$ on the emission radius, and $\Gamma_0^{2(1+\beta)}$ (i.e. $\Gamma_0^{-2.4}$ for $\beta=-2.2$) coming from the dependency of $\tau_{\gamma\gamma}$ on the photon field. In the internal shock model, the radius is mainly fixed by the minimum Lorentz factor $\Gamma_\mathrm{inf}$ within the outflow. Indeed, in a two shell collision with a ratio $\kappa=\Gamma_\mathrm{sup}/\Gamma_\mathrm{inf}$ between the Lorentz factors and a variability timescale $\Delta t_\mathrm{var}$, the internal shock radius depends only on $\Gamma_\mathrm{inf}$ for efficient collisions. It equals $R\simeq 2 \Gamma_\mathrm{inf}^2 c \Delta t_\mathrm{var}$ with a precision better than 10 \%
for $\kappa\ga 3.2$.  On the other hand the Lorentz factor relevant for the radiation field should be the Lorentz factor in the shocked regions, which is not too different from 
 the mean Lorentz factor $\overline{\Gamma}$ in the outflow. 
 This leads to
\begin{equation}
\tau_{\gamma\gamma}\propto \Gamma_\mathrm{inf}^{-4}\overline{\Gamma}^{2(1+\beta)}\, .
\label{eq_correctscaling}
\end{equation}
All the examples presented in this paper agree well with this scaling law. We find that a good approximate formula for the $\gamma\gamma$ opacity can be deduced from the comparison of \refeq{equation_tgg_flash_approximate} and the detailed numerical calculation presented in this paper and is given by
\begin{equation}
\tau_{\gamma\gamma}(E_\mathrm{HE}) 
\simeq
K'_0 \frac{A_0 \sigma_\mathrm{T} \mathcal{E}_\mathrm{rad}}{4\pi \left(c \Delta t_\mathrm{var}\right)^2 E_\mathrm{p,0}}
\left(\frac{\left(m_\mathrm{e}c^2\right)^2}{E_\mathrm{HE}E_\mathrm{p,0}}\right)^{1+\beta} 
\!\!\!\!\!\!\!
\Gamma_\mathrm{inf}^{-4}\, \overline{\Gamma}^{2\left(1+\beta\right)}\, ,
\label{eq:C1}
\end{equation}
with $K'_0= C_1 K_0 = C_1 \left( 2^{1+2\beta}\mathcal{I}(\beta)\right)$. The numerical constant $C_1\simeq 4\cdot 10^{-2}$ has been obtained from the comparison of \refeq{eq:C1} with the result of the full numerical calculation in a series of runs based on the synthetic single pulse GRB shown in  \reffig{fig:proxy}. 
In this example,
we have $\Gamma_\mathrm{inf}=100$, $\overline{\Gamma}=290$, $E_\mathrm{p,0}=1\, \mathrm{MeV}$, $\beta=-2.2$,  $\Delta t_\mathrm{var}\simeq 0.4\,\mathrm{s}$
and the radiated energy above $E_\mathrm{p,0}$ is $1.3\cdot 10^{53}\, \mathrm{erg}$, i.e. $\mathcal{E}_\mathrm{rad}=5.2\cdot 10^{52}\, \mathrm{erg}$ after a correction by a factor $0.4\, \mathrm{s}/1\, \mathrm{s}$ to estimate $\mathcal{E}_\mathrm{rad}$ over $\Delta t_\mathrm{var}$. Then each of these parameters has been varied in different series of runs, to allow for the calibration of $C_1$. 
As seen in \reffig{fig:proxy}, it is not relevant to compare the approximate value of the opacity given by \refeq{eq:C1}
 with the result of the time-dependent calculation, where the opacity varies by several orders of magnitude during the evolution of the pulse. Therefore, the comparison is based on the time-integrated spectrum over the full duration of the burst. A
 ``time-averaged'' 
   opacity is computed from the ratio of the unabsorbed over the absorbed spectrum, and then compared to 
 \refeq{eq:C1}. As the single zone model predicts an exponential cutoff at high-energy, whereas the detailed calculation shows a more complex shape (see \refpar{subsection_scattering}), the comparison can be done only in the relevant energy range, i.e. in the region where the opacity is close to unity. At higher energy, the $\gamma\gamma$ opacity is overestimated by the single zone model. At lower energy, we find that the discrepancy on the $\gamma\gamma$ opacity is less than a factor of 2, which would lead to a factor $2^{1/(2(1-\beta))}$ for the minimum Lorentz factor, i.e. about 10\, \%.  In the series of runs which were used to calibrate $C_1$, we have varied the mean Lorentz factor $\overline{\Gamma}$ from 100 to 1000, the ratio $\kappa=\Gamma_\mathrm{sup}/\Gamma_\mathrm{inf}$ from 2 to 8, the kinetic power $\dot{E}_\mathrm{kin}$ from $10^{50}$ to $10^{56}\, \mathrm{erg.s^{-1}}$, the duration $t_\mathrm{w}$ from 0.1 to 10 s, and the peak energy $E_\mathrm{p,0}$ from 100 keV to 10 MeV and we have found  that -- after calibration -- the discrepancy for the $\gamma\gamma$ opacity remains less than a factor of 2, except for low $\kappa$ where it can increase to a factor of $10$. This is because the simple scaling law between the internal shock radius and the Lorentz factor $\Gamma_\mathrm{inf}$ is not valid anymore for $\kappa \la 3$.  We have checked a posteriori that the discrepancy for the opacity is less than a factor of 2  for all the cases presented in this paper.\\

To compare to the estimates by \citet{abdo:2009b} (resp. \citealt{lithwick:2001}) where the details of the distribution of the Lorentz factor are unknown, we use \refeq{eq:C1} with $\Gamma_\mathrm{inf}=\overline{\Gamma}$. This leads to a minimum Lorentz factor $\Gamma_\mathrm{min}$ which is reduced by a factor $\left(K_0^\mathrm{A09}/K'_0\right)^{\frac{1}{2(1-\beta)}}\simeq 2.4$ 
(resp. $\left(K_0^\mathrm{LS01}/K'_0\right)^{\frac{1}{2(1-\beta)}}\simeq 2.2$)
 for $\beta=-2.3$. This is in reasonable agreement with 
the pre-\textit{Fermi} study presented by \citet{granot:2008}. In this earlier work, a semi-analytical calculation of $\tau_{\gamma\gamma}$ using a simple prescription for the propagation of the shock wave  responsible for the pulse (constant Lorentz factor) but taking into account the exact radiation field led to a normalization in the equivalent of 
\refeq{eq:C1} given by (see Eq.(126) in \citealt{granot:2008}):
\begin{equation}
{K'_0}^\mathrm{G08}\simeq \frac{1.86\cdot 10^{-4} }{c_2} \left(\frac{-\beta}{2}\right)^{-5/3}\, , 
\end{equation}
where $c_2\simeq 1$. For $\beta=-2.3$, this leads to a reduction factor $\left({K_0^\mathrm{A09}}/{{K'_0}^\mathrm{G08}}\right)^{\frac{1}{2(1-\beta)}}\simeq 2.6$ for the estimate of the minimum Lorentz factor. \\

\begin{figure*}
\begin{center}
\begin{tabular}{cc}
\includegraphics[scale=0.4]{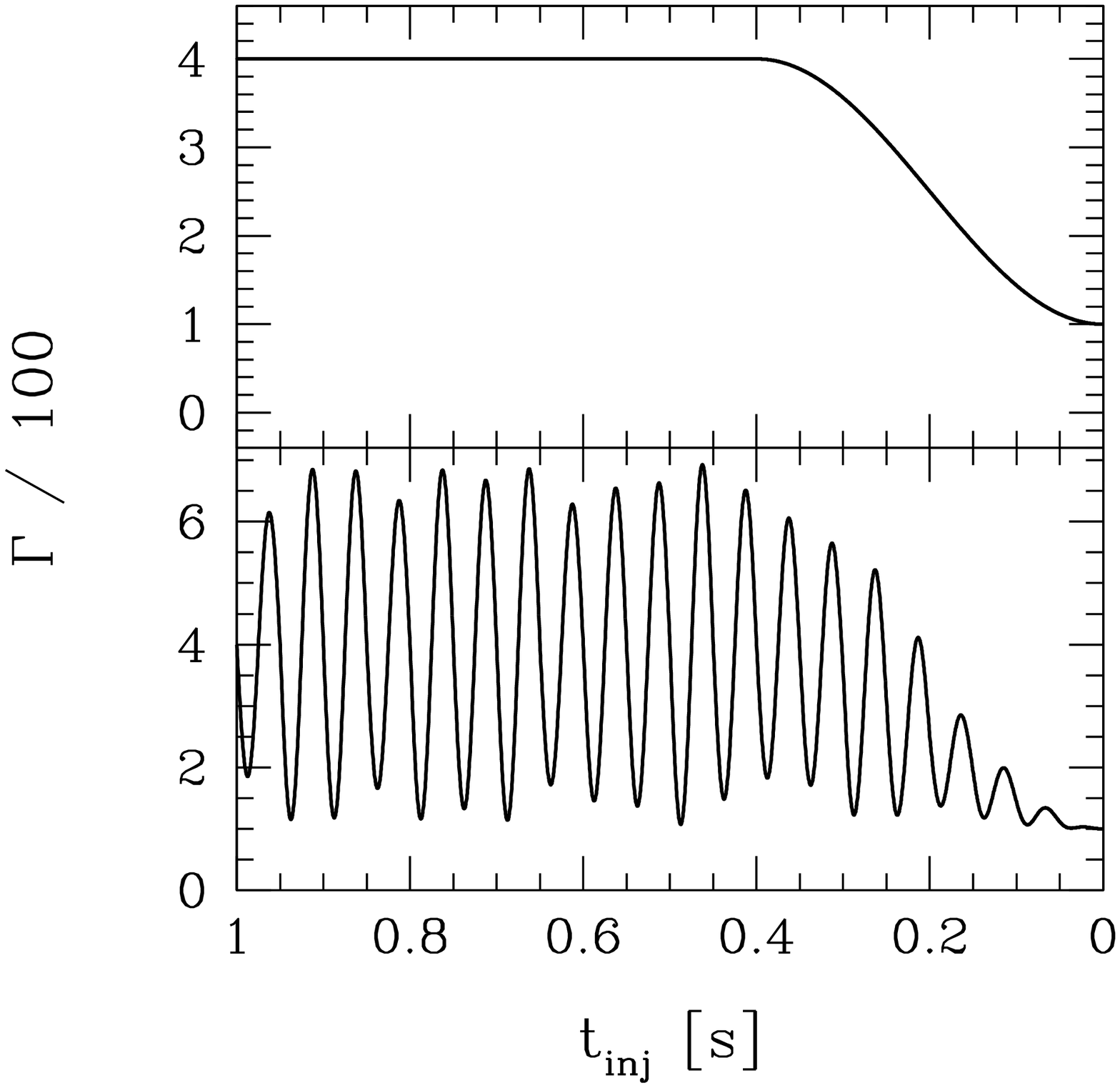} & \includegraphics[scale=0.4]{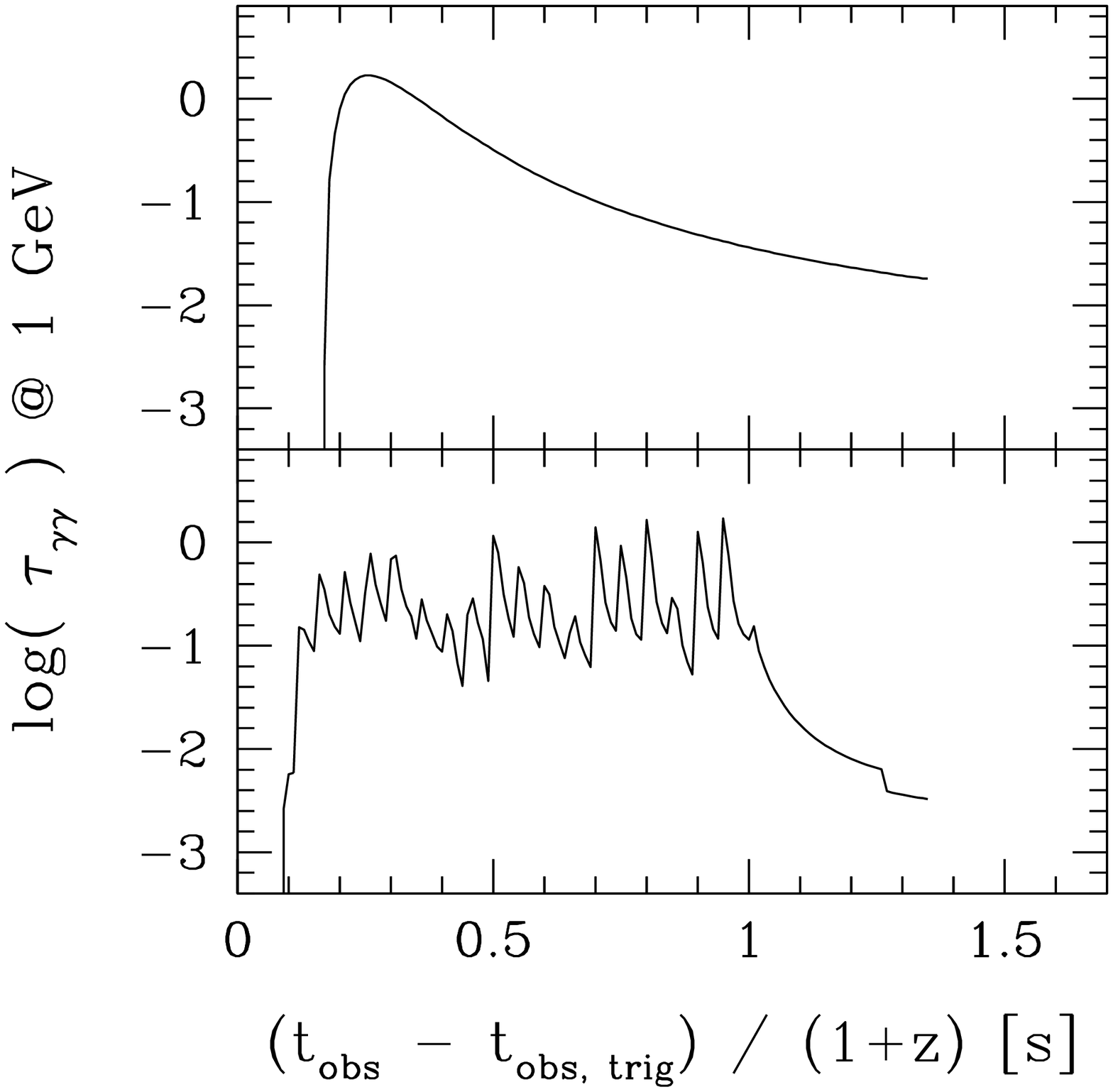} \\
\includegraphics[scale=0.4]{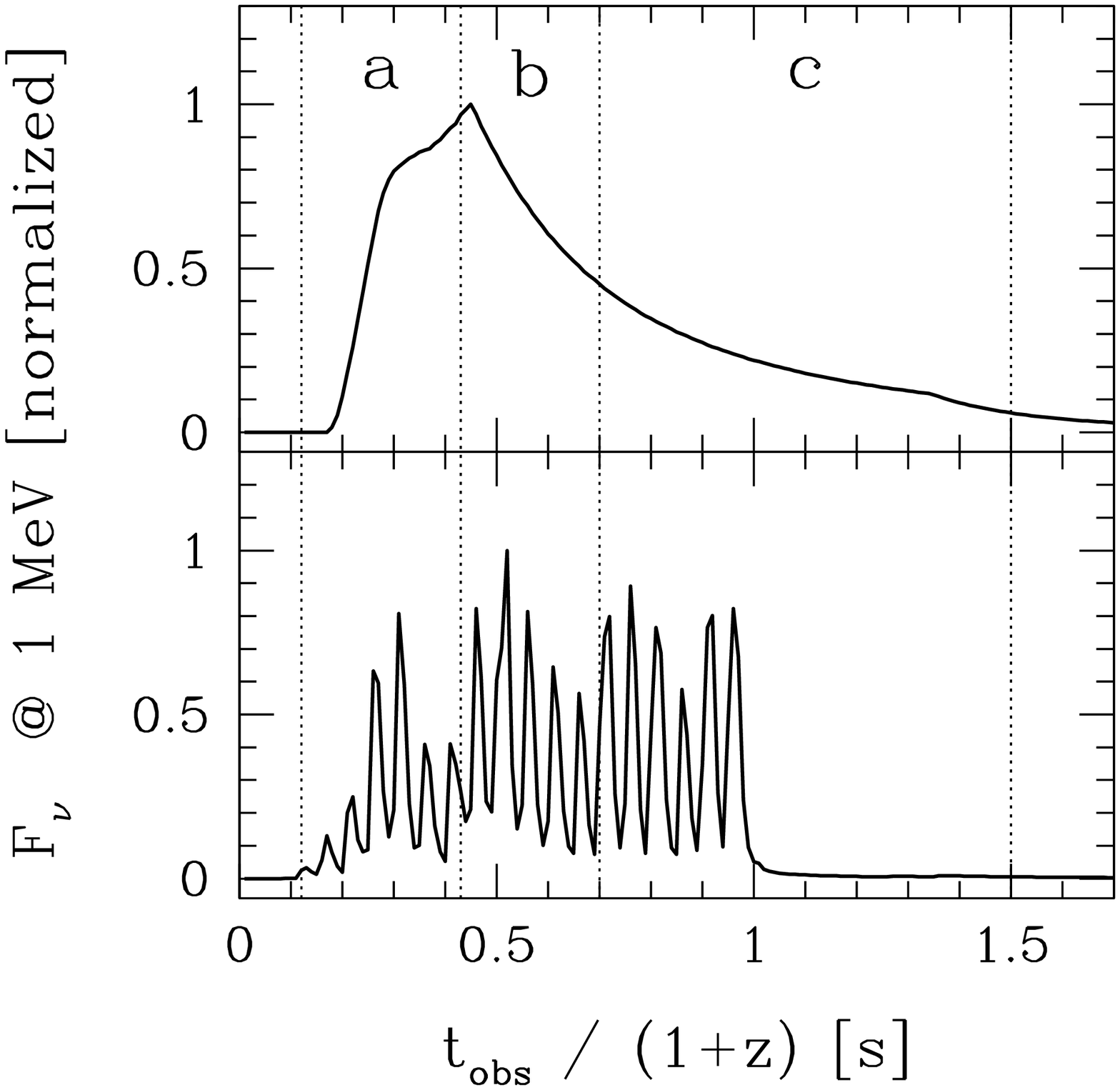} & \includegraphics[scale=0.4]{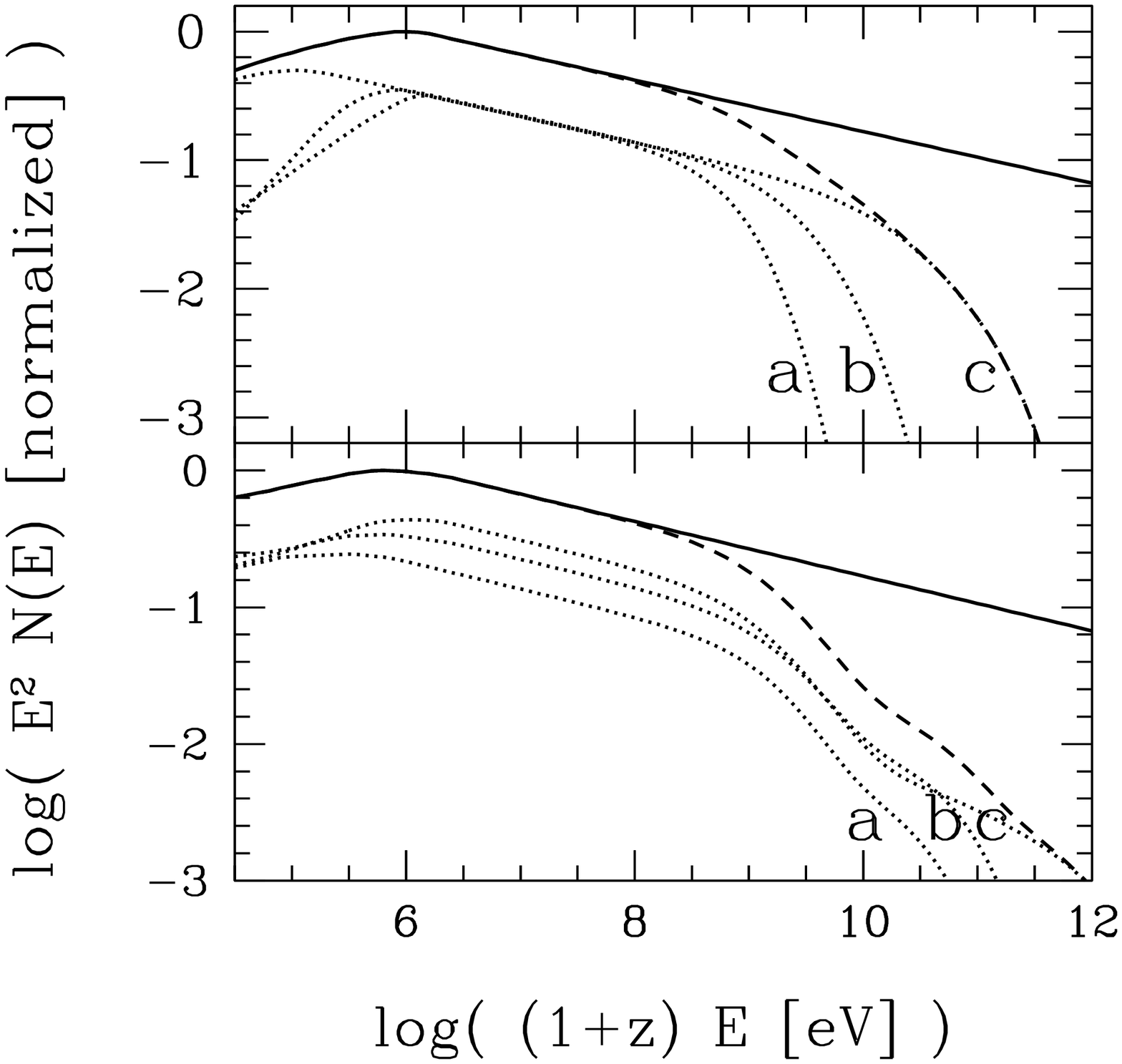} 
\end{tabular}
\end{center}

\caption{\textbf{Temporal and spectral evolution of the $\gamma\gamma$ opacity - Example of two synthetic GRBs.} The upper part of the four figures illustrates the case of a mono-pulse GRB with no additional temporal substructures whereas the bottom part show the case of a complex multiple-pulse GRB. 
The Lorentz factor 
during the ejection
rises monotonously from 100 to 400 in the first case, leading to a collision between the slow and the fast parts of the outflow, and from 100 to 700 in the second case with the addition of variability on a shorter time-scales.
The kinetic power of the outflow is assumed to be constant, with $\dot{E}_\mathrm{kin}=10^{54}\, \mathrm{erg.s^{-1}}$ in the first case and $5\cdot10^{53}\, \mathrm{erg.s^{-1}}$ in the second case (so that the radiated energy $2\cdot10^{52}$ ergs is the same in both cases). \textit{Upper left panel:} initial Lorentz factor distribution of the outflow as a function of the injection time $t_\mathrm{inj}$. 
\textit{Upper right panel:}  the evolution of the approximate 
opacity $\tilde{\tau}_{\gamma\gamma}$, as defined in \S\ref{sec:proxy},  seen by 1 GeV photons (source frame) as a function of the observer time $t_\mathrm{obs}-t_\mathrm{obs,trig}$. The curve is plotted only as long as the on-axis emission is active.
\textit{Lower left panel:} 
the $\gamma$-ray light curve at 1 MeV (source frame) as a function of the observer time $t_\mathrm{obs}-t_\mathrm{obs,trig}$, where $t_\mathrm{obs,trig}$ is the observer time of the first detected photons in the complex GRB case. Three integration time bins for the simulated spectra (see lower right panel) are delimited by vertical dotted lines and labeled 'a', 'b' and 'c'.
\textit{Lower right panel:} 
Time-integrated spectra corresponding to time bins 'a', 'b' and 'c' (dotted line) and 'a+b+c' (dashed line). In the latter case, the time-integrated spectrum without $\gamma\gamma$ absorption is plotted in solid line. 
}
\label{fig_scattering}
\end{figure*}

\subsection{Shape of the $\gamma \gamma$ attenuated time-integrated spectrum} 
\label{subsection_scattering}

As pointed out by \citet{granot:2008}, due to the temporal evolution of $\tau_{\gamma \gamma}$, the opacity cutoff in a time-integrated spectrum will be smoother than a sharp exponential decay: the cutoff transition will be close to 
a power-law steepening. The detailed model presented here is appropriate to characterize from an observer point of view this time evolution effect.
The smoother $\tau_{\gamma \gamma}$ transition is due to the fact that the time integrated spectrum is a superposition of instant spectra which can have different $\gamma\gamma$ cutoff photon energies. This time evolution takes place within a given $\gamma$-ray pulse, and can be even stronger in a complex burst where the light curve is made of many pulses \citep{aoi:2010}. In addition, in the latter case,  cross-interactions between different pulses can become important and strongly influence the time evolution of  the opacity: such 
cross-interactions are fully taken into account for the first time by our approach.\\

To illustrate this effect, two examples of synthetic GRBs are shown in \reffig{fig_scattering}. In the case of the ``mono-pulse'' GRB the $\gamma\gamma$ opacity seen by high energy photons evolves smoothly, with a regular decay 
(except at the very beginning when the interacting photon field progressively builds up, see also \citealt{granot:2008}): as a result, 
even for time intervals corresponding to a noticeable fraction of the burst duration, 
the $\gamma\gamma$ cutoff in the time-integrated spectrum remains close to an exponential cutoff. In the case of the ``complex'' multiple-pulse GRB, the $\gamma\gamma$ opacity shows much stronger variations, due to the intrinsic variability of the outflow, and  consequently, 
the  $\gamma\gamma$ cutoff is much closer to a power-law steepening. \\

To our knowledge, GRB 090926A is the best case where a cutoff in the spectrum at high energy has been clearly identified at about 1.4 GeV \citep{ackermann:2011}. However the shape of this cutoff is weakly constrained due to the poor photon statistics at high energy. As a result, it is difficult to further discuss if this break is due to the $\gamma \gamma$ annihilation or some other spectral effect. For example in a leptonic scenario, the main Band component could be produced by synchrotron emission whereas a second sub-dominant component at higher energy could be due to inverse Compton scatterings in Klein Nishina regime (see e.g. \citealt{bosnjak:2009}). In this scenario, it is difficult to distinguish the cutoff due to $\gamma\gamma$ annihilation from the 
expected cutoff
 of the inverse Compton component at high energy,
 without a good characterization of the spectral shape above 1 GeV. 
 Even in the absence of clear spectral signatures of a cutoff for most \textit{Fermi} GRBs, an
 evidence for an attenuation at high-energy in a large fraction of  \textit{Fermi} bursts is given by the comparison of the LAT detection rate and the theoretical rate obtained by extrapolating the GRB spectrum measured by the GBM  \citep{nakar:2011}. Unfortunately, this method does not allow any comparison of  the spectral shape of the cutoff with theoretical predictions.

\begin{figure*}
\begin{center}
\begin{tabular}{cc}
\includegraphics[scale=0.4]{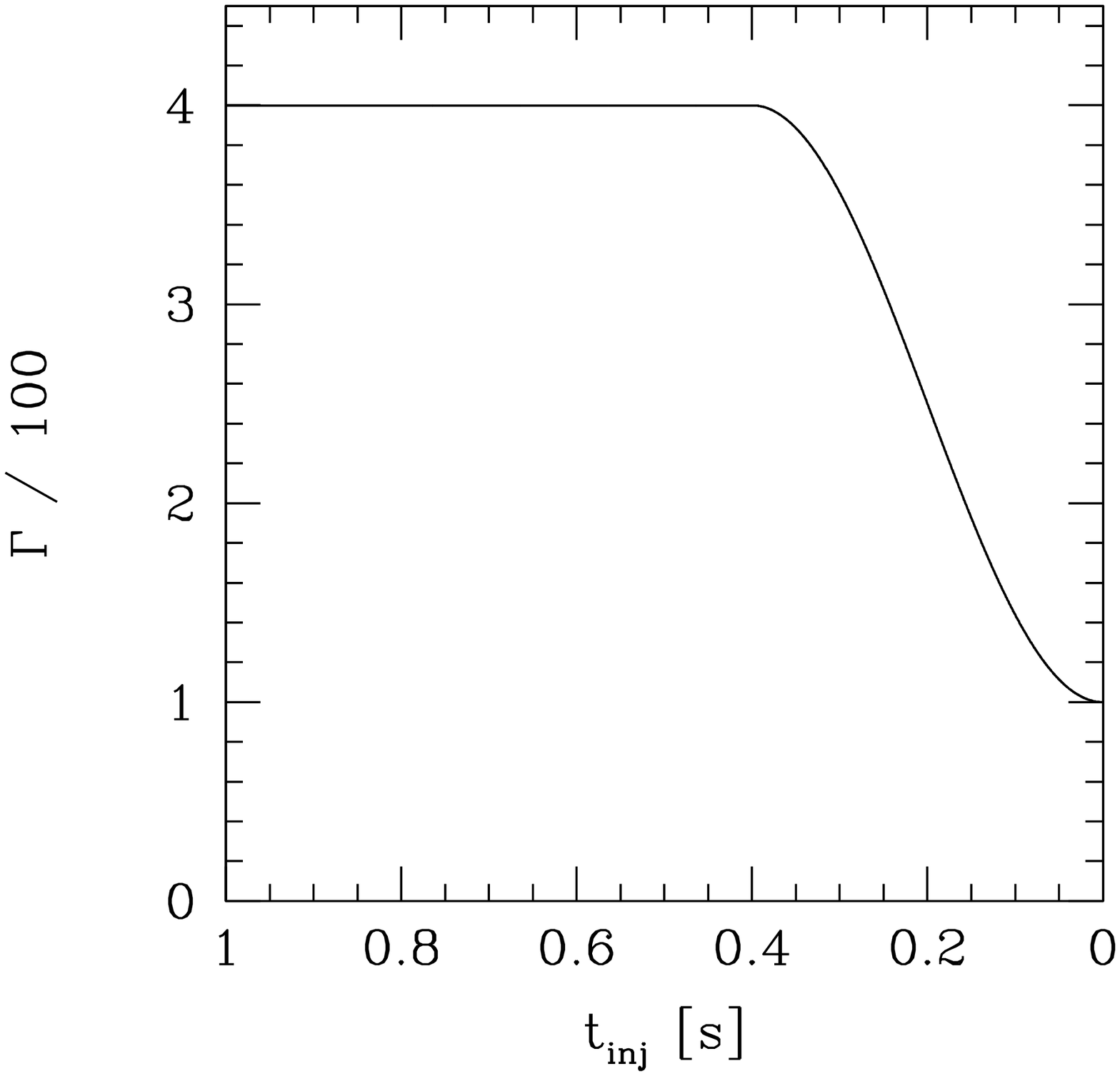} & \includegraphics[scale=0.4]{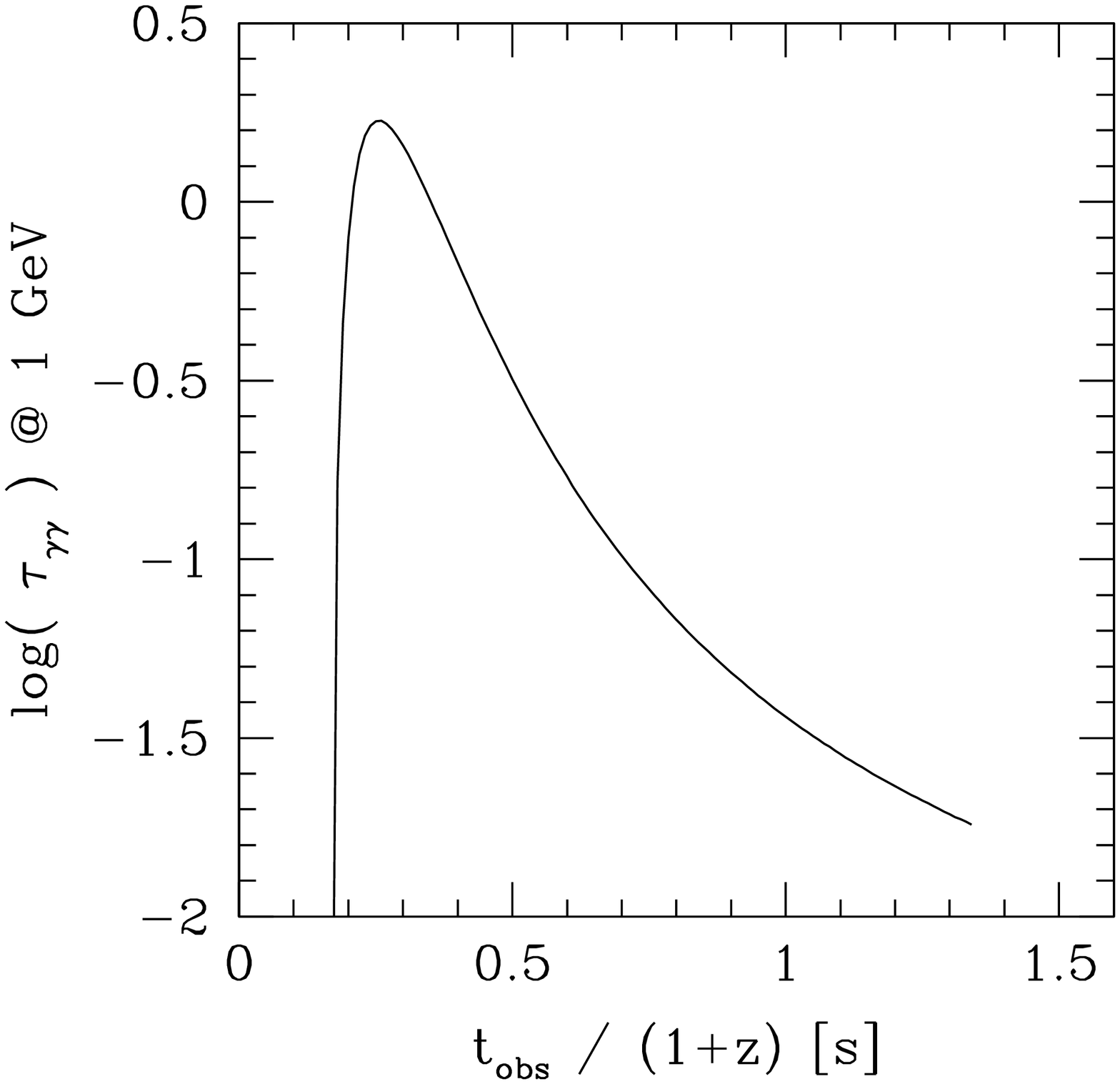} \\ 
\includegraphics[scale=0.4]{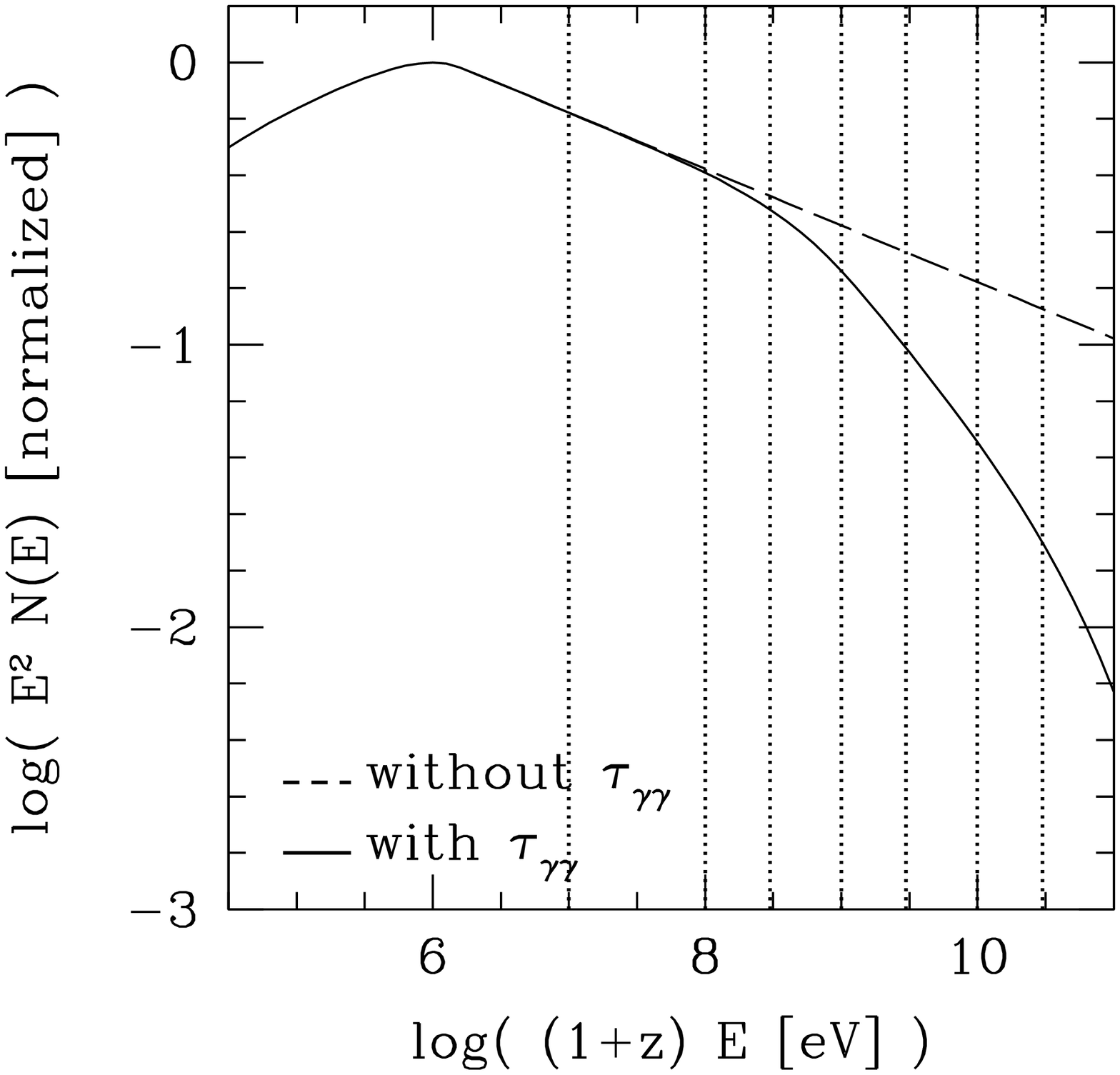} & \includegraphics[scale=0.4]{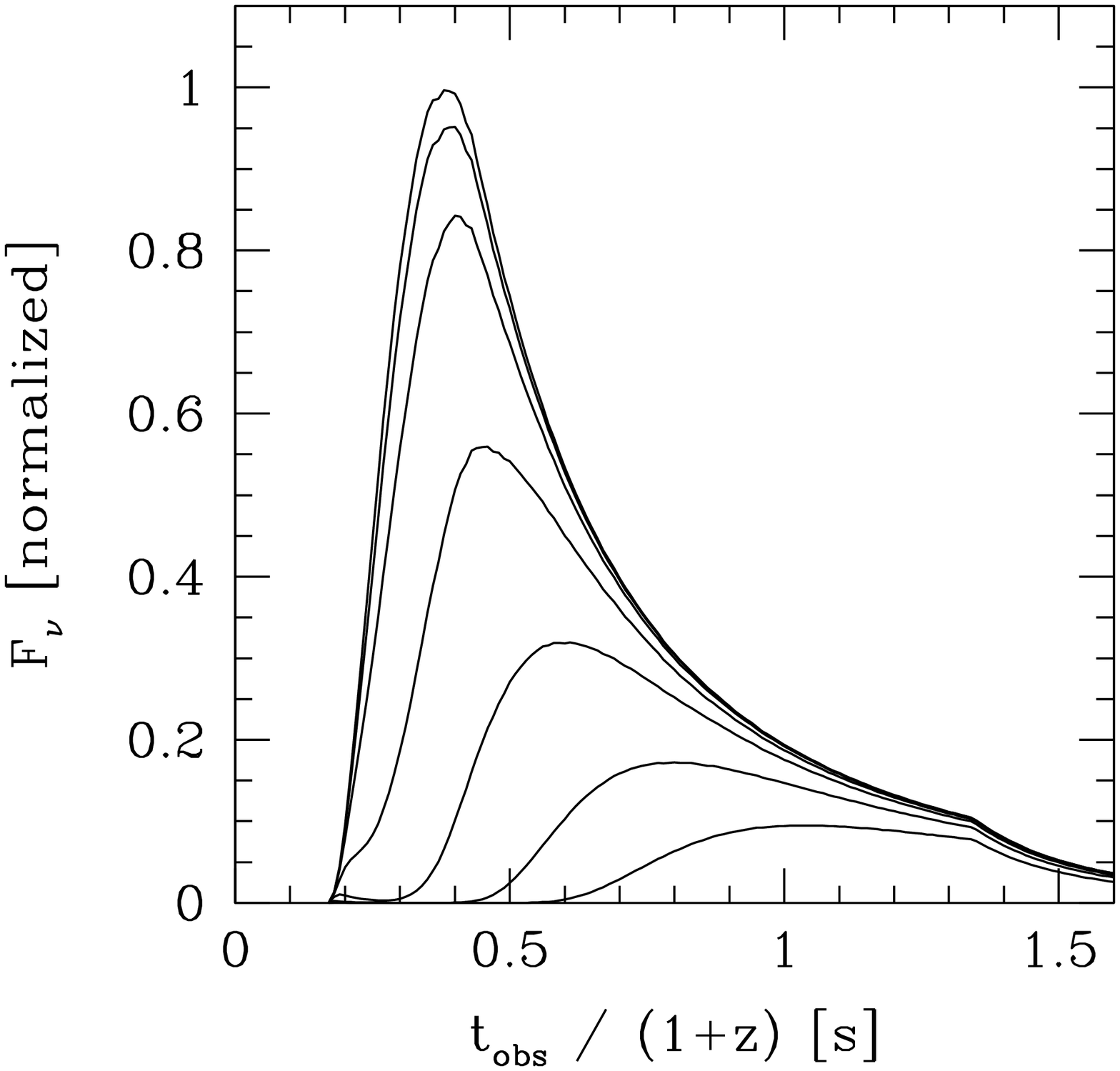}
\end{tabular}
\end{center}

\caption{\textbf{Delayed onset produced by the $\gamma\gamma$ opacity.} We consider a single pulse burst generated by internal shocks in a relativistic outflow.
\textit{Upper left panel:} initial Lorentz factor distribution of the outflow as a function of the injection time $t_\mathrm{inj}$. The kinetic power is constant and equals  $10^{55}\, \mathrm{erg.s^{-1}}$.
\textit{Upper right panel:}
the approximate opacity $\tilde{\tau}_{\gamma\gamma}$, as defined in \S\ref{sec:proxy},  seen by
1 GeV photons (source frame) 
is plotted
as a function of the observer time $t_\mathrm{obs}$. 
\textit{Lower right panel:} high-energy light curves calculated for $E_\mathrm{HE}=$ 10, 100, 300 MeV, 1 , 3 , 10, 30 GeV  (source frame). The delayed onset begins to show up 
above 1 GeV, which would correspond to 300 MeV for a GRB at $z=2$ (such as GRB 090926A) or 200 MeV for a GRB at $z=4$ (such as GRB 080916C).
\textit{Lower left panel:} time-integrated spectrum over the whole duration of the burst. The unabsorbed spectrum is plotted in dashed line. The vertical dotted lines mark the photon energies at which light curves are calculated.}
\label{fig_delay}
\end{figure*}

\subsection{Is the delayed onset of the GeV emission a signature of the $\gamma\gamma$ opacity ?}
\label{subsection_delay}

\subsubsection{Effect on the high-energy lightcurve of a time evolving $\gamma\gamma$ opacity}

The high energy emission (above 100 MeV) detected by \textit{Fermi} in a few bright GRBs often shows a delayed onset compared to the softer $\gamma$-ray emission (below 5 MeV). 
The analysis by \citet{bbzhang:2011} indicates that such a delayed onset is present in at least 7 in a sample of 17 GRBs detected by \textit{Fermi}-LAT. This feature seems to be common to long and short GRBs and its origin is debated \citep{granot:2010}. 
Among the proposed explanations \citep[see e.g.][]{zou:2009,li:2010,toma:2009}, the possibility that this delayed onset is induced by a $\gamma\gamma$ opacity temporal evolution effect has already been discussed by \citet{abdo:2009a}: 
as the shock wave producing the $\gamma$-ray emission expands to larger radii, the opacity seen by the
high energy photons evolve from an optically thick to an optically thin regime. The model developed in the present study is well appropriate to investigate this possibility in more details. 
\reffig{fig_delay} illustrates a simple example of a synthetic burst displaying a delay between the high-energy and soft $\gamma$-ray lightcurves due to $\gamma \gamma$ annihilation. The initial Lorentz factor distribution of the relativistic outflow is simple, with the formation of two shocks when the slow part of the ejecta catches up with the fast part: an internal ``forward'' shock sweeping the slow part and an internal ``reverse'' shock sweeping the fast part. The internal ``forward'' shock quickly disappears  and contributes to the observed $\gamma$-ray emission only at early times ($t_\mathrm{obs}/(1+z)<0.21$ s). On the other hand the internal ``reverse'' shock has a longer duration: it forms at 
$R=8\cdot 10^{13}$ cm and 
propagates until $R=3\cdot 10^{15}$ cm, leading to most of the 
prompt $\gamma$-ray emission. 
During the propagation of the shock, two effects contribute to a reduction of the $\gamma\gamma$ opacity for high-energy photons detected at later times: both the radius and the Lorentz factor of the shocked material are increasing. \\

This example shows that the temporal evolution of the $\gamma\gamma$ opacity can actually induce a significant delay between the high energy and the soft $\gamma$-ray emission. The synthetic burst used in \reffig{fig_gmin} to model bins 'a' and 'b' of GRB 080916C gives another example of a delayed onset at 3 GeV induced by an evolving $\gamma\gamma$ opacity. The first pulse is produced at lower radii and in lower Lorentz factor material and is therefore strongly absorbed. For this reason, it is almost suppressed in the 3 GeV lightcurve, whereas the second pulse is well visible.

\begin{figure*}
\begin{tabular}{ccc}
\includegraphics[width=0.32\textwidth]{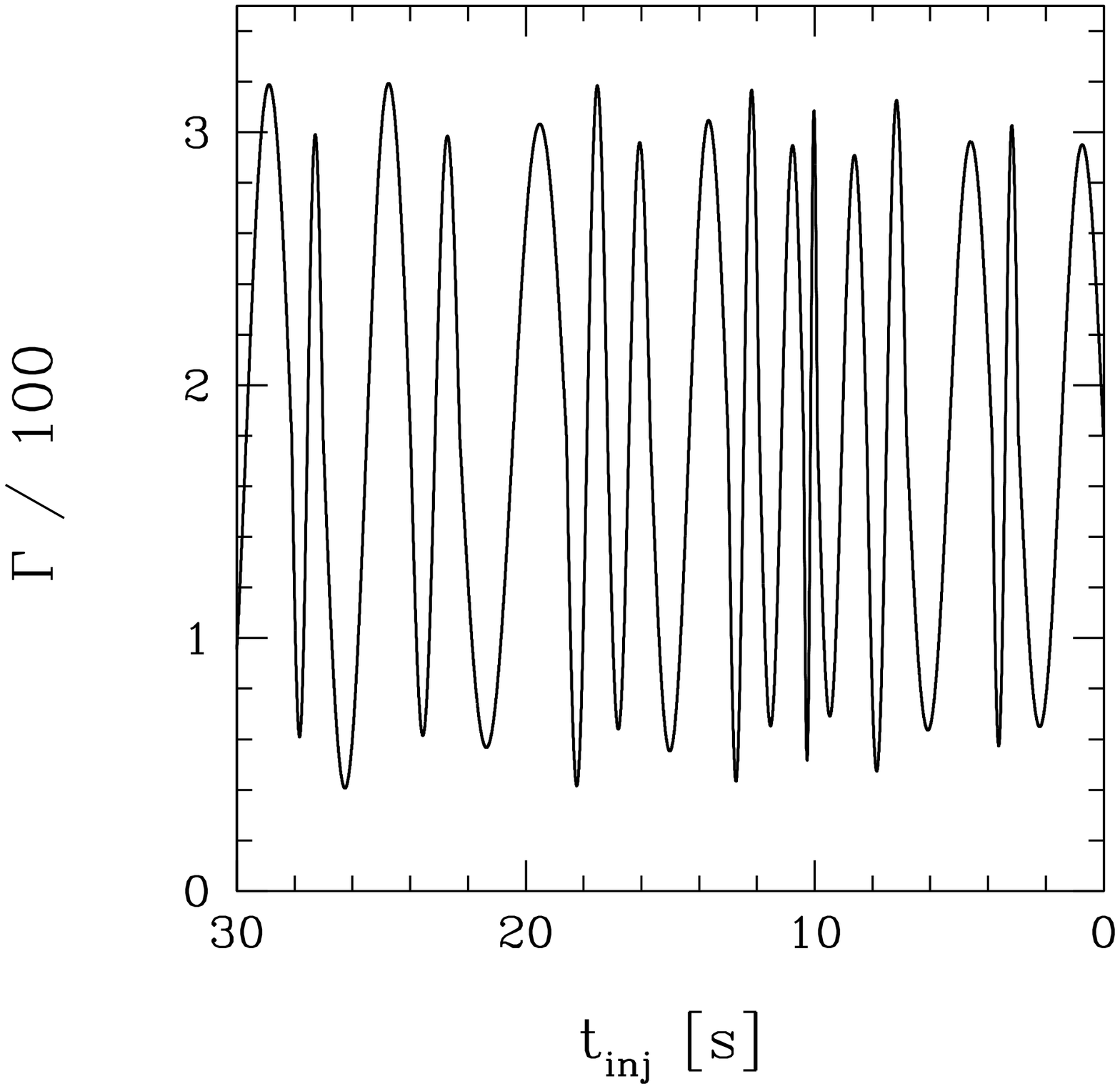} &
\includegraphics[width=0.32\textwidth]{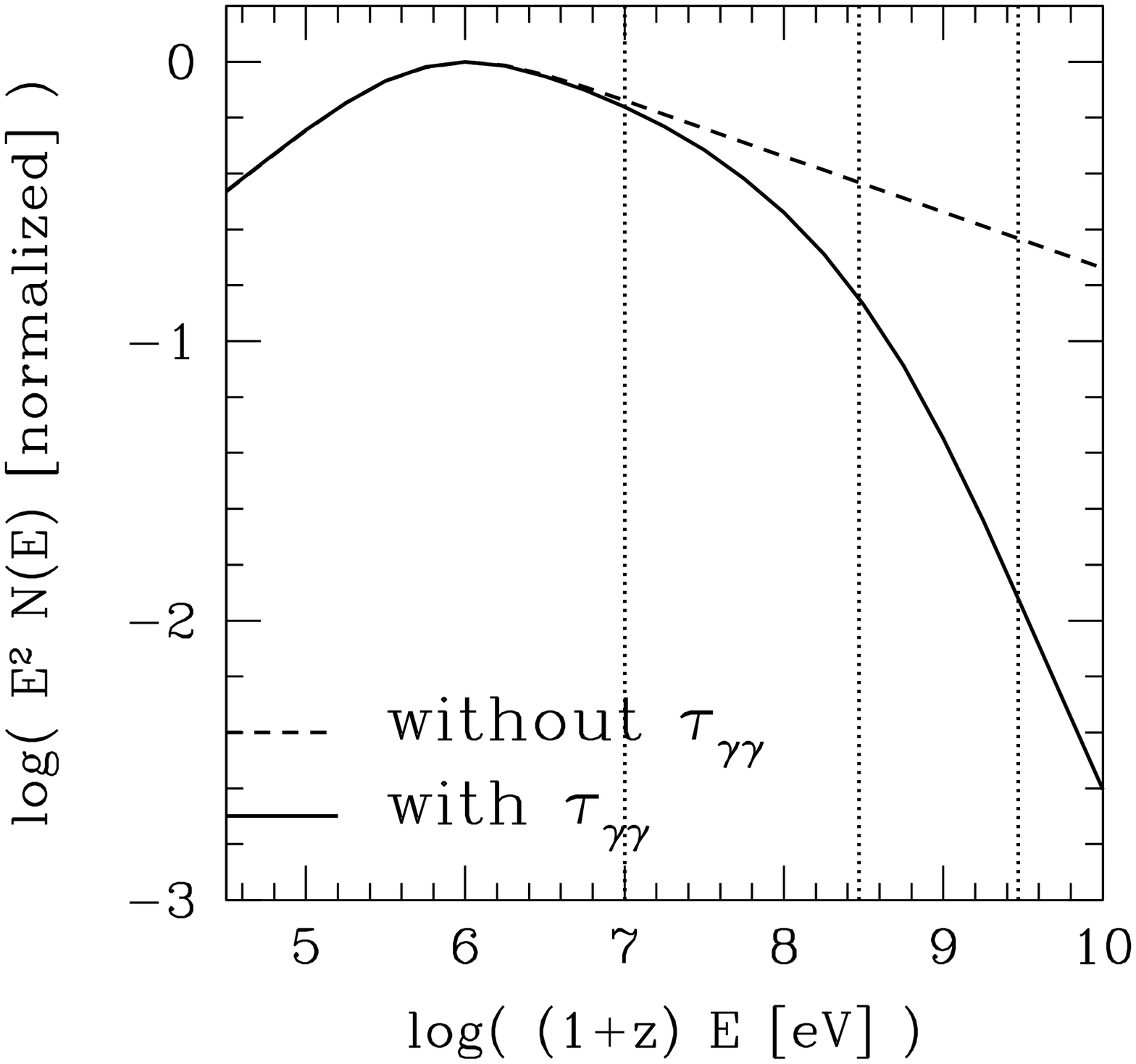} &
\includegraphics[width=0.32\textwidth]{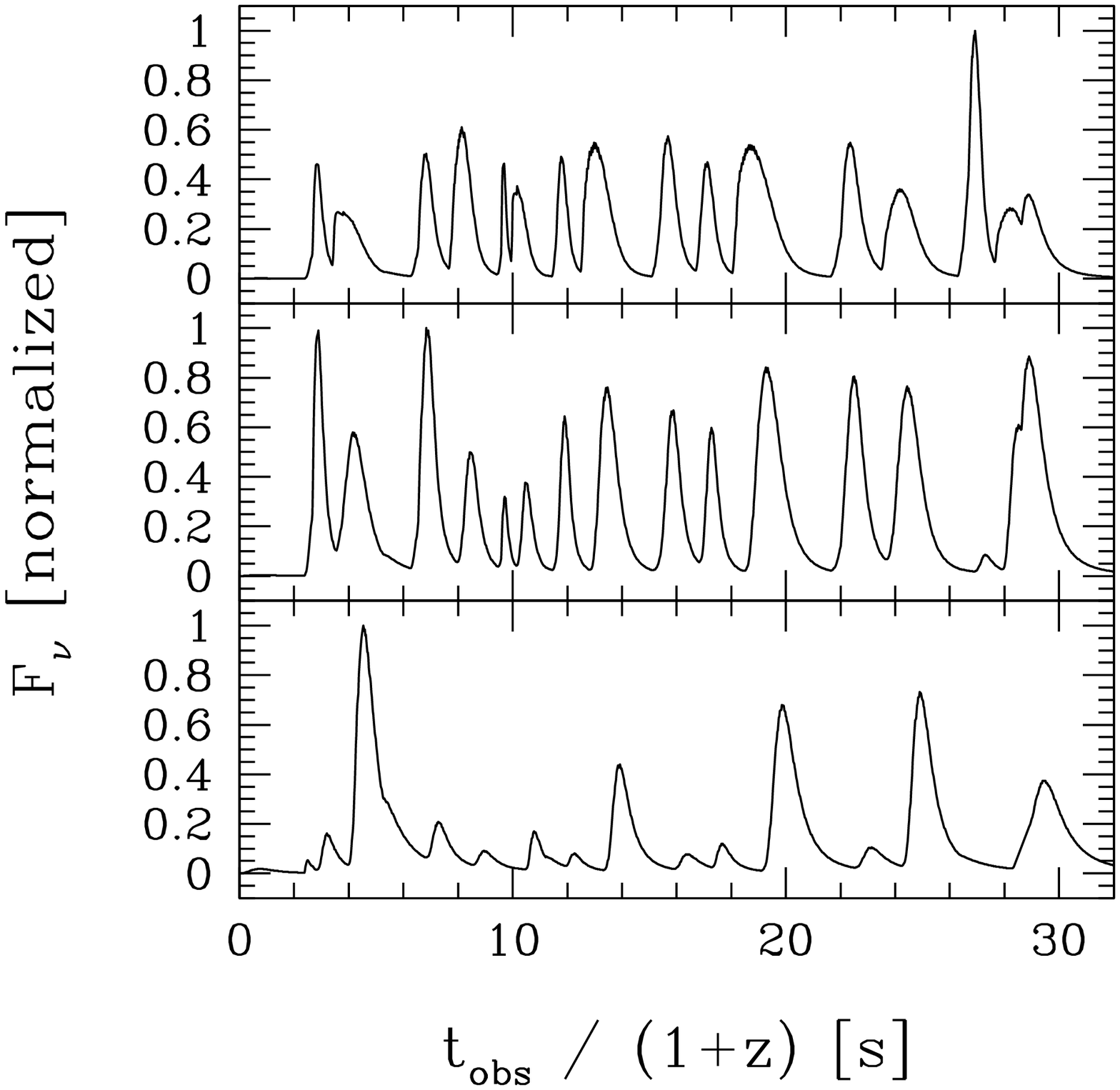} \\
\end{tabular}
\caption{\textbf{Temporal smoothing effect of the $\gamma\gamma$ opacity at high energy.} A complex multi-pulse burst has been generated by assuming an outflow ejected for 30 s with a constant kinetic power $2\cdot 10^{54}\, \mathrm{erg.s^{-1}}$ and an initial Lorentz factor varying between 50 and 350 with modulations on timescales varying from 100 ms to 4 s. 
\textit{Left:} initial distribution of the Lorentz factor in the outflow.
\textit{Center:} time-integrated spectrum. The unabsorbed spectrum is plotted in dashed line. The three vertical dotted lines indicate the energies corresponding to the three lightcurves in the next panel. 
\textit{Right:} lightcurves at 10 MeV (top), 300 MeV (middle) and 3 GeV (bottom) (source frame). Note the attenuation of the short duration pulses in the 3 GeV lightcurve.
}
\label{fig_smoothing}
\end{figure*}

\subsubsection{Characteristics of the $\gamma\gamma$ induced delay}
\textbf{Consistency with the spectral analysis.}
Testing this theoretical idea for the observed delay remains difficult as the precise specifications of the high-energy $\gamma$-ray spectrum in \textit{Fermi}-LAT GRBs are not fully understood. In some cases, the observed time integrated spectra are consistent with a unique Band function component over the GBM and LAT energy ranges (see e.g. GRB 080916C, \citealt{abdo:2009a}), but in some other cases, they show an additional component at high energy  (e.g. GRB 090510, \citealt{ackermann:2010}, GRB 090902B, \citealt{abdo:2009b} and GRB 090926A, \citealt{ackermann:2011}. 
It is then unclear whether the observed evolution of the high-energy slope $\beta$, such as the steepening from time bin 'a' to time bin 'b' in GRB 080916C, is real or related to the evolution of an additional high-energy weak component. In the former case, this spectral evolution could also contribute to the observed delay. Note that even if $\beta$ is intrinsically constant, the time evolution of the $\gamma\gamma$ opacity discussed here to explain the delayed onset of the GeV emission would also affect the measured value of the high-energy slope, due to the power-law steepening of the spectrum (see \refpar{subsection_scattering}).
Clearly, despite a very significant improvement compared to previous instruments, \textit{Fermi}-LAT cannot yet provide a detailed picture of the high energy part of the spectrum where several effects are expected to superimpose: shape of the high-energy tail of the main component, location and shape of the additional weak high energy component, location and shape of the $\gamma\gamma$ attenuation. \\

\noindent\textbf{Predicted duration of the delay.} An interesting prediction of the $\gamma\gamma$ induced delayed onset of the \textit{Fermi}-LAT emission is that
the delay should increase when observing at higher energy, 
since the transparency regime is reached at larger radii (and so at larger observer times) for photons of higher energies. However the paucity of photons above 1 GeV
 does not allow to strongly constrain this potential behavior.\\

In a simplified picture one could also expect that
the delay could not be larger that the duration of the
$\gamma$-ray pulse. Within the internal shock model, 
this would be true if each $\gamma$-ray pulse was due to a distinct 
shock wave. This is not necessarily the case as the observed lightcurves often show many superimposed pulses, which suggests that another, more complex, situation is possible, where a single propagating shock wave is responsible both for the envelope of the lightcurve and the superimposed short duration pulses, due to modulations in the kinetic power and/or the Lorentz factor of the outflow.
For instance, the synthetic GRB shown in \reffig{fig_gmin} to model bins 'a' and 'b' of GRB 080916C exhibits  two main pulses with a superimposed short time-scale variability and is nevertheless due to a single shock wave with variations in the shocked region due to the modulation in the Lorentz factor (see top left panel of \reffig{fig_gmin}). The delayed onset in the GeV lightcurve induced by the evolving $\gamma\gamma$ opacity is of the order $\simeq 5$ s, i.e. much larger that the typical duration of short duration pulses in this burst ($\simeq 0.5$ s). Such a delay is in good agreement with \textit{Fermi}-LAT observations. \\

\noindent\textbf{A high-energy precursor ?} It has been argued by \citet{granot:2008, abdo:2009a} that  a high energy emission should also be detected at the very beginning of the burst while the photon field inducing the $\gamma\gamma$ opacity is still building up. However this early transparency phase is very short (of the order of $\Delta t_\mathrm{obs} \simeq 100\, \mathrm{ms}$ for the synthetic burst presented here) and occurs when the $\gamma$-ray emission is still at the beginning of its rise. As seen in the bottom right panel of \reffig{fig_delay}, the high-energy precursor is visible at 1 GeV (source frame) but 
remains extremely weak. The relative amplitude of the precursor and the main pulse increases at higher energy but the flux becomes very low. It is then unlikely that such a high-energy precursor can be detected by current instruments.

\subsection{Temporal smoothing  at high energy}
\label{sec_smoothing}
Within the internal shock model, another expected signature of the $\gamma\gamma$ annihilation process is the smoothing of the short time-scale variability  
in the high energy light curves. Indeed, 
in complex bursts where the $\gamma$-ray emission is made of several shock waves, the opacity at a given frequency will be larger for photons emitted at smaller radii where the smallest time-scales of the observed light curves are produced.
This effect is illustrated in \reffig{fig_smoothing} where we have simulated a complex multi-pulses GRB, using a highly variable distribution of the initial Lorentz factor in the outflow. The variability has been produced on different timescales from 100 ms to 4 s, with a total duration of the ejection of 30 s. Lightcurves are plotted at three energies 10 MeV, 300 MeV and 3 GeV (source frame). The attenuation due to $\gamma\gamma$ annihilation becomes strong above a cutoff energy which evolves strongly during the burst. In particular, short pulses tend to have a lower cutoff energy, because they are produced at smaller radii. The cutoff remains however always above 10 MeV, so that the first lightcurve is un-attenuated and shows all the variability initially introduced in the Lorentz factor distribution. At higher energies, the lightcurves show less and less variability, the suppression affecting mainly the short duration pulses. 
It has been proposed that the study of the variability in the GeV lightcurve could be used to test whether GeV photons observed by \textit{Fermi}-LAT have an internal origin, or are produced by the external shock as proposed by \citet{kumar:2010,ghisellini:2010}. 
In practice, the temporal smoothing effect described here could make such a test difficult.

\subsection{Other sources of opacities in the outflow}
\label{sec_opacity_others}
Thomson scatterings by leptons in the outflow are an additional source of opacity to take into account in GRBs. Both primary electrons injected in the outflow by the central source with baryons, and secondary electron-positron pairs produced by $\gamma\gamma$ annihilation should be considered \citep[see e.g][]{lithwick:2001}.

\subsubsection{Thomson opacity from primary electrons}
\begin{figure*}
\begin{center}
\begin{tabular}{cc}
\includegraphics[scale=0.4]{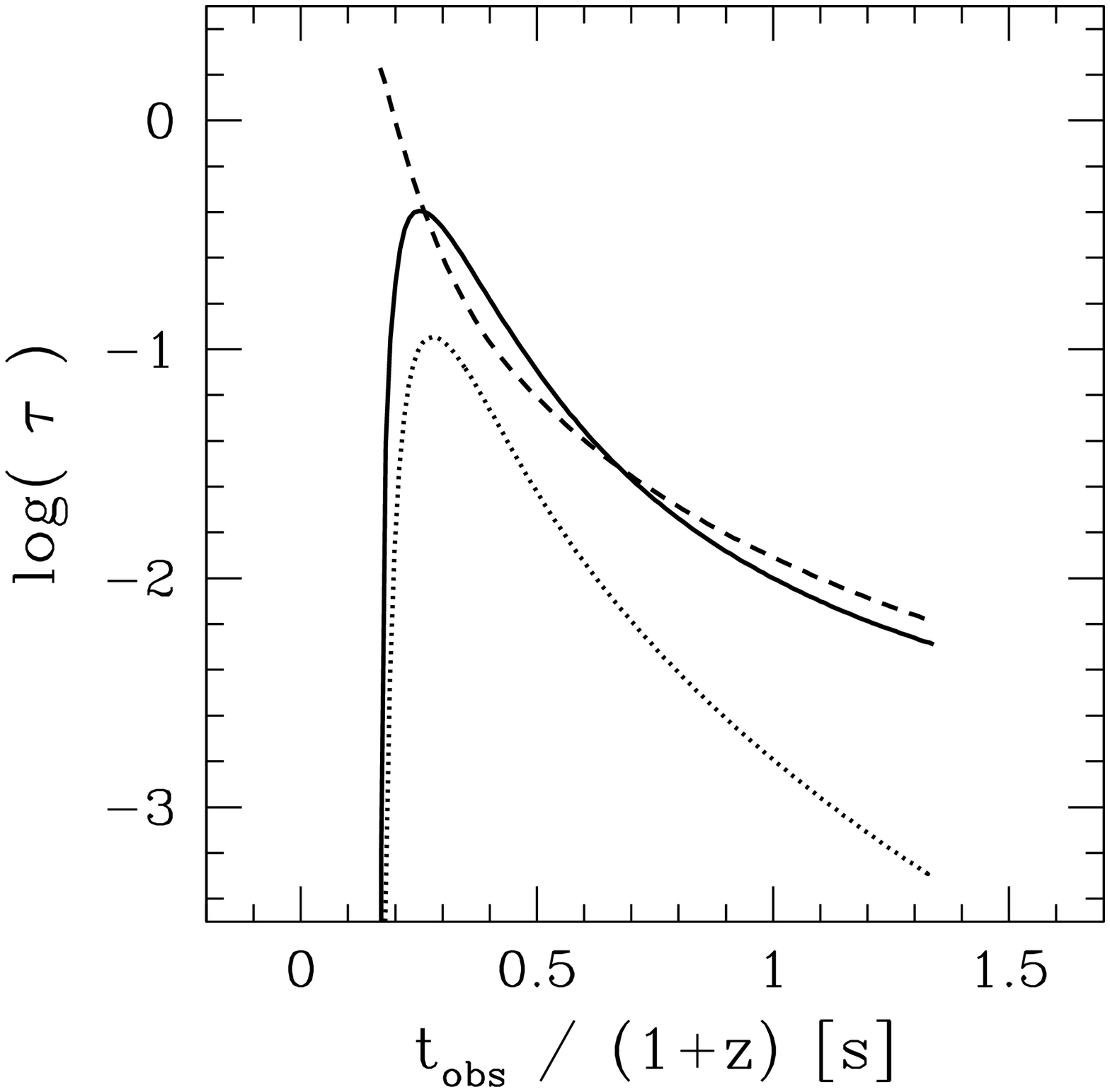} & \includegraphics[scale=0.4]{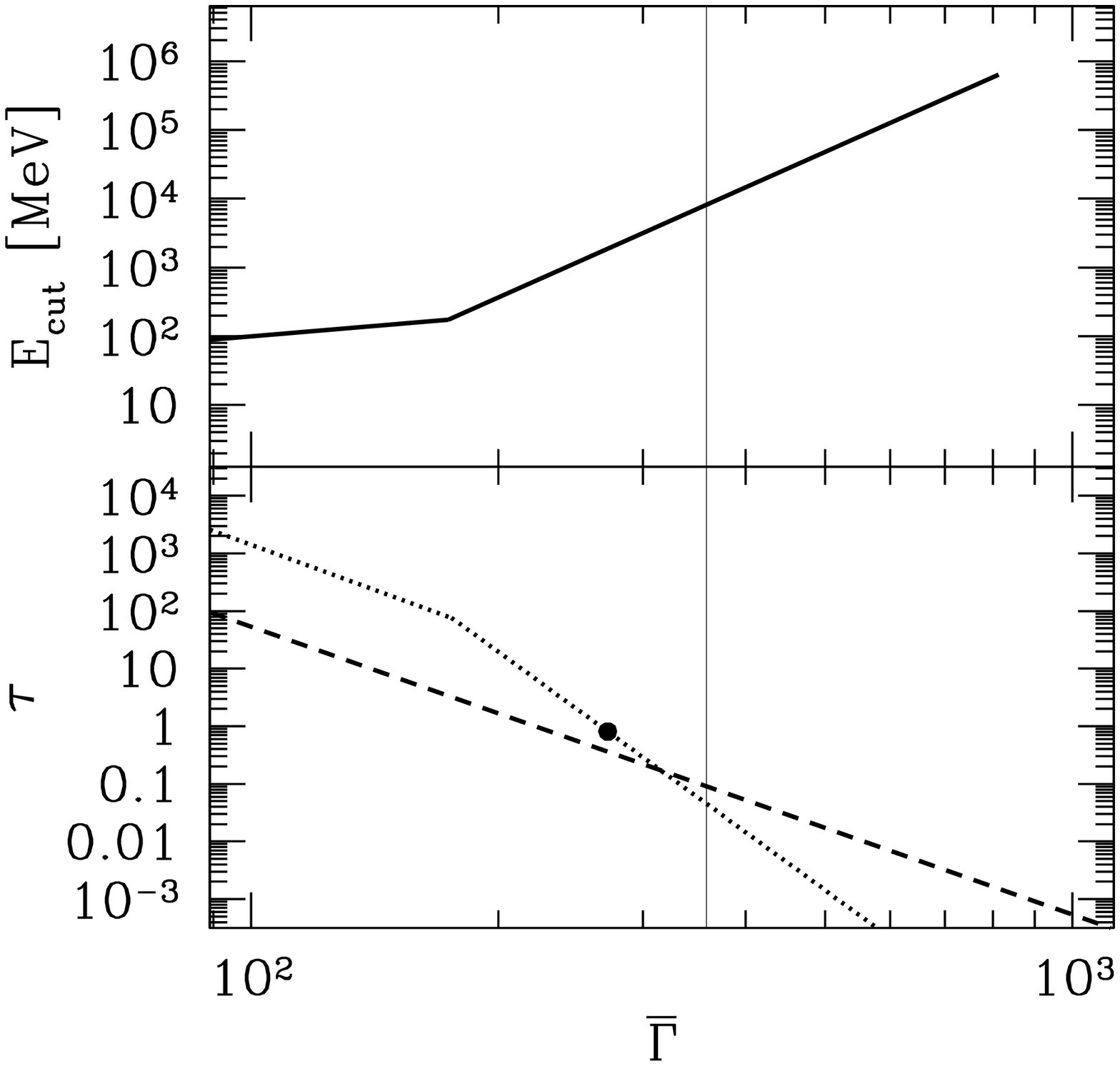} 
\end{tabular}
\end{center}

\caption{\textbf{Comparison between the $\gamma\gamma$ and the Thomson opacities.} 
We consider a single pulse burst generated by an outflow with a distribution of Lorentz factor from 125 to 500 with a similar shape as  in \reffig{fig_delay}, a total duration of 1 s and a kinetic power of $\dot{E}_\mathrm{kin}=10^{55}\, \mathrm{erg.s^{-1}}$.
 \textit{Left panel:}  opacities seen by photons emitted on the line of sight as a function of the observer time $t_\mathrm{obs}$:
$\gamma\gamma$ opacity $\tau_{\gamma\gamma}$ seen by 1 GeV photons (source frame) in solid line; Thomson opacity $\tau_\mathrm{e}$ due to the primary electrons in dashed line; Thomson opacity $\tau_{\pm}$ due to pair-produced $e^{\pm}$ in dotted line.  These three opacities are plotted only as long as the on-axis emission is active.  
\textit{Right panel:} the cutoff energy due to $\gamma\gamma$ (source frame, top) and the Thomson opacity due to primary electrons (dashed line, bottom) and secondary leptons (dotted line, bottom) are plotted as a function of the mean Lorentz factor $\overline\Gamma$, keeping the same relative shape for the initial distribution of the Lorentz factor   in the outflow. The evolutions of $E_\mathrm{cut}$ and $\tau_\pm$ show a break at $\overline{\Gamma}\simeq 170$ when $E_\mathrm{cut}=E_\mathrm{self}$ (see \refeq{tau_pairs}). 
In addition, the evolution of $E_\mathrm{cut}$ is interrupted at $\bar{\Gamma}\simeq 810$ when $\tilde{E}_\mathrm{cut}=E_\mathrm{p,0}$. 
The location of the case shown in the left panel with $\overline{\Gamma}=360$ is indicated with a vertical thin solid line. The minimum value of the Lorentz factor for the outflow to remain optically thin for the Thomson opacity (here dominated by the contribution of secondary leptons) is indicated with a big dot and corresponds to $\overline\Gamma=270$.}
\label{fig_thomson}
\end{figure*}

The Thomson opacity $\tau_\mathrm{e}$ due to the primary electrons  
can be computed in a variable outflow following the procedure described in \citet{daigne:2002} to estimate the photospheric emission expected in the internal shock model.
 \reffig{fig_thomson} presents the result of such a calculation for a synthetic GRB very close to the example used in
\reffig{fig_delay}, except for a slightly higher Lorentz factor (mean value of 360 instead of 290 chosen to have well separated contributions to the opacity in \reffig{fig_thomson}). The $\gamma\gamma$ opacity at 1 GeV (source frame) is plotted as a function of the observer time and follows the evolution which has already been described earlier. 
The Thomson opacity due to primary electrons continously decreases as a function of the observer time, due to the dilution associated to the radial expansion of the radiating shocked region.
As expected, if we exclude the very beginning of the internal shock phase, most of the internal shock propagation occurs well above the photosphere of the outflow ($\tau_\mathrm{e}\ll 1$). Note that this calculation is not as detailed as the calculation of the $\gamma\gamma$ opacity presented here, as we limit it to on-axis photons (for a detailed discussion of the off-axis photosphere, see e.g. \citealt{peer:08}). Another source of uncertainty is the initial number $Y_\mathrm{e}$ of electrons per nucleons in the outflow. It has however a limited impact, at most a factor 2 in $\tau_\mathrm{e}$. Everywhere in this paper, we adopt $Y_\mathrm{e}=0.5$.

\subsubsection{Thomson opacity from pair-produced $e^{\pm}$}
\label{subsubsection_opacity_pair}
The detailed calculation of the opacity $\tau_{\pm}$ due to secondary pairs of leptons produced by $\gamma\gamma$ annihilation is beyond the scope of this paper. We estimate $\tau_\pm$ with several simplifying assumptions : (i) we assume that 
 the Thomson regime is valid\footnote{If the Klein Nishina corrections were not negligible, the interaction cross-section would be smaller, reducing $\tau_{\pm}$.} 
which is reasonable since even if the pairs $e^{\pm}$ 
could be highly energetic right after their production, they will radiate rapidly their energy.
In practice, pairs will be mainly produced by the annihilation of photons at energy $E_\mathrm{cut}$ defined by  $\tau_{\gamma\gamma}\left(E_\mathrm{cut}\right)\simeq 1$, with seed photons interacting with a typical interaction angle $\Psi\simeq 1/\Gamma$ at the threshold of pair production, i.e. at energy 
\begin{equation}
\tilde{E}_\mathrm{cut}=\frac{\left(2\Gamma m_\mathrm{e}c^2\right)^2}{E_\mathrm{cut}}\, .
\end{equation} 
The two pair-produced leptons in this case have an energy (lab frame)
\begin{equation}
\Gamma \gamma_\pm m_\mathrm{e}c^2 \simeq \frac{1}{2}\left(E_\mathrm{cut}+\tilde{E}_\mathrm{cut}
\right)\, .
\end{equation}
For $E_\mathrm{cut}\simeq 300$ MeV to 1 GeV, and $\Gamma \simeq 300$, this leads to low values of the Lorentz factor of the leptons, $\gamma_\pm \simeq 2-4$. 
The scattering of typical $\gamma$-ray photons at energy $E_\mathrm{p,0}$ (source frame) by these leptons will occur in the Thomson regime if the photon energy in the proper frame of the lepton is small compared to the rest mass energy of the lepton, i.e. if 
\begin{equation}
w_\pm = \frac{\gamma_\pm E_\mathrm{p,0}}{\Gamma m_\mathrm{e}c^2} \ll 1\, .
\end{equation}
This condition is clearly fulfilled for the typical values of $\gamma_\pm$ estimated above;
 (ii)
it is assumed that pairs are produced at the location where the photons eventually annihilating were emitted. 
This 
approximation is 
excellent
for simple GRBs where $\gamma\gamma$ cross-interactions between different shock waves are limited, since in such simple cases most of the high energy photons annihilate very close to their emission location with seed photons produced by the same shock wave. 
In more complex situations, this approximation is less justified.\\

The value of $\tau_\pm$ is plotted as a function of the observer time in \reffig{fig_thomson}. As expected, it decreases with time, due both to the decrease of the $\gamma\gamma$ opacity leading to a decrease of the pair production, and to dilution due to the radial expansion. \\

Note that we do not include in this study the contribution to radiation of secondary leptons, such as pairs produced by $\gamma\gamma$ annihilation, or additional pairs produced by cascades if the first generation of pairs produce new high energy photons that can annihilate 
(the cascade process would stop when the secondary photons have too low energies to produce new pairs). As the fraction of the radiated energy in annihilating photons is low in all examples presented here (the $\gamma\gamma$ cutoff being at high energy), the neglected contribution remains weak.

\subsubsection{What is the dominant opacity ?}
\label{sec_otheropacities}
In the example shown in \reffig{fig_thomson}, all opacities ($\gamma\gamma$ opacity at 1 GeV and Thomson opacities due to primary electrons and pair-produced leptons) are decreasing with time. For most of the burst, especially at the maximum of the lightcurve at $t_\mathrm{obs}\simeq 0.4$ s (see \reffig{fig_delay}), the dominant opacity is the $\gamma\gamma$ opacity at 1 GeV, which means that, assuming that a GeV photons is detected in this burst,  the most constraining limit on the minimum Lorentz factor in the outflow will be provided by the $\gamma\gamma$ opacity limit discussed earlier. However, the dependency of these three opacities with the Lorentz factor of the outflow is not the same. Then, depending on the maximum energy of detected photons, the limit could be provided by any of the three constraints, as discussed in \citet{lithwick:2001}.\\

In the right panel of \reffig{fig_thomson}, we have plotted the value of the Thomson opacity due to primary electrons and secondary leptons and the cutoff energy due to the $\gamma\gamma$ opacity as a function of the mean Lorentz factor in the outflow (other parameters are kept constant: see caption). We find that these values are in good agreement with the following approximate formulae:
\begin{enumerate}
\item \textit{Opacity due to $\gamma\gamma$ annihilation:} 
\refeq{eq:C1}; 

\item \textit{Thomson opacity due to primary electrons:} from \citet{daigne:2002}, 
this opacity
is approximatively given by 
\begin{equation}
\tau_\mathrm{e}\simeq C_2 \frac{Y_\mathrm{e} \sigma_\mathrm{T} \mathcal{E}_\mathrm{rad}}{8\pi m_\mathrm{p} c^4 \Delta t_\mathrm{var}^2 f_\gamma}  
\Gamma_\mathrm{inf}^{-2}\overline{\Gamma}^{-3}\, ,
\label{tau_thomson}
\end{equation}
where  $C_2\simeq 0.2$ is a correction factor obtained by comparing the approximate formula and the numerical calculation, and  $f_\gamma$ is the efficiency of the prompt emission phase, so that the kinetic energy of the flow is given by $\mathcal{E}_\mathrm{rad}/f_\gamma$. For the example used in \reffig{fig_thomson}, this efficiency equals $f_\gamma\simeq 0.02$;
\item \textit{Thomson opacity due to pair-produced leptons:} 
this opacity
can be estimated in a similar way as in \citet{lithwick:2001}, taking into account the correction factors for the $\gamma\gamma$ opacity: we define $E_\mathrm{self}=2\overline{\Gamma} m_\mathrm{e}c^2$ as the energy (source frame) of photons that self-annihilate with an interaction angle $\Psi=1/\overline{\Gamma}$. Then the Thomson opacity due to pairs is approximatively given by
\begin{equation}
\tau_\pm \simeq C_3 \times \left\lbrace
\begin{array}{cl}
-\frac{K'_0 }{1+\beta}
\left(\tau_* \Gamma_\mathrm{inf}^{-3}\overline{\Gamma}^\beta\right)^2
& \mathrm{if}\, E_\mathrm{cut}>E_\mathrm{self}\\
-\frac{2^{1+\beta}}{1+\beta}\tau_*
\Gamma_\mathrm{inf}^{-2}\overline{\Gamma}^{\beta-1}
 & \mathrm{if}\, E_\mathrm{cut}<E_\mathrm{self}
\end{array}
\right.\, ,
\label{tau_pairs}
\end{equation}
with
\begin{equation}
\tau_* = \frac{A_0 \sigma_\mathrm{T}\mathcal{E}_\mathrm{rad}}{4\pi \left(c \Delta t_\mathrm{var}\right)^2 E_\mathrm{p,0}}\left(\frac{m_\mathrm{e}c^2}{E_\mathrm{p,0}}\right)^{1+\beta}\, ,
\end{equation}
and where $C_3\simeq 3$ is a correction factor introduced from the comparison with the detailed calculation. The cutoff energy due to $\gamma\gamma$ annihilation can be estimated from \refeq{eq:C1}:
\begin{equation}
E_\mathrm{cut}\simeq m_\mathrm{e}c^2 \left( K'_0
\tau_*\right)^{\frac{1}{1+\beta}} 
\Gamma_\mathrm{inf}^{\frac{-4}{1+\beta}}\overline{\Gamma}^{2}
\, .
\end{equation}
The condition $E_\mathrm{cut}>E_\mathrm{self}$ is equivalent to 
\begin{equation}
\frac{K'_0}{2^{1+\beta}}
\tau_* 
\Gamma_\mathrm{inf}^{-4} \overline{\Gamma}^{1+\beta}
< 1 \, .
\end{equation}
\end{enumerate} 
The approximate formulae for $\tau_{\gamma\gamma}$ and $\tau_\pm$ assume a power-law spectrum, as in \S\ref{sec_approx_formula}. Therefore, these equations are valid if $E_\mathrm{p,0}$ is understood as the energy above which the power-law with slope $\beta$ is observed, $\mathcal{E}_\mathrm{rad}$ is the radiated energy above $E_\mathrm{p,0}$ over a timescale $\Delta t_\mathrm{var}$ and $E_\mathrm{cut}$ and $E_\mathrm{self}$ are above $E_\mathrm{p,0}$. For a comparison with \textit{Fermi observations}, $\mathcal{E}_\mathrm{rad}$ can be estimated from the measured photon fluence above $E_\mathrm{p,0}$ (see \refeq{eq:erad_fluence}). The variability timescale can be estimated from the observed variability timescale in the GBM lightcurve.  
In addition, corrections due to the redshift $z$ of the source should also be taken into account. \\
 
\begin{figure*}
\begin{center}
\begin{tabular}{ccc}
(a) $\beta=-2.2$ and $E_\mathrm{p,0}=1\, \mathrm{MeV}$
& (b) $\beta=-2.5$ and $E_\mathrm{p,0}=1\, \mathrm{MeV}$
& (c) $\beta=-2.5$ and $E_\mathrm{p,0}=100\, \mathrm{keV}$\\
\\
\includegraphics[width=0.32\textwidth]{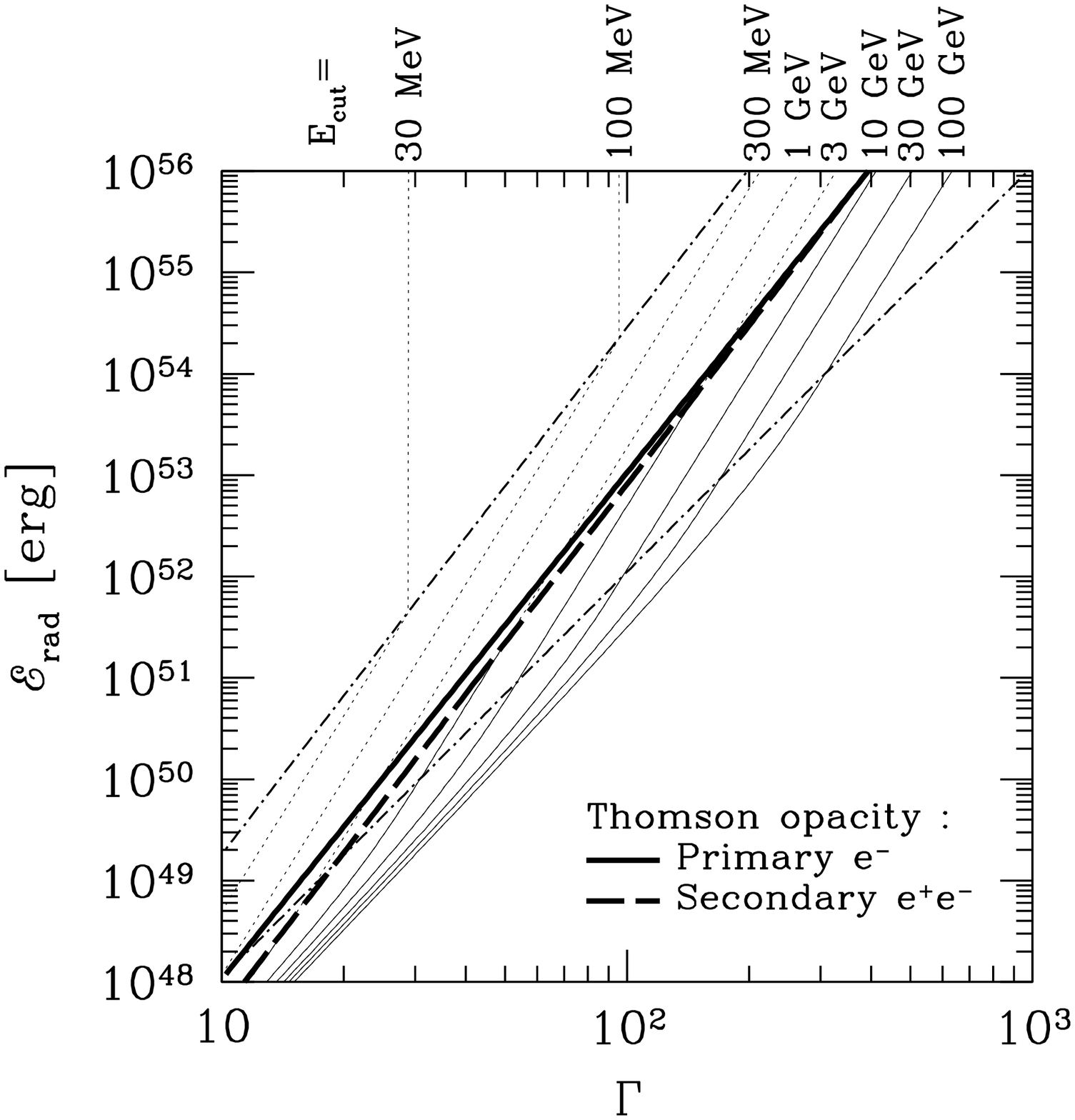} &
\includegraphics[width=0.32\textwidth]{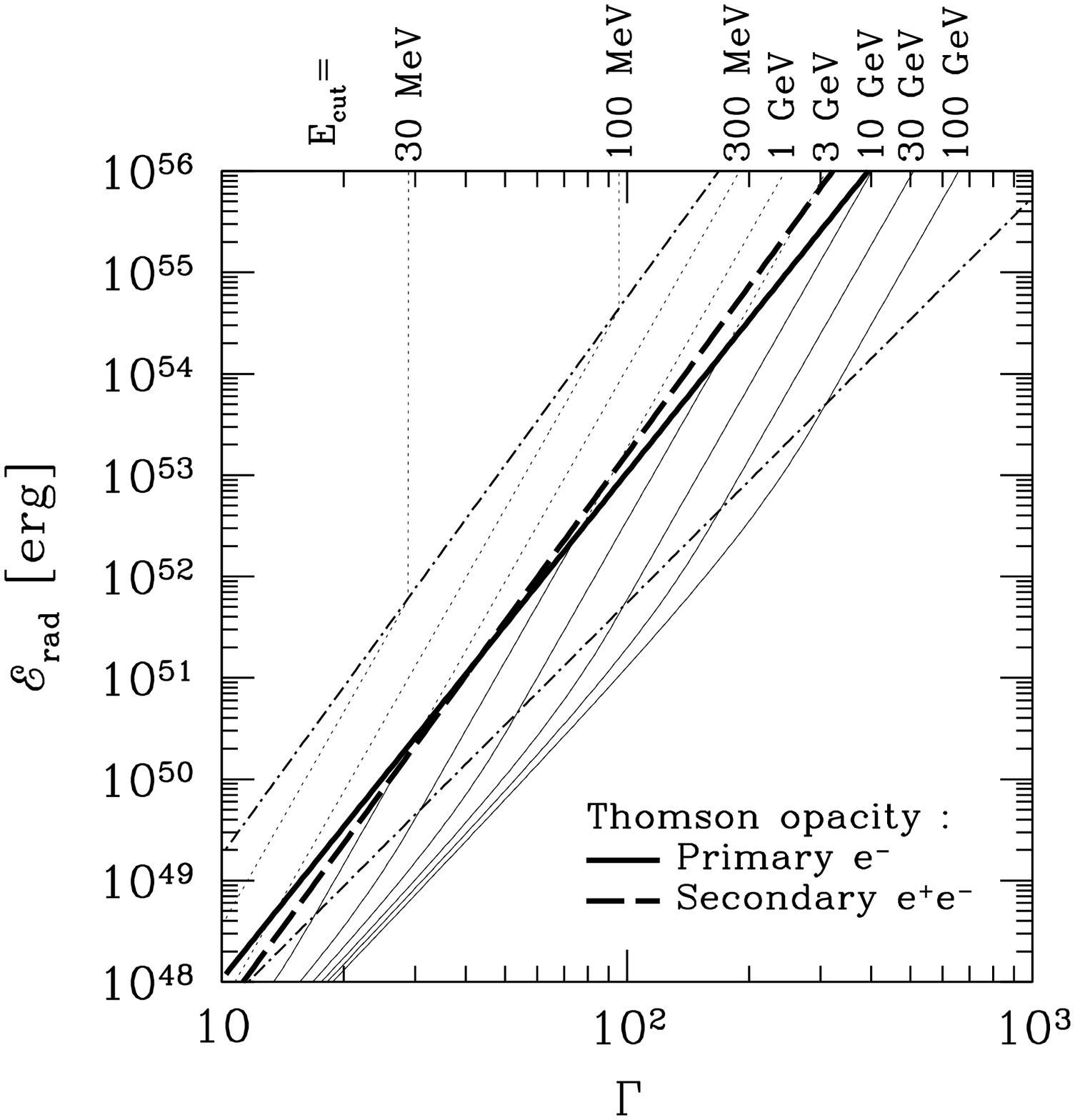} &
\includegraphics[width=0.32\textwidth]{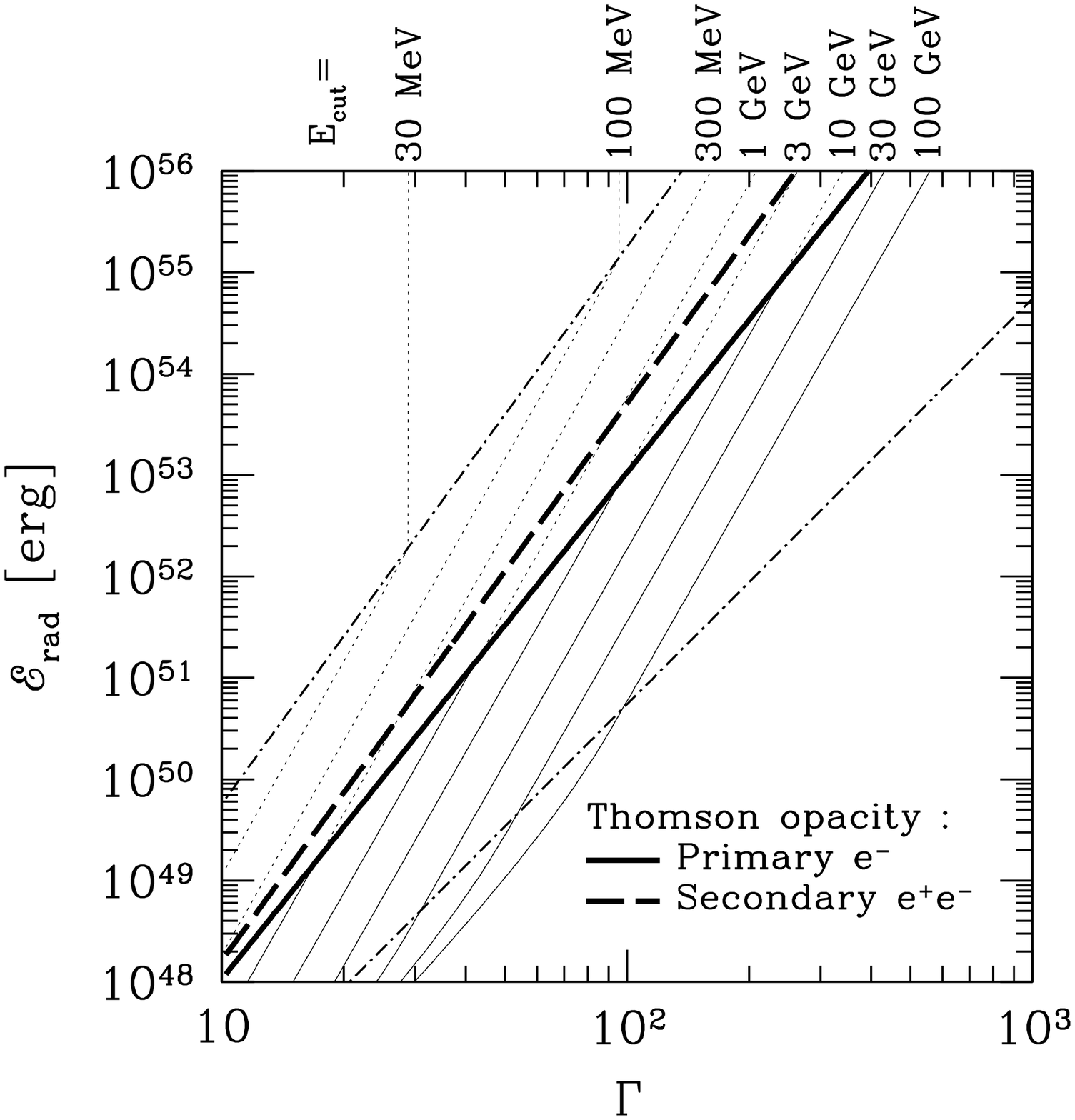} \\
\end{tabular}
\end{center}
\vspace*{-2ex}

\caption{\textbf{Opacity constraints on the minimum Lorentz factor in GRB outflows.} The limits for the outflow to be optically thin for the Thomson opacity due to primary electrons (solid thick line) and secondary leptons (dashed thick line) are plotted in the radiated energy--Lorentz factor plane, using \refeq{gmin_e} and \refeq{gmin_ep} with $\Delta t_\mathrm{var}=1\, \mathrm{s}$ and $f_\gamma=0.02$. Note that $\mathcal{E}_\mathrm{rad}$ in the y-axis is the total radiated energy on a timescale $\Delta t_\mathrm{var}$. When necessary, the radiated energy above $E_\mathrm{p,0}$ is deduced from $\mathcal{E}_\mathrm{rad}$, $\alpha$ and $\beta$. Lines of constant cutoff energy due to $\gamma\gamma$ annihilation are also plotted in thin line for different values of $E_\mathrm{cut}$ using \refeq{gmin_gg}: the dotted part of the line corresponds to the region where the outflow is optically thick for the Thomson opacity. Two dot-dashed lines indicate the limits where $E_\mathrm{cut}=E_\mathrm{self}$ and $\tilde{E}_\mathrm{cut}=E_\mathrm{p,0}$.
For $\tilde{E}_\mathrm{cut}<E_\mathrm{p,0}$, the lines have been extended using a low-energy slope $\alpha=-1.1$. For $E_\mathrm{cut}>E_\mathrm{self}$, we consider that the cutoff saturates at $E_\mathrm{self}=2\Gamma m_\mathrm{e}c^2$. The three panels correspond to three different sets of values of $\beta$ and $E_\mathrm{p,0}$.}
\label{fig_opacity_gmin}
\end{figure*}

We have checked that all simulations presented in this paper (including the model of GRB 080916C in \reffig{fig_gmin}) corresponds to situations where the GRB outflow during the emission is optically thin for Thomson scatterings by primary electrons or pair-produced leptons. However, from the scaling laws listed above, it appears that 
the limit on the minimum Lorentz factor in the outflow could be constrained by any of the three opacities, depending on the highest energy photon detected in a given burst \citep{lithwick:2001}. With our corrected estimates of the three opacities, we obtain the following limits, assuming $\Gamma\simeq\Gamma_\mathrm{inf}\simeq\overline\Gamma$ to allow a comparison with previous single-zone studies:
\begin{enumerate}
\item \textit{Opacity due to $\gamma\gamma$ annihilation.} From \refeq{eq:C1}, if the highest energy photon has an energy $E_\mathrm{max}$ (source frame), the minimum Lorentz factor is given by
\begin{equation}
\Gamma_\mathrm{min,\gamma\gamma}\simeq\left[
C_1 K_0 \frac{A_0 \sigma_\mathrm{T} \mathcal{E}_\mathrm{rad}}{4\pi\left(c \Delta t_\mathrm{var}\right)^2 E_\mathrm{p,0}}
\right]^{\frac{1}{2(1-\beta)}} \left(\frac{\left(m_\mathrm{e}c^2\right)^2}{E_\mathrm{max} E_\mathrm{p,0}}\right)^{\frac{1}{2}\frac{1+\beta}{1-\beta}}\, .
\label{gmin_gg}
\end{equation}
If the cutoff energy $E_\mathrm{cut}$ is identified, the previous equation can be used with $E_\mathrm{cut}$ instead of $E_\mathrm{max}$ to estimate the Lorentz factor;
\item \textit{Thomson opacity due to primary electrons.} From \refeq{tau_thomson}, the minimum Lorentz factor for transparency is given by
\begin{equation}
\Gamma_\mathrm{min,e}\simeq\left(C_2 \frac{Y_\mathrm{e} \sigma_\mathrm{T} \mathcal{E}_\mathrm{rad}}{8\pi m_\mathrm{p} c^2 \left(c\Delta t_\mathrm{var}\right)^2 f_\gamma}\right)^{\frac{1}{5}}\, ;
\label{gmin_e}
\end{equation}
\item \textit{Thomson opacity due to pair-produced leptons.} From \refeq{tau_pairs}, the minimum Lorentz factor for transparency is given by
\begin{equation}
\Gamma_\mathrm{min,\pm}\simeq\left\lbrace\begin{array}{cl}
\left(C_3 C_1 \frac{-K_0}{1+\beta}\right)^{\frac{1}{2(3-\beta)}} \tau_*^{\frac{1}{3-\beta}} & \mathrm{if}\, \frac{C_1}{C_3}K_0 < -\frac{2^{2+2\beta}}{1+\beta}\\
\left(C_3 \frac{-2^{1+\beta}}{1+\beta}\right)^{\frac{1}{3-\beta}} \tau_*^{\frac{1}{3-\beta}} & \mathrm{if}
\frac{C_1}{C_3}K_0 > -\frac{2^{2+2\beta}}{1+\beta}\\
\end{array}\right.\, .
\label{gmin_ep}
\end{equation}

\end{enumerate}

We have plotted in \reffig{fig_opacity_gmin} the three constraints in a plane Lorentz factor -- radiated energy,  based on \refeq{gmin_gg}, \refeq{gmin_e} and \refeq{gmin_ep}. In panel (a), for $\beta=-2.2$ and $E_\mathrm{p,0}=1\, \mathrm{MeV}$ (source frame), the minimum value of the Lorentz factor is determined by the $\gamma\gamma$ opacity as long as the cutoff energy is above $\simeq$1--3 GeV (the precise limit depends on the radiated energy $\mathcal{E}_\mathrm{rad}$). For a reasonable range in the radiated energy, $\mathcal{E}_\mathrm{rad}=10^{49}\to 10^{55}$ erg, and for cutoff energies below 100 GeV, this leads to minimum Lorentz factors from 21 to 440. If the cutoff energy due to $\gamma\gamma$ annihilation is below 1 GeV, then the minimum Lorentz factor is determined by the Thomson opacity due to leptons. In the present case, it is dominated by the opacity from pair-produced leptons. This limit goes from $\Gamma=18$ for $\mathcal{E}_\mathrm{rad}=10^{49}\, \mathrm{erg}$ to 250 for $\mathcal{E}_\mathrm{rad}=10^{55}\, \mathrm{erg}$. Notice that the cutoff energies listed here are measured in the source frame. In the observer frame, the limit at 1-3 GeV is located in the low-energy part of the LAT spectral range (100 MeV--1 GeV) for usual GRB redshifts.  In panel (b), we study the impact of the high-energy spectral slope $\beta$ by using
 $\beta=-2.5$ and $E_\mathrm{p,0}=1\, \mathrm{MeV}$ (source frame). The main effect is to increase the region where the $\gamma\gamma$ opacity is dominant: for a cutoff energy above 1 GeV for large $\mathcal{E}_\mathrm{rad}$ and above  300 MeV for lower $\mathcal{E}_\mathrm{rad}$. For lower cutoff energies, the Thomson opacity is now dominated by secondary leptons at low $\mathcal{E}_\mathrm{rad}$ and primary electrons at high $\mathcal{E}_\mathrm{rad}$. The corresponding minimum Lorentz factor is not very different from the previous case. In panel (c), we study the influence of the peak energy by using $\beta=-2.5$ and $E_\mathrm{p,0}=100\, \mathrm{keV}$. The main effect is that the Thomson opacity is now always dominated by the contribution of primary electrons. We have also studied the effect of a change in the variability timescale: it mainly moves the limits towards higher Lorentz factors when the variability timescale decreases.

\subsubsection{$\gamma\gamma$ opacity due to the interaction with the circumburst medium} 
\label{sec_belo}
Another mechanism which could be a cause of $\gamma\gamma$ annihilation in GRBs are scatterings of $\gamma$-ray photons by electrons in the circumburst medium \citep{thompson:2000}. This process has been investigated in details by \citet{beloborodov:2002}. After their emission in the jet, $\gamma$-ray photons
will progressively overtake the relativistic outflow due to a difference $\sim c/2\Gamma^2$ in velocity, and 
sweep the ambient medium which has not been shocked yet.
Under appropriate conditions, this can lead to an efficient pre-acceleration of the circumburst medium, as photons deposit energy and momentum via Compton scatterings. 
The small fraction of scattered photons tends to be isotropized by the scatterings, which results in efficient $\gamma\gamma$ annihilation with unscattered photons (due to the possibility of head-on interactions), and therefore to the pair-enrichment of the circumburst medium. Both Compton scatterings and $\gamma\gamma$ annihilations are strongly reduced when the radius increases (dilution) and when the motion of the medium becomes relativistic (beaming).
The dynamics of the deceleration (which is delayed) and the emission of the shocked external medium (which is lepton rich) are modified by this mechanism : first the relativistic outflow is expanding in an empty cavity, as the circumburst medium is moving faster; then the external shock forms but is weaker due to the relativistic motion of the pre-accelerated material ; at larger distances there is no  pre-acceleration any more but the circumburst medium remains pair-enriched; finally the standard deceleration is recovered.\\

Such a mechanism may affect the prompt high-energy spectrum of the GRB, due to the $\gamma\gamma$ annihilations between scattered and unscattered $\gamma$-ray photons. As shown by \citet{beloborodov:2002}, when effective, these annihilations lead to a break at 5--50 MeV in the prompt GRB spectrum.  
For a
typical high density wind-like medium, the annihilations rate is large enough to produce a break if
the prompt GRB photons enter the ambient medium at a radius $R_\mathrm{esc}$ lower than a characteristic radius $R_{\gamma\gamma, \mathrm{ext}}$ given by (see equation (70) of \citealt{beloborodov:2002}):
\begin{equation}
R_{\gamma \gamma,\mathrm{ext}} \simeq \left(6\cdot10^{14}\to 10^{15}\right) \left( \frac{ E_{\mathrm{rad},53}}{\mu_e} \right)^{\frac{3}{6-2\beta}} 
\left( \frac{A}{A_*} \right)^{-\frac{2\beta}{6-2\beta}} \mathrm{cm}\, ,
\label{eqn_wind_andrei}
\end{equation}
where $E_{\mathrm{rad},53}$ 
is the radiated prompt $\gamma$-ray energy of the GRB in unit of $10^{53}$ erg, $\mu_\mathrm{e}$ is the mean particle mass per electron in the ambient medium ($\mu_\mathrm{e}=1$ for ionized hydrogen),  
$A$ is the normalization of the wind density profile (see \refeq{eqn_external_density}) and $A_* = 5 \cdot 10^{11}\,\mathrm{ g \cdot cm^{-1}} $ is the value of $A$ expected for 
a typical Wolf-Rayet star wind. \\

We note that the escape radius $R_\mathrm{esc}$ of the prompt photons
 is usually larger than their emission radius $R_\mathrm{e}$, except for photons radiated at the very front of the jet. Typically, $R_\mathrm{esc}\simeq R_\mathrm{e}+2\overline{\Gamma}^2 \Delta_\mathrm{e}$, where $\Delta_\mathrm{e}$ is the width of the outflow located in front of the emission region. Taking into account this effect, we compute the escape radius  
of all emitted photons in the synthetic burst used to model GRB 080916C in  \reffig{fig_gmin}.
We obtain a minimum value of $R_\mathrm{esc} \simeq 10^{15}$ cm. 
For  $\beta = -2.2$, $E_{\mathrm{rad},53} = 88$ (as in GRB 080916C), and $\mu_\mathrm{e}=1$, we get $R_{\gamma \gamma , \mathrm{ext}} \simeq \left(2.2\to3.6\right)\cdot 10^{15} \left( A/A_* \right)^{0.423}$ cm. Therefore, the condition $R_\mathrm{esc} \ga R_{\gamma\gamma,\mathrm{ext}}$ to avoid a break in the prompt spectrum is fulfilled as long as $A/A_* \la 0.2\to0.05$. 

\section{Consequences of distinct emission regions for MeV and GeV photons}
\label{sec_2zones_model}
\subsection{Are GeV and MeV photons produced in the same region ?}
It has been proposed in several recent studies that the delayed onset and/or the long-lasting tail of the high energy emission could be an evidence of two different regions for the emission of MeV and GeV photons.
An extreme version is the scenario proposed by \citet{kumar:2010} and \citet{ghisellini:2010}
where the whole GeV emission (prompt and long lasting) is produced by the external shock during the early deceleration of the relativistic outflow. Note that this scenario leads to strong constraints on the density and magnetization of the external medium \citep{piran:2010} and that the observed temporal slope of the long-lasting high-energy emission would imply a strongly pair-enriched medium as discussed in \refpar{sec_belo} \citep{ghisellini:2010}.\\

 Even in scenarios where the prompt GeV emission has an internal origin, a partially distinct emission region could be due to  a spectral evolution of the prompt mechanism. For instance, in the framework of internal shocks, the evolution of the physical conditions in the shocked region during the propagation of a shock wave leads to an evolving efficiency of the IC scatterings, depending on the importance of Klein-Nishina corrections. This naturally leads to a variable high-energy component following with a delay the main (Band) component in the MeV range \citep{wang:2009,bosnjak:2009,daigne:2011}. Successive generations of collisions in a variable outflow can also lead naturally to different emission regions \citep{li:2010}.
  An evolution in the microphysics of the acceleration process could also be responsible for some spectral evolution in scenarios where there is a dominant hadronic component at high energy \citep[see e.g.][]{asano:2009}. Finally, two emission regions are naturally expected in photospheric models, as it is often assumed that the main (Band) component has a photospheric origin and that internal shocks or magnetic dissipation occurring at larger distance produce an additional component at high energy \citep[see e.g.][]{toma:2010,vurm:2011}.

\subsection{Loosening the constraint on $\Gamma_\mathrm{min}$}
\label{sec_gmin_2zones}
As discussed in \citet{zhao:2011,zou:2011,racusin:2011}, the possibility for the GeV photons to be produced in a different region than the MeV photons can loosen the constraint on the minimum Lorentz factor in GRB outflows. To investigate this effect, 
we consider the same synthetic GRB as used in \refpar{subsection_lfmin} to model time bins 'a' and 'b' of GRB 080916C. We focus on the onset of the GeV component, which occurs at $t_\mathrm{obs,onset}=t_\mathrm{obs,trig}+0.67\left(1+z\right)\,\mathrm{s}$, where $t_\mathrm{obs,trig}$ corresponds to the observer time of the first MeV photons. The MeV photons observed at $t_\mathrm{obs,onset}$ are emitted at radius $R_\mathrm{MeV}$ and it is assumed that the emerging GeV photons observed at the same time were emitted 
 by material moving with Lorentz factor $\Gamma_\mathrm{GeV}$ (velocity $\beta_\mathrm{GeV} c$)
at radius $R_\mathrm{GeV}$ and time $t_\mathrm{GeV}$ (source frame) with $t_\mathrm{GeV}-R_\mathrm{GeV}/c =  t_\mathrm{obs,onset}/(1+z)$. The flash of GeV photons emitted at $R_\mathrm{GeV}$ is assumed to have a power-law spectrum with photon slope $\beta=-2.2$. We define a latitude-averaged $\gamma\gamma$ opacity for GeV photons of energy $E_\mathrm{GeV}$ by
\begin{equation}
e^{-\overline{\tau}_{\gamma\gamma}\left(E_\mathrm{GeV}\right)} = \frac{ 
\int e^{- \tau_{\gamma\gamma}(E_\mathrm{GeV},\Theta_\mathrm{e})} \mathcal{D}(\Theta_\mathrm{e})^{1-\beta}  \sin{\Theta_\mathrm{e}} d \Theta_\mathrm{e}
}{ 
\int \mathcal{D}(\Theta_\mathrm{e})^{1-\beta}  \sin{\Theta_\mathrm{e}} d \Theta_\mathrm{e}
}\, ,
\label{eqn_mean_tgg}
\end{equation}
where $\tau_{\gamma\gamma}(E_\mathrm{GeV},\Theta_\mathrm{e})$ is the opacity seen  
photons emitted at colatitude $\Theta_\mathrm{e}$ and $\mathcal{D}\left(\Theta_\mathrm{e}\right)=\left( \Gamma_\mathrm{GeV}\left(1-\beta_\mathrm{GeV}\cos{\Theta_\mathrm{e}}\right)\right)^{-1}$ is the corresponding Doppler factor. The contribution of each colatitude to the mean value is weighted by the corresponding fluence, leading to the $1-\beta$ exponent.\\

\begin{figure}
\begin{center}
\begin{tabular}{c}
\includegraphics[scale=0.4]{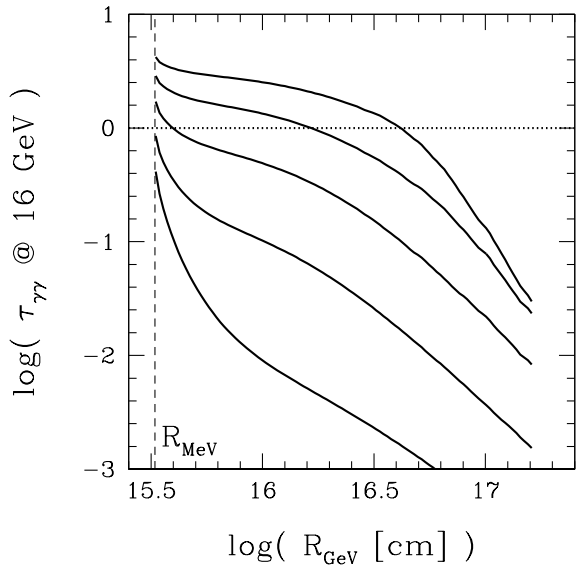} 
\end{tabular}
\end{center}

\caption{\textbf{Two emitting region scenario.} The 
opacity $\tau_{\gamma \gamma}$ seen by 16 GeV  photons (source frame) observed at $t_\mathrm{obs,onset}-t_\mathrm{obs,trig}\simeq 0.67~(1+z)$ s (see text) is plotted as a function of their emission radius $R_\mathrm{GeV}$ for different values of the Lorentz factor  $\Gamma_\mathrm{GeV}=50$, $100$, $200$, $400$ and $800$ from top to bottom. }
\label{fig_2zones}
\end{figure}

We plot in \reffig{fig_2zones} the evolution of the latitude averaged $\gamma\gamma$ opacity $\overline{\tau}_{\gamma\gamma}$ at 16 GeV (source frame) as a function of $R_\mathrm{GeV}$ for $R_\mathrm{GeV}>R_\mathrm{MeV}$ and for different values of the Lorentz factor $\Gamma_\mathrm{GeV}$. When GeV and MeV photons are emitted at the same location, we find that $\overline{\tau}_{\gamma\gamma}\la1$ for $\Gamma_\mathrm{GeV} \ga \Gamma_\mathrm{GeV,min,same\,zone}\simeq 340$, i.e. the same limite as in \refpar{subsection_lfmin}. 
When $R_\mathrm{GeV}$ increases, the opacity $\overline{\tau}_{\gamma\gamma}$ decreases as expected, which loosen the constraint on the minimum Lorentz factor $\Gamma_\mathrm{min,GeV}$ of the material emitting GeV photons :
\begin{center}
\begin{tabular}{l||cccc}
$R_\mathrm{GeV}/R_\mathrm{MeV}$ & 1 & 1.2 & 5.1 & 13\\
\hline
$\Gamma_\mathrm{min,GeV}/\Gamma_\mathrm{min,GeV,same\,zone}$ & 1 & 0.59 & 0.29 & 0.15
\end{tabular}
\end{center}
This follows approximatively the dependency on $R_\mathrm{e}/R_0$ found in \refpar{sec_approx_formula}, i.e. 
 $\tau_{\gamma\gamma}\propto \left(R_\mathrm{e}/R_0\right)^{2(\beta-1)}$ when $R_\mathrm{e}\gg R_0$, leading to $\Gamma_\mathrm{min,GeV}\propto \left(R_\mathrm{e}/R_0\right)^{-1}$.
As shown in \refpar{subsection_lfmin} the detailed modeling of the $\gamma\gamma$ opacity in a scenario where GeV and MeV photons are emitted in the same regions leads to a reduction of the minimum Lorentz factor by a factor $\simeq 2$--$3$ compared to single zone models. The calculation presented here shows in addition that the minimum Lorentz factor can be reduced further more by another factor $\simeq2-8$ for $\Gamma_\mathrm{GeV}$ if GeV emission becomes efficient at a radius larger than for MeV photons. Assuming that the radiated energy at $R_\mathrm{GeV}$ is not larger than the radiated energy at $R_\mathrm{MeV}$, we have checked that the outflow remains optically thin for the Thomson opacity due to primary electrons and secondary leptons at $R_\mathrm{GeV}$ in the case shown in \reffig{fig_2zones}.\\

 Note that 
  the loosening of the constraint on the minimum Lorentz factor does not apply to models where GeV photons are entirely due to the external shock. Indeed, the small value of $t_\mathrm{obs,onset}$ implies an early deceleration. As the isotropic equivalent energy of GRB 080916C is huge, this leads to a minimum Lorentz factor $\overline{\Gamma} > 10^3$ in the outflow, which is more constraining that the $\gamma\gamma$ opacity limit.\\ 

The discussion of the effect of a distinct GeV emission region presented here is quite simplified and some limitations should be kept in mind. If an additional GeV component could be firmly identified in GRB 080916C, the maximum energy $E_\mathrm{MeV,max}$ of photons associated with the main component should be taken into account to derive a new constraint $\Gamma_\mathrm{min}$ on the Lorentz factor of the outflow during the MeV emission phase. We have assumed here $\overline{\Gamma}=340$ as derived in \refpar{subsection_lfmin} using $E_\mathrm{MeV,max}=16$ GeV (source frame) but $\Gamma_\mathrm{min}$ will be reduced by a factor $\left(E_\mathrm{MeV,max}/16\, \mathrm{GeV}\right)^{\frac{\beta+1}{2(\beta-1)}}$ if $E_\mathrm{MeV,max}$ is lower. There is one further complication: the component produced at $R_\mathrm{GeV}$ extends probably in the soft gamma-ray range, as suggested by the observation of a soft excess correlated with the high energy component in some GRBs such as  GRB 090926 \citep{ackermann:2011}, GRB 090926 \citep{abdo:2009b} and GRB 090510 \citep{ackermann:2010}. It has been assumed here that the annihilation rate of GeV photons with the seed photons produced at $R_\mathrm{GeV}$ is negligible compared to the annihilation
rate with MeV photons produced earlier. This is however not necessarily the case depending on the relative intensity of the two components. 
Clearly, a detailed modeling of the emitted spectrum is necessary to investigate such effects. This is  
 beyond the scope of this paper and we leave to a forthcoming study the coupling of the formalism presented here to compute the $\gamma\gamma$ opacity with a detailed radiative model such as developed by  \citet{bosnjak:2009}.

\section{Conclusions}
\label{sec_conclusions}
Photon--photon annihilation is an important process in GRBs. It has been known for a long time that it provides one of the main constraints on GRB models, i.e. the minimum Lorentz factor in the jet. In most GRBs, this leads to $\Gamma_\mathrm{min}\simeq 10^{2}$ \citep{lithwick:2001}. The detection of GeV photons in a few GRBs by \textit{Fermi}-LAT has led to much stronger limits, of the order of $\Gamma_\mathrm{min}\simeq 10^3$ \citep{abdo:2009a}. However these estimates are based on highly simplified single zone formulae, which assume an isotropic radiation field in the coming frame of the outflow. 
We present here a detailed calculation of the $\gamma\gamma$ opacity in GRB outflows taking into account the exact time-dependent anisotropic photon field produced by multiple relativistically geometrically thin emitting regions. This calculation can be implemented numerically, for any model prescription of the prompt emission mechanism, as long as the emitting regions can be considered as expanding thin spherical shells. We present results obtained in the framework of the internal shock model, where the emitting regions are shock waves propagating within the outflow. Our calculation has been validated by a comparison with the previous semi-analytical study made by \citet{granot:2008}. We have also estimated the opacity due to Thomson scatterings by ambient electrons in the outflow, to pairs created by $\gamma\gamma$ annihilation and to $\gamma\gamma$ annihilation with prompt photons back-scattered by the external medium. For all the examples presented in the paper, the $\gamma\gamma$ opacity is the dominant term (see \refpar{sec_opacity_others}). We have obtained the following results:
\begin{enumerate}
\item \textit{Spectral shape of the $\gamma\gamma$ cutoff:} as described by \citet{granot:2008}, due to a time evolution of the $\gamma\gamma$ opacity, the spectral shape of the $\gamma\gamma$ cutoff in time-integrated spectra is not expected to be exponential, but rather power-law like. 
We illustrate this effect in a synthetic burst in \refpar{subsection_scattering}. This spectral shape makes more difficult the identification of the cutoff in \textit{Fermi}-LAT spectra. The main candidate is GRB 090926A \citep{ackermann:2011}, where the number of photons detected above the cutoff at $\simeq 1.4$ GeV is too low to allow for a precise characterization of the spectral shape. Nevertheless, in most \textit{Fermi}-LAT GRBs, there is no evidence for an observed cutoff, which puts interesting constraints on the outflow.\\

\item \textit{Minimum Lorentz factor in GRB outflows:} when taking into account the exact photon field, the $\gamma\gamma$ opacity for GeV photons is reduced compared to single zone isotropic estimates, leading to minimum Lorentz factors which are a factor of $\simeq2-3$ lower. As an illustration, we present in \refpar{subsection_lfmin} a synthetic burst that reproduces well the observed features of time bins 'a' and 'b' of GRB 080916C with a mean Lorentz factor of $\simeq 340$. This minimum Lorentz factor is obtained assuming that there is only one component in the emitted spectrum, responsible both for the MeV and the GeV photons. This is not necessarily the case as several \textit{Fermi}-LAT burst show evidence of an additional high-energy component. We show in \refpar{sec_gmin_2zones} that the minimum Lorentz factor can be reduced by another factor $2$ to $8$ if the GeV photons are emitted at larger radius than the main component. This study clearly illustrates the need for a detailed modeling to constrain the Lorentz factor in GRB outflows. When it is not possible, a reasonably accurate estimate of $\Gamma_\mathrm{min}$ can be obtained from the following formula
\begin{eqnarray}
\Gamma_\mathrm{min} & \simeq &  \frac{\left[C_1 2^{1+2\beta} \mathcal{I}(\beta)\right]^{\frac{1}{2(1-\beta)}}}{\left[\frac{1}{2}\left(1+\frac{R_\mathrm{GeV}}{R_\mathrm{MeV}}\right)\left(\frac{R_\mathrm{GeV}}{R_\mathrm{MeV}}\right)\right]^{1/2}}\, \left(1+z\right)^{-\frac{1+\beta}{1-\beta}}
\nonumber\\ 
& &\!\!\!\! \!\!\!\!\!\!\!\!\!\!\!\!\!\!\! \times 
\left[ \sigma_\mathrm{T} \left(\frac{D_\mathrm{L}(z)}{c \Delta t_\mathrm{var}}\right)^2 \!\!\! E_\mathrm{c} F(E_\mathrm{c})\right]^{\frac{1}{2(1-\beta)}} \left(\frac{E_\mathrm{max}E_\mathrm{c}}{(m_\mathrm{e}c^2)^2}\right)^{\frac{\beta+1}{2(\beta-1)}}
\!\!\!\!
\, ,\nonumber\\ 
\label{eq_newapprox}
\end{eqnarray}
where $C_1 \simeq 4\cdot 10^{-2}$, 
$z$ and $D_\mathrm{L}(z)$ are the redshift and the luminosity distance of the source, $\Delta t_\mathrm{var}$ is the observed variability timescale,
$R_\mathrm{GeV}/R_\mathrm{MeV}$ is the ratio of the radii where the GeV and MeV components are emitted, 
and where 
the high energy spectrum (over a duration $\sim \Delta t_\mathrm{var}$) is assumed to follow a power-law with photon index $\beta$ above an observed characteristic energy $E_\mathrm{c}$ : $\overline{F}(E)=\overline{F}(E_\mathrm{c})(E/E_\mathrm{c})^\beta$ ($\mathrm{ph.cm^{-2}.keV^{-1}}$). Energy $E_\mathrm{max}$ is the observed energy of the most  energetic detected photons. 
The integral $\mathcal{I}(\beta)$ is defined by \refeq{eqn_i_beta} in \refpar{sec:flash}.
As usually the spectrum is measured over a time interval $\Delta t_\mathrm{spec}$ which is larger than the variability timescale $\Delta t_\mathrm{var}$, the normalization $F(E_\mathrm{c})$ entering in \refeq{eq_newapprox} (fluence at energy $E_\mathrm{c}$ in $\mathrm{ph.cm^{-2}.keV^{-1}}$) must be corrected by a factor $F(E_\mathrm{c})=\overline{F}(E_\mathrm{c})\times \left(\Delta t_\mathrm{var}/\Delta t_\mathrm{spec}\right)$.
This equation can be directly applied to \textit{Fermi}-LAT observations and generalizes the usual formula given by \citet{abdo:2009a} by introducing two corrections: (1) a more accurate normalization taking into account the anisotropy of the radiation field and including a numerical factor $C_1$ obtained from the comparison with numerical simulations in \refpar{subsection_lfmin}; (2) the possibility to take into account two different emitting regions for MeV and GeV photons. The standard limit is obtained with $R_\mathrm{GeV}/R_\mathrm{MeV}=1$ (same region) : then the denominator in \refeq{eq_newapprox} equals $1$. The radius $R_\mathrm{MeV}$ is estimated from the variability timescale by $R_\mathrm{MeV}\simeq \Gamma^2 c\Delta t_\mathrm{var}/(1+z)$, which is valid for most models of the prompt emission. The radius $R_\mathrm{GeV}$ is difficult to constrain without a detailed model of the high-energy emission mechanism. 
If GeV photons have an internal origin, an upper limit for $R_\mathrm{GeV}$ is given by the deceleration radius. In the future, a measurement of the variability timescale in the GeV lightcurve could provide a better estimate of this radius.\\ 
 
\item \textit{Delayed onset of the GeV emission:} due to the variable nature of the GRB phenomenon, the $\gamma\gamma$ opacity is expected to be strongly time dependent. This can lead naturally to a delayed onset of the high-energy component, if 
it is initially highly absorbed. For instance, in internal shocks, the emission radius increases during the propagation of a given shock wave, which favors such an evolution. We show several examples of synthetic bursts where a delayed onset of the GeV component is observed. In the case of internal shocks, the delay before the onset is comparable with the observed duration associated with the propagation of the shock wave that produces the seed photons for $\gamma\gamma$ annihilation. This duration has to be distinguished from the shortest timescale of variability. In the synthetic burst which models 
time bins 'a' and 'b' of GRB 080916C, we obtain a GeV onset delayed by $\simeq 5\, \mathrm{s}$, comparable with the observed value, while reproducing variability on shorter timescales ($\simeq 0.5$ s) in the MeV lightcurve. A time evolving $\gamma\gamma$ opacity appears as a good candidate to explain the observed delayed onset of the GeV emission. Note that this effect is obtained here assuming a single component in the emitted spectrum. More complex spectral evolution, such as a varying inverse Compton component \citep{bosnjak:2009}, may be an additional source of delay.\\

\item \textit{Smoothing the variability in the GeV lightcurve:} in highly variable outflows, the short timescale variability in the GRB lightcurve is expected to be associated with emission regions at low radius. Therefore, the associated high-energy emission may be highly absorbed by $\gamma\gamma$ annihilation due to a denser photon field. This predicts a smoothing of the lightcurve in the GeV  
range. We illustrate this effect in the framework of the internal shock model in \refpar{sec_smoothing}. For this reason, the variability of the GeV lightcurve is an indicator to use carefully if one wants to distinguish between an internal and an external origin for the GeV emission.
\end{enumerate}
Our results loosen the constraints on models of  GRB central engines as Lorentz factor of $10^3$ do not seem to be required, even in the most extreme bursts observed by \textit{Fermi}-LAT. In addition, they solve another contradiction. For Lorentz factors of $10^3$, the deceleration by the external medium occurs very early. Then, it becomes difficult to interpret \textit{Fermi}-LAT GRBs such as GRB 080916C if the prompt emission is radiated well above the photosphere (see also \citealt{zhang:2009}). On the other hand, the early steep decay observed in X-ray afterglows by Swift-XRT is a strong evidence that the prompt emission phase ends at a large radius \citep{lyutikov:2006,lazzati:2006,kumar:2007,genet:2009}. 
 With lower limits on the Lorentz factor as obtained in the present paper, bright \textit{Fermi}-LAT GRBs such as GRB 080916C are consistent with models, including internal shocks, where the prompt emission is produced between the photospheric and the deceleration radii.


\appendix
\section{Photon field created by a spherical flash}
\label{sec_calc}

Let us consider the interaction between a spherical flash and a high energy photon (the geometry  is illustrated in \reffig{fig_flash_photon}, see also the caption of the figure for the adopted notations). The emissivity in the source frame is given :
\begin{equation}
j_{\nu} = \mathcal{D}^2 j'_{\nu'}\, ,
\label{eqn_emissivity_chframe}
\end{equation}
where the Doppler factor $\mathcal{D}$ is defined by
\begin{equation}
\mathcal{D} = \frac{\nu}{\nu'}=\frac{1}{\Gamma\left(1-\beta \cos{\delta}\right)}\, .
\end{equation}
 The specific intensity $I_\nu$ $[\mathrm{erg}\cdot \mathrm{cm}^{-2} \cdot \mathrm{s}^{-1} \cdot \mathrm{Hz}^{-1} \cdot \mathrm{sr}^{-1}]$ created by the spherical flash at the interaction point I (radius $R_\mathrm{I}$ and time $t_\mathrm{I}$) in direction $\alpha$ is given by
\begin{equation}
I_{\nu}\left(R_\mathrm{I},t_\mathrm{I},\alpha\right) = \int  j_\nu\left(\tilde{s},t_\mathrm{I}-\frac{\tilde{s}}{c}\right)\, d\tilde{s}\, ,
\label{eqn_specific_intensity_gen}
\end{equation}
assuming negligible absorption at frequency $\nu$. Indeed,
high energy photons interact preferentially with seed low energy photons close to the threshold with $\psi \simeq 1/\Gamma$, i.e. 
\begin{equation}
\frac{E}{10\, \mathrm{MeV}}\frac{E_\mathrm{HE}}{1\, \mathrm{GeV}}\simeq 1.04 \left(\frac{\Gamma}{100}\right)^2\, .
\end{equation}
As the number of emitted photons at energy $E$, $N(E)$, decreases as a power-law with energy (typically $N(E)\propto E^{\beta}$ with $\beta\simeq -2$ to $-3$ above 1 MeV),  the ratio $N(E_\mathrm{HE})/N(E)$ is always small and the fraction of absorbed seed photons negligible. The radius $r$ is related to $\tilde{s}$ and $\alpha$ by 
\begin{equation}
r^2 = R_\mathrm{I}^2+\tilde{s}^2-2 R_\mathrm{I}\tilde{s}\cos{\alpha}\, ,
\end{equation}
which leads to
\begin{equation}
\int \delta\left(r-R_0\right) d\tilde{s} = \frac{R_0}{s-R_\mathrm{I} \cos{\alpha}}\, . 
\label{eqn_dirac_radius}
\end{equation}
The same relation applied at $r=R_0$ ($\tilde{s}=s$) leads to
\begin{equation}
\delta\left(t_\mathrm{I}-\frac{s}{c}-t_0\right) = c\,\, \frac{s-R_\mathrm{I}\cos{\alpha}}{s R_\mathrm{I}\sin{\alpha}}\, \delta\left(\alpha-\alpha_\mathrm{I}\right)\, .
\label{eqn_dirac_time}
\end{equation}
Therefore, from \refeq{eqn_emissivity_comov} and \refeq{eqn_specific_intensity_gen}, we obtain
\begin{eqnarray}
I_{\nu}\left(R_\mathrm{I},t_\mathrm{I},\alpha\right) & = &  \frac{1}{4\pi} \frac{\mathcal{E}_\mathrm{rad}}{4\pi R_0^2 \Gamma_0}\, \frac{\mathcal{D}^2}{\nu'_\mathrm{p,0}} \mathcal{B}\left(\frac{\nu}{\mathcal{D}\nu'_\mathrm{p,0}}\right)
\nonumber\\ 
& & \times 
\frac{c R_0}{s R_\mathrm{I}\sin{\alpha}}\, \delta\left(\alpha-\alpha_\mathrm{I}\right)\, .
\label{eqn_specific_intensity_spec}
\end{eqnarray}
As expected from the flash geometry, the intensity is positive for a unique direction $\alpha=\alpha_\mathrm{I}$.
Then, the photon density $[\mathrm{ph}\cdot \mathrm{cm}^{-3}  \cdot \mathrm{keV}^{-1} \cdot \mathrm{sr}^{-1}]$ 
is given by 
\begin{equation}
n_{\Omega}(E) = \frac{1}{h c} \frac{I_{\nu}}{h \nu }\, .
\label{eqn_photon_density}
\end{equation}

\section{Approximate formula for the $\gamma\gamma$ opacity}
\label{sec_calc_approx}

In this appendix, we derive the approximate formula given by \refeq{equation_tgg_flash_approximate} for the $\gamma\gamma$ opacity seen by a on-axis high energy photon emitted at $R_\mathrm{e}$ interacting with a spherical flash radiated at $R_0$ and having a power-law spectrum. Using $R_\mathrm{e}=R_0+\beta_0 c \left(t_\mathrm{e}-t_0\right)$ and introducing normalized coordinates $X = \left(R_\mathrm{e}-R_0\right)/R_0$ et $x=\ell/R_0$, the opacity given by \refeq{eqn_taugamma_flash_pl_onaxis} reads
\begin{eqnarray}
\tau_{\gamma\gamma}(E_\mathrm{HE}) & = & \mathcal{I}(\beta) \tau_0 A_0 \left(\frac{2\left(m_\mathrm{e}c^2\right)^2}{E_\mathrm{HE}E_\mathrm{p,0}}\right)^{1+\beta} \Gamma_0^{1+\beta}
\nonumber\\ 
& & \!\!\!\!\!\!\!\!\!\!\!\!\!\!\!\!\!\!\! \times 
\int_0^{+\infty}\frac{dx}{X}\,\frac{1}{1+X+x}\,\frac{X}{\frac{X}{\beta_0}+x}\,\mathcal{D}^{1-\beta}\,\left(1-\cos{\psi}\right)^{-\beta}
\nonumber\\ 
\end{eqnarray}
Several contributions appear in the integral :
\begin{itemize}
\item \textit{Dilution term :} the first factor, $\left(1+X+x\right)^{-1}$, expresses the dilution of the radiation from the flash since its emission ;
\item \textit{Propagation term :} the second factor, $\left(X/\beta_0+x\right)^{-1}$, takes into account the propagation of the high energy photon up  to location of the interaction ;
\item \textit{Doppler term :} the third factor, $\mathcal{D}^{1-\beta}$, is due to the Doppler effect applied to the seed photons. 
\item \textit{Interaction angle term :} the fourth term, $\left(1-\cos{\psi}\right)^{-\beta}$, expresses the effect of the interaction angle $\psi$ on the effective cross section. 
\end{itemize}
The product of these four terms is a strictly decreasing function, so that the integral is dominated by the contributions at low $x$. For $x\ll X$ and $\Gamma_0\gg 1$, we have 
$\left(1+X+x\right)^{-1} \simeq \left(1+X\right)^{-1}$ (dilution), $X/\left(X/\beta_0+x\right)\simeq \beta_0\simeq 1$ (propagation), 
$\mathcal{D}^{1-\beta} \simeq \left({2\Gamma_0}/{(2+X)}\right)^{1-\beta}$ (Doppler)
and $\left(1-\cos{\psi}\right)^{-\beta} \simeq  \left( 2\Gamma_0^2(1+X)\right)^\beta$ (interaction angle),
where we have used the Taylor expansion of the Doppler angle $\delta$ and the interaction angle $\psi$ :
\begin{eqnarray}
\delta & \simeq & \frac{\sqrt{1+X}}{\Gamma_0}\, ,\nonumber\\
\psi  & \simeq & \frac{1}{\sqrt{1+X}\Gamma_0}\, .\nonumber
\end{eqnarray}
The product of the four terms is therefore approximatively constant for $x\ll X$, before starting a steep decay.
Keeping only the contribution for $0\le x\la X$ to the integral gives the following approximate formula:
\begin{eqnarray}
\tau_{\gamma\gamma}(E_\mathrm{HE}) & \simeq & \mathcal{I}(\beta) \tau_0 A_0 \left(\frac{2\left(m_\mathrm{e}c^2\right)^2}{E_\mathrm{HE}E_\mathrm{p,0}}\right)^{1+\beta} \Gamma_0^{1+\beta}
\nonumber\\ 
& & \!\!\!\!\!\times 
\int_0^{X}\frac{dx}{X}
\frac{1}{1+X}\left(\frac{2\Gamma_0}{2+X}\right)^{1-\beta}\left( 2\Gamma_0^2(1+X)\right)^\beta\nonumber\\
& \simeq & 2^{1+2\beta} \mathcal{I}(\beta) \tau_0 A_0 \left(\frac{\left(m_\mathrm{e}c^2\right)^2}{E_\mathrm{HE}E_\mathrm{p,0}}\right)^{1+\beta} \Gamma_0^{2(1+\beta)}
\nonumber\\ 
& & \times 
\frac{1}{\left[\left( 1+X\right)\left(1+\frac{X}{2}\right)\right]^{1-\beta}}
\nonumber\\ 
\end{eqnarray}
which leads to \refeq{equation_tgg_flash_approximate}. This approximate formula is compared to the exact calculation in the top-left panel of \reffig{fig_case_flash}. The behaviour of the opacity with the emission radius of the high-energy photon can be simply understood from the Taylor expansions above : the integral is dominated by the product of the dilution, Doppler et interaction angle terms. When the high energy photon is emitted just after the flash ($R_\mathrm{e}\to R_0$, i.e. $X\to 0$), the dilution term is negligible, the Doppler term is optimal as $\delta \simeq 1/\Gamma_0$ and the interaction angle term is also optimal as $\psi \simeq 1/\Gamma_0$. Therefore, $\gamma\gamma$ annihilation is efficient. At larger emission radii ($R_\mathrm{e}\gg R_0$, i.e. $X\to +\infty$), the dilution factor ($\propto X^{-1}$) reduces the seed radiation field, the Doppler term ($\propto X^{\beta-1}$) also reduces strongly the seed radiation field as $\delta$ becomes large compared to $1/\Gamma_0$, and the interaction angle term ($\propto X^{\beta}$) strongly reduces the probablity of interaction as $\psi\to 0$. This explains the steep decay of $\tau_{\gamma\gamma}\propto X^{2(\beta-1)}$ for $X\ga 1$ .

\section*{Acknowledgments}
The authors thank J. Granot for useful discussions on $\gamma\gamma$ opacity in GRBs, and F. Piron, V. Pelassa and S. Guiriec for stimulating discussions on \textit{Fermi} observations.
This work is partially supported by the French Space Agency (CNES).
R.H.'s PhD work is funded by a Fondation CFM-JP Aguilar grant. 
\bibliographystyle{mn2e}   
\bibliography{opacity}

\label{lastpage}

\end{document}